\documentclass[secnumarabic,amssymb,amsmath,natbib,graphicx,aps,twocolumn]{revtex4-2}
\usepackage{lipsum}
\usepackage{subcaption}
\usepackage{tikz}       
\usepackage{hyperref,comment}
\usepackage{booktabs}
\usepackage{float}
\usepackage{dcolumn}

\def\bea{\begin{eqnarray}}
\def\eea{\end{eqnarray}}

\begin{document}

\title{Circular orbits and observational features of the rotating Simpson–Visser black hole surrounded by a thin accretion disk}
\author{Ziyang Li}
\author{Shou-Qi Liu}
\author{Jia-Hui Huang}
\email{huangjh@m.scnu.edu.cn}
\affiliation{Key Laboratory of Atomic and Subatomic Structure and Quantum Control (Ministry of Education), Guangdong Basic Research Center of Excellence for Structure and Fundamental Interactions of Matter, School of Physics, South China Normal University, Guangzhou 510006, China}
\affiliation{Guangdong Provincial Key Laboratory of Quantum Engineering and Quantum Materials, Guangdong-Hong Kong Joint Laboratory of Quantum Matter, South China Normal University, Guangzhou 510006, China}

\date{\today}

	\begin{abstract}
	Since the Event Horizon Telescope (EHT) collaboration released horizon-scale images of the supermassive black holes Sgr~A* and M87*, a new observational window for probing black hole spacetimes in the strong-gravity regime has opened.
	As an important class of Kerr black hole mimickers, rotating Simpson--Visser (SV) black holes exhibit a degeneracy with Kerr black holes at the level of shadow size, making it difficult to distinguish them using shadow observations alone.
	Motivated by this issue, we present a systematic investigation of the radiative properties and optical appearance of rotating SV black holes surrounded by a thin accretion disks, and mainly analyze the influence of the regularization parameter $g$ on related observables.   
	The results show that although the kinetic quantities and the location of the innermost stable circular orbit (ISCO) depend on the regularization parameter $g$, 
the radiative efficiency of the rotating SV black hole is the same as its Kerr counterpart.
	Within the Novikov--Thorne thin-disk model, the radiative flux, effective temperature, and spectral luminosity are studied, and by adopting observational parameters relevant to Sgr~A* and M87*, concrete examples of the rotating SV black holes are calculated and compared with that of the Kerr black holes. The results show that the parameter $g$
suppresses the maximum values of these quantities. 
    In addition, using a backward ray-tracing technique, we numerically simulate the optical appearance of rotating SV black holes and analyze the corresponding intensity images,
redshift and observed flux distributions. Our results show that these quantities are affected by $g$. In particular, as $g$ increases, the observed intensity is significantly suppressed and the photon ring region has remarkable increase in its width.
	Our findings suggest that accretion-disk-related observables may provide important avenues to distinguish rotating SV black holes and Kerr black holes, and offer theoretical guidance for future high-resolution observations.
	\end{abstract}

\maketitle	

	\section{Introduction}
	The recent detection of gravitational waves by the LIGO/Virgo collaboration \cite{Abbott:2016blz} and the groundbreaking horizon-scale imaging of supermassive black holes M87* \cite{EventHorizonTelescope:2019dse} and Sgr A* \cite{EventHorizonTelescope:2022wkp} have confirmed the existence of black holes and verified the validity of general relativity (GR) in the strong-gravity regime. Nevertheless, GR is plagued by the prediction of singularities, as demonstrated by the celebrated Penrose-Hawking singularity theorems \cite{Penrose:1965,Hawking:1970}. At the singularity, the spacetime curvature diverges, leading to the breakdown of physical laws and the loss of predictability \cite{Marolf:2017jkr}. To resolve this fundamental problem, the concept of regular black holes was proposed \cite{Bardeen1968, Hayward:2006part,Bambi:2023try,Lan:2023cvz}. These solutions avoid the central singularity by replacing it with a regular core, which ensures the spacetime to be geodesically complete.
	
	Historically, the regular black hole solutions, such as the Bardeen \cite{Bardeen1968} and Hayward \cite{Hayward:2006part} metrics, were often constructed phenomenologically and interpreted as arising from the coupling of general relativity to nonlinear electrodynamics. While these models successfully removed the singularity, questions regarding their dynamical origins persist.
	Recently, a significant theoretical progress has been made \cite{Zhang:2025ccx}, which demonstrates that static, spherically symmetric regular geometries can naturally emerge as vacuum solutions to a broad class of generally covariant gravity theories. The work also suggests a powerful inverse problem approach: if high-precision observations could determine the spacetime metric, one can, in principle, reconstruct the underlying gravitational theory. Although the work has not yet been fully generalized to the rotating case, it profoundly enhances the motivation for investigating the observational signatures of regular black holes. Studying their images is thus not merely a test of a specific metric, but a crucial step toward revealing or constraining possible modifications to GR.
	
	Among various nonsingular spacetimes, Simpson and Visser \cite{Simpson:2018tsi} proposed a novel class of black bounce geometries. This model introduces a regularization parameter $g$, which effectively smears out the central singularity by replacing it with a minimal surface, or a throat. The spacetime structure smoothly interpolates between a regular black hole (when $g$ is small) and a traversable wormhole (when $g$ is large). It is also demonstrated that such geometries can be sourced dynamically by a combination of a minimally coupled phantom scalar field and a nonlinear electromagnetic field \cite{Bronnikov:2021uta}. From an astrophysical perspective, since realistic compact objects are expected to possess angular momentum \cite{Kerr:1963ud}, the static SV solution was subsequently generalized to the rotating case \cite{Mazza:2021rgq}. This rotating SV metric serves as an excellent Kerr mimicker, which preserves the Kerr-like exterior while modifying the interior structure to resolve the singularity, and provides an ideal laboratory for testing strong-field gravity.

The static and rotating SV metrics have been studied from various aspects in the literature, including thermodynamics \cite{Nosirov:2023ism,Ahmed:2026bwm,Joshi:2025ozt,Kumar:2025nio}, quasinormal modes \cite{Churilova:2019cyt,Khoo:2025qjc,Franzin:2022iai,Jha:2023wzo}, shadows \cite{KumarWalia:2022aop,Lima:2021las,Shaikh:2021yux,Shaikh:2022ivr,Chaudhary:2025ssf,Jha:2023wzo,Jafarzade:2021umv},  gravitational lensing \cite{Tsukamoto:2020bjm,Islam:2021ful,Nascimento:2020ime,He:2024ozb}, imagings \cite{Bambhaniya:2021ugr,Chen:2024tss,Combi:2024ehi}, and others \cite{Dasgupta:2025qwq,Viththani:2024map,Patel:2022jbk,Jiang:2021ajk,Lobo:2020kxn,Bargueno:2020ais,Jumaniyozov:2025irx,Calza:2025yfm,Yang:2025uaa,Yang:2024mro,Hadi:2024lxg,Vrba:2023byq,Arora:2023ltv,Rodrigues:2022rfj,LimaJunior:2022zvu,Vagnozzi:2022moj}. Despite these theoretical advances, a critical challenge arises when attempting to identify these metrics via shadow imaging. Recent investigations \cite{KumarWalia:2022aop,Shaikh:2021yux} have revealed an intrinsic degeneracy between rotating SV black holes and the Kerr metric. Specifically, for the black hole branch of the SV solution (where the regularization parameter $g$ is sufficiently small), the central bounce throat is hidden deep within the photon sphere. Since the shadow boundary is determined by the photon capture region, which remains unaffected by the internal modification of the spacetime, the resulting shadow size and shape are theoretically identical to those of a Kerr black hole with the same mass and spin. This exact degeneracy implies that shadow imaging alone is blind to the regularization parameter $g$ in this regime. Consequently, to break this degeneracy and probe the nonsingular nature of the core, it is necessary to explore other observational signatures \cite{Bambi:2025wjx} that are more sensitive to the near-horizon geometry, such as the radiative and observational properties of an accretion disk around the rotating SV black hole.

In this paper, we perform a comprehensive study of the radiative properties and optical appearance of rotating SV black holes surrounded by a thin accretion disk and focus on the influence of the regularization parameter $g$ on the related physical quantities.
This paper is organized as follows.
In Sec.~\ref{sec:metric}, we introduce the rotating SV black hole metric and analyze the dependence of the event horizon radius on the spin parameter and the regularization parameter. We then study the circular motion of massive test particles on the equatorial plane and present the specific energy, angular momentum, and angular velocity outside the ISCO for various parameters. After determining the radius of the ISCO numerically, the radiative efficiency of the rotating SV black hole is subsequently evaluated.
In Sec.~\ref{sec:disk}, we numerically investigate several radiative properties of the thin accretion disk around rotating SV black holes. In particular, we analyze the radiative flux, effective temperature, and spectral luminosity, and discuss the observational differences between rotating SV black holes and Kerr black holes.
In Sec.~\ref{sec:raytracing}, we adopt ray-tracing method to simulate the optical appearance of rotating SV black holes. The intensity profiles on the observer’s screen are compared with those of the Kerr black hole. In addition, the distributions of the redshift factor and the observed flux are presented and analyzed for different model parameters and observational inclination angles.
 Sec.~\ref{sec:conclusion} is devoted to the conclusion.

	\section{The Rotating Simpson–Visser Metric and Equatorial Circular Orbits}
	\label{sec:metric}
The spherically symmetric SV black hole was originally proposed as a regular black hole mimicker obtained by introducing a regularization parameter that removes the central curvature singularity while preserving asymptotic flatness \cite{Simpson:2018tsi,Bronnikov:2021uta}. 
Using a modified Newman--Janis algorithm, the static SV spacetime can be generalized to a stationary and axisymmetric rotating geometry, referred as the rotating SV black hole \cite{Mazza:2021rgq,Shaikh:2021yux}. 
Working in natural units with $G=c=1$, the line element of the rotating SV black hole in Boyer--Lindquist coordinates $(t,r,\theta,\phi)$ is given by
\begin{align}
	\label{eq:metric}
	ds^{2} ={}&
	-\left(1-\frac{2 M \sqrt{r^{2}+g^{2}}}{\rho^2}\right) dt^{2}
+ \frac{\rho^2}{\Delta} \, dr^{2}
	+ \rho^2 \, d\theta^{2} \nonumber\\
	&- \frac{4 a M \sqrt{r^{2}+g^{2}} \sin^{2}\theta}{\rho^2} \, dt \, d\phi
	+ \frac{\Sigma \sin^2\theta}{\rho^2} \, d\phi^{2}, 
\end{align}
where
 	\begin{equation}
 	\begin{split}
    &\rho^2 = r^2 + g^2 + a^2\cos^2\theta, \\
    &\Delta = r^2 + g^2 + a^2 - 2M\sqrt{r^2 + g^2}, \\
    &\Sigma = (r^2 + g^2 + a^2)^2 - \Delta a^2 \sin^2\theta.
    \end{split}
    \end{equation}
Here $M$ denotes the ADM mass, $a$ is the spin parameter associated with the black hole angular momentum, and $g$ is the regularization parameter characterizing deviations from the Kerr geometry and ensuring spacetime regularity. 
In the limit $g\to0$, the metric reduces to the Kerr black hole, while for $a\to0$ it further degenerates to the Schwarzschild solution.

   \begin{table*}[th]
	\centering
	\caption{Classification of rotating SV spacetime according to the parameter $g$ \cite{Mazza:2021rgq}.}
	\label{tab:horizon_classification}
	\begin{tabular}{c c c}
		\hline
		Range of $g$ & Horizons & Spacetime \\
		\hline
		$[0,M-\sqrt{M^2-a^2})$ 
		& $r_-,\,r_+$ 
		& Regular black hole \\
		
		$[M-\sqrt{M^2-a^2}  , M+\sqrt{M^2-a^2})$ 
		& $r_+$
		& Single-horizon regular black hole \\
		
		$[M+\sqrt{M^2-a^2},\infty)$ 
		& No horizon 
		& Wormhole \\
		\hline
	\end{tabular}
\end{table*}
 
The condition $\Delta(r)=0$ defines the event horizon of the rotating regular spacetime.
This equation admits analytical solutions, yielding two positive roots that are the inner and outer horizons,
$r_-$ and $r_+$, respectively,
\begin{equation}
	r_{\pm} = \sqrt{(M \pm \sqrt{M^2-a^2})^2 - g^2}.
\end{equation}
The existence and number of horizons are jointly controlled by the parameter $a$ and $g$.
According to the value of $g$, the rotating SV spacetime can be classified into three distinct regimes,
as summarized in Table~\ref{tab:horizon_classification}.
A comprehensive discussion of the global parameter space can be found in \cite{Mazza:2021rgq}.

\begin{figure}[H]
	\centering
	\includegraphics[width=0.4\textwidth]{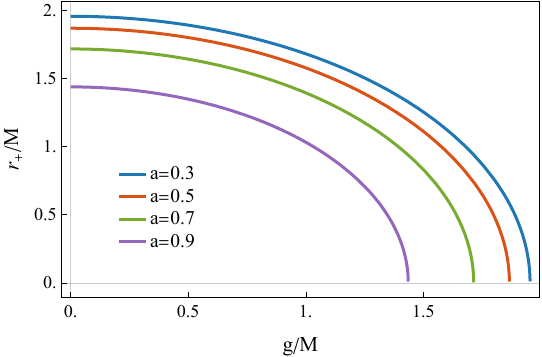}
	\caption{Variation of the event horizon radius $r_+$ with respect to the parameter $g$ for several representative values of the parameter $a$.
	}
	\label{fig:horizon}
\end{figure}

In the present work, we focus exclusively on regular black hole configurations possessing an outer event horizon.
Accordingly, the parameter $g$ is restricted to the range $0 \le g < M+\sqrt{M^2-a^2}$ in the subsequent analyses.
To illustrate the impact of $g$ on the horizon geometry,
Fig.~\ref{fig:horizon} shows the variation of the outer horizon radius $r_+$ with $g$ for several representative values of the parameter $a$.
As $g$ increases, the outer horizon radius decreases monotonically and can become significantly smaller than its Kerr counterpart,
highlighting the strong influence of spacetime regularization on the near-horizon geometry.

The dynamics of test particles in the equatorial plane is of fundamental importance for modeling astrophysical accretion disks \cite{Bambi:2016sac}. In the standard thin-disk model, the accreting matter is assumed to follow nearly circular, equatorial orbits prior to its plunge into the black hole. Therefore, a detailed analysis of equatorial circular orbits is essential for understanding the radiative properties and observational signatures of accretion disks around the rotating SV black hole, which will be explored. Restricting the motion to the equatorial plane, we impose the conditions $\theta=\pi/2$ and $\dot{\theta}=0$. The particle dynamics is then governed by the Lagrangian
\begin{equation}
	\mathcal{L}
	=\frac{1}{2} g_{\mu\nu}\dot{x}^{\mu}\dot{x}^{\nu}
	=\frac{1}{2}\left(
	g_{tt}\dot{t}^{2}
	+2g_{t\phi}\dot{t}\dot{\phi}
	+g_{rr}\dot{r}^{2}
	+g_{\phi\phi}\dot{\phi}^{2}
	\right), 
\end{equation}
where the dot denotes differentiation with respect to the proper time. Owing to the stationarity and axial symmetry of the spacetime, the coordinates $t$ and $\phi$ are cyclic. The conjugate momenta defined from the Lagrangian,
$p_{\mu}=\frac{\partial\mathcal{L}}{\partial\dot{x}^{\mu}},$
lead to two conserved quantities along geodesics, namely the energy $E$ and the angular momentum $L$, defined as
\begin{equation}
	E=-p_{t}=-g_{tt}\dot{t} - g_{t\phi}\dot{\phi},
\end{equation}
\begin{equation}
	L=p_{\phi}=g_{t\phi}\dot{t} + g_{\phi\phi}\dot{\phi}.
\end{equation}
Under the assumption of equatorial circular motion, the dynamics of test particles can be derived following the standard procedure presented in \cite{Bambi:2016sac}. We impose the circular orbit constraints $\dot{r}=0$, and the angular velocity $\Omega=d\phi/dt=\dot{\phi}/\dot{t}$  can be written in the general form
\begin{equation}
	\Omega_{\pm}
	=
	\frac{-\partial_r g_{t\phi}
		\pm
		\sqrt{(\partial_r g_{t\phi})^2
			-
			(\partial_r g_{tt})(\partial_r g_{\phi\phi})}}
	{\partial_r g_{\phi\phi}},
\end{equation}
where the plus and minus signs correspond to co-rotating and counter-rotating circular orbits, respectively. Unless otherwise stated, we shall focus on the co-rotating branch $\Omega=\Omega_{+}$ in the following.
For a massive test particle, its motion is subject to the normalization condition of the four velocity $g_{\mu\nu}x^\mu x^\nu=-1$. Then,  the conserved energy and angular momentum can be written as
\begin{align}
	E &= \frac{-g_{tt}-g_{t\phi}\Omega}
	{\sqrt{-g_{tt}-2g_{t\phi}\Omega-g_{\phi\phi}\Omega^2}}, \\
	L &= \frac{g_{t\phi}+g_{\phi\phi}\Omega}
	{\sqrt{-g_{tt}-2g_{t\phi}\Omega-g_{\phi\phi}\Omega^2}} .
\end{align}
\begin{figure*}[tb]
	\centering
	\begin{subfigure}[b]{0.32\textwidth}
		\centering
		\includegraphics[width=\textwidth]{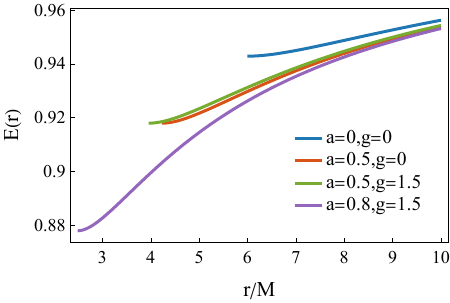}
	\end{subfigure}
	\begin{subfigure}[b]{0.32\textwidth}
		\centering
		\includegraphics[width=\textwidth]{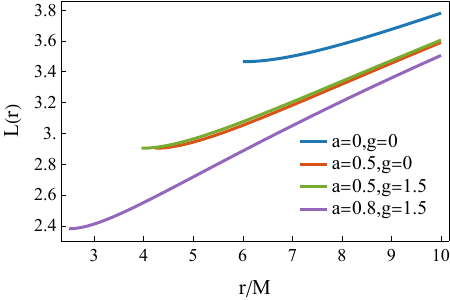}
	\end{subfigure}
	\begin{subfigure}[b]{0.32\textwidth}
		\centering
		\includegraphics[width=\textwidth]{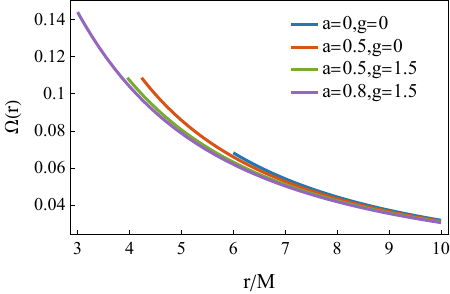}
	\end{subfigure}
	\caption{Radial profiles of the specific energy $E$, specific angular momentum $L$, and angular velocity $\Omega$ for particles moving on prograde equatorial circular orbits.}
	\label{fig:ELOmega}
\end{figure*}

Having obtained the analytical expressions for the specific energy $E$, angular momentum $L$,
and angular velocity $\Omega$ of test particles moving on equatorial circular orbits, we now investigate their numerical behavior
as functions of the radial coordinate $r$ for different choices of the model parameters.
The results are summarized in Fig.~\ref{fig:ELOmega}, where four representative parameter sets are considered.
The Schwarzschild case corresponds to $a=0$ and $g=0$, the Kerr spacetime is recovered for $g=0$ and $a\neq0$,
while the remaining curves describe rotating regular SV black holes with nonvanishing $g$.
For each curve, the leftmost endpoint marks the location of the ISCO,
which sets the natural inner edge of stable particle motion.
Accordingly, all quantities are shown only for radii larger than the corresponding ISCO radius.
From Fig.~\ref{fig:ELOmega}, several characteristic trends can be identified.
For a fixed parameter $a$, increasing parameter $g$ leads to an increase in the
specific energy $E$ and angular momentum $L$ at a given radius,
while the angular velocity $\Omega$ correspondingly decreases.
This indicates that spacetime regularization effectively allows particles to orbit with higher energy
and angular momentum, and to rotate slower. 
In contrast, for fixed $g$, an increase in $a$ results in a systematic reduction of
$E$, $L$, and $\Omega$, reflecting the well-known effect of frame dragging in rotating spacetimes.
Notably, while parameter $g$ affects the radial dependence of these quantities,
the values of $E$, $L$, and $\Omega$ evaluated at the ISCO remain unchanged when $g$ varies and $a$ is fixed.
This demonstrates that, although the location of the ISCO depends on the spacetime regularization parameter $g$,
the corresponding orbital properties of particles at the ISCO are insensitive to $g$.

For the motion of massive particles confined to the equatorial plane,
the radial equation of motion can be written as
\bea
&&g_{rr}\dot r^{2}+V_{\mathrm{eff}}(r)=0,\\	
&&V_{\mathrm{eff}}(r)\equiv 1-\frac{E^{2}g_{\phi\phi}+2ELg_{t\phi}+L^{2}g_{tt}}{g_{t\phi}^{2}-g_{tt}g_{\phi\phi}},\nonumber
\eea
where $V_{\mathrm{eff}}(r)$ denotes the radial effective potential.
The location of the ISCO is determined by the following equation
\begin{equation}
\frac{d^{2}V_{\mathrm{eff}}}{dr^{2}}=0.
\end{equation}
Since this condition generally does not admit closed-form solutions,
the ISCO radius is obtained numerically.

\begin{figure}[H]
	\centering
	\begin{subfigure}[t]{0.4\textwidth}
		\centering
		\includegraphics[width=\textwidth]{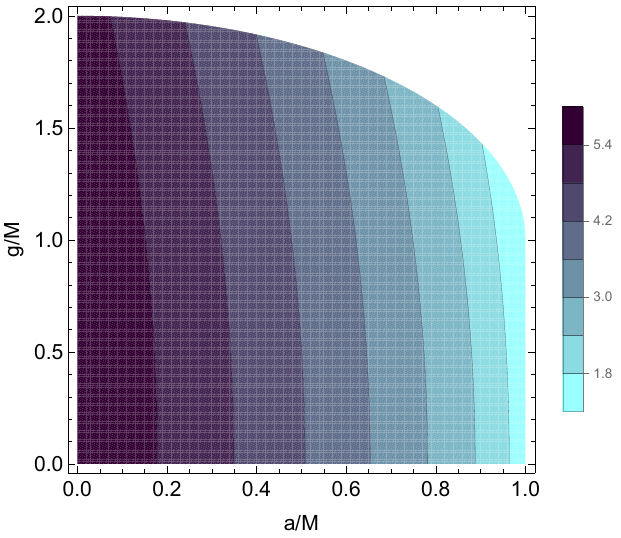}
		\caption{Prograde case}
	\end{subfigure}
	\hfill
	\begin{subfigure}[t]{0.4\textwidth}
		\centering
		\includegraphics[width=\textwidth]{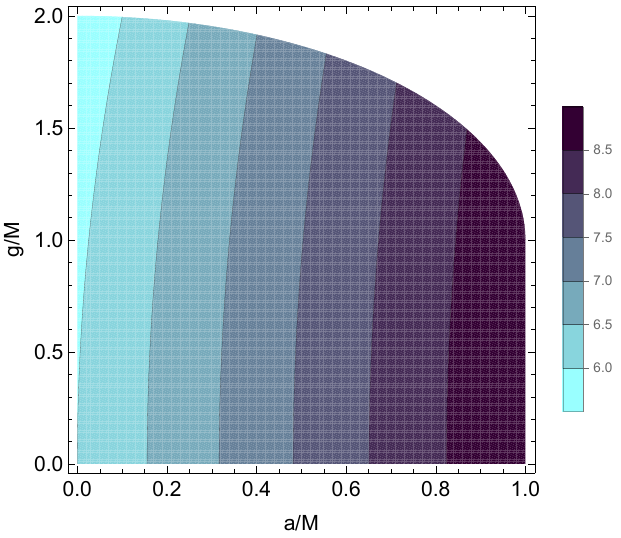}
		\caption{Retrograde case}
	\end{subfigure}
	\caption{ISCO radius as a function of the model parameters $a$ and $g$.
		}
	\label{fig:ISCO}
\end{figure}

Figure~\ref{fig:ISCO} displays the ISCO radius as a function of parameters $a$
and $g$ for both prograde and retrograde circular orbits.
For prograde motion, the ISCO radius decreases monotonically with increasing $a$ and $g$.
In contrast, for retrograde motion, the ISCO radius increases with $a$
while still decreasing as $g$ increases.
Moreover, in both prograde and retrograde cases, the dependence of the ISCO radius on $a$
is significantly stronger than that on $g$,
indicating that spacetime rotation plays a dominant role in determining the ISCO location.

\begin{figure}[H]
	\centering
	\begin{subfigure}[t]{0.4\textwidth}
		\centering
		\includegraphics[width=\textwidth]{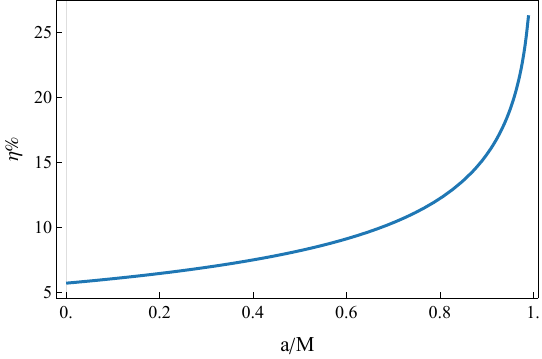}
	\end{subfigure}
	\hfill
	\begin{subfigure}[t]{0.4\textwidth}
		\centering
		\includegraphics[width=\textwidth]{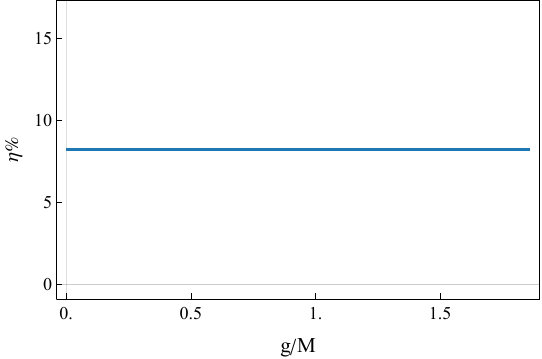}
	\end{subfigure}
	\caption{Radiative efficiency of the rotating SV black hole.
		The top panel shows the dependence on spin parameter $a$ for a fixed parameter $g=1.5$,
		while the bottom panel displays the dependence on $g$ for a fixed $a=0.5$.}
	\label{fig:efficiency}
\end{figure}

The radiative efficiency is a key quantity characterizing the conversion of gravitational energy
into radiation during the accretion process and is widely used to quantify the energy release
of the black hole accretion disk \cite{Page:1974he,Kurmanov:2024hpn,Collodel:2021gxu,Wu:2024sng,Liu:2024brf,Lee:2022rtg,Olmo:2023lil,Boshkayev:2023fft}.
Within the thin-disk approximation, assuming that particles slowly fall from infinity
and terminate their inspiral at the ISCO,
the radiative efficiency can be defined as
\begin{equation}
	\eta=\frac{E_{\infty}-E_{\mathrm{ISCO}}}{E_{\infty}},
\end{equation}
where $E_{\infty}$ denotes the specific energy of the particle at infinity.
For particles starting from rest at infinity, one may approximate $E_{\infty}\simeq 1$,
leading to
\begin{equation}
	\eta \simeq 1 - E_{\mathrm{ISCO}}.
\end{equation}
Figure~\ref{fig:efficiency} presents the dependence of the radiative efficiency on the parameters $a$ and $g$.
As the parameter $a$ increases, the radiative efficiency grows significantly,
in agreement with the well-known behavior in the Kerr black hole case.
In contrast, variation of the parameter $g$ does not affect the radiative efficiency,
which is a direct consequence of the fact that the specific energy evaluated at the ISCO
is independent of $g$. 
This is a typical feature of the SV black hole that spacetime regularization does not modify the efficiency of energy release.

	\section{Radiative Properties of Thin Disks around the Black Holes}
	\label{sec:disk}
	
	In the previous section, we analyzed the dynamical properties of equatorial circular
	orbits and the structure of the ISCO
	in the rotating SV black hole spacetime.
	Based on these results, we now investigate the  radiative properties of thin
	accretion disks in this background, with particular emphasis on the radial distribution
	of the radiative flux.
	Within the Novikov--Thorne thin-disk model, the radiative flux emitted from the disk
	surface is given by \cite{Page:1974he,Novikov1973,Li:2026ogy,Shu:2024tut}
	\begin{equation}
		\mathcal{F}(r)=\frac{\dot{M}}{4\pi\sqrt{-\tilde{g}/g_{\theta\theta}}}
		\frac{-\Omega_{,r}}{(E-\Omega L)^2}
		\int_{r_{\mathrm{ISCO}}}^{r}(E-\Omega L)L_{,r}\,dr,
	\end{equation}
	where $\dot{M}$ is the mass accretion rate, which is assumed as a
	constant here, and $\tilde{g}$ denotes the determinant of the spacetime metric. 	
	
Once the radiative flux from the accretion disk surface is obtained,
one can further define the local effective temperature of the disk.
Assuming that the disk emits as a  blackbody locally,
the radiative flux and the effective temperature are related
through the Stefan--Boltzmann law \cite{Nampalliwar:2018tup}. 
Accordingly, the local effective temperature of the accretion disk
at radius $r$ is given by
\begin{equation}
	T_{\mathrm{eff}}(r)
	=\left(\frac{\mathcal{F}(r)}{\sigma}\right)^{1/4},
\end{equation}
where $\sigma$ is the Stefan--Boltzmann constant,
whose numerical value is
\begin{equation}
	\sigma = 5.67 \times 10^{-5}\;
	\mathrm{erg\,cm^{-2}\,s^{-1}\,K^{-4}}.
\end{equation}

	To eliminate the overall scaling effects introduced by the black hole mass and the mass accretion rate, it is convenient to introduce dimensionless radiative flux and effective temperature \cite{Liu:2025hhg}. Accordingly, the dimensionless radiative flux $\mathcal{F}^*$ and the dimensionless effective temperature $T_{\rm eff}^*$ are defined as
	\begin{equation}
		\mathcal{F}^*=\frac{G^2 M^2}{c^6 \dot{M}}\,\mathcal{F}, \qquad
		T_{\rm eff}^* = \mathcal{F}^{*1/4},
	\end{equation}
	where $G$ is the gravitational constant, $c$ is the speed of light in vacuum, and $M$ and $\dot{M}$ denote the black hole mass and the mass accretion rate, respectively. With these dimensionless quantities, the radiative flux and effective temperature distributions of accretion disks around black holes with different masses and accretion rates can be directly compared, and the corresponding numerical results are presented and discussed in the following figures.
	
	Figure~\ref{fig:flux} presents the numerical results of the dimensionless radiative flux $\mathcal{F}^*$ in two representative cases: varying the parameter $g$ with fixed parameter $a$, and varying $a$ with fixed $g$. In all cases, the radial profiles of the radiative flux exhibit a well-defined peak, whose magnitude and radial location depend sensitively on the parameters.
	The upper panel shows four representative curves obtained for fixed $g=1$ and different values of the parameter $a$. As $a$ increases, the radiative flux is enhanced at all radial locations, the peak value becomes significantly larger, and the position of the peak shifts inward. This indicates that the spin parameter plays a dominant role in shaping the radiative properties of the accretion disk. In particular, the maximum radiative flux for $a=0.8$ is approximately three times larger than that for $a=0.5$, highlighting the strong impact of black hole spin.
	The lower panel corresponds to the case with fixed $a=0.5$, with the parameter $g$ taking values $g=0$ and $g=1.86$, the latter being close to the upper bound allowed for regular black hole configurations. This choice is intended to maximize the visibility of the effects induced by $g$. It is evident that, within the parameter range permitted by regular black hole solutions, the influence of $g$ on the radiative flux is much weaker than that of the parameter $a$. As $g$ increases, the peak value of the radiative flux decreases and its location shifts slightly inward. However, this trend is not monotonic for the whole disk: in some regions of the disk, the radiative flux is instead slightly enhanced. 

	\begin{figure}[H]
		\centering
		\begin{subfigure}[t]{0.45\textwidth}
			\centering
			\includegraphics[width=\textwidth]{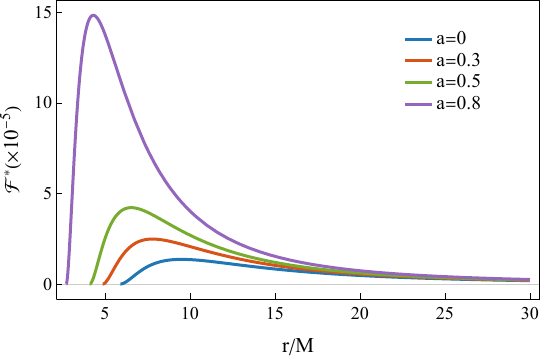}
		\end{subfigure}
		\hfill
	\begin{subfigure}[t]{0.45\textwidth}
		    \centering
			\includegraphics[width=\textwidth]{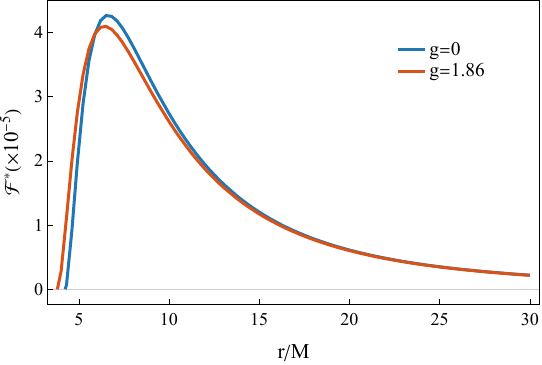}
		\end{subfigure}
		\caption{Dimensionless radiative flux of a thin accretion disk surrounding the rotating SV black hole.
			The top panel shows the dependence of the flux on the parameter $a$ for a fixed parameter $g=1$.
			The bottom panel displays the dependence on the parameter $g$ for a fixed parameter $a=0.5$.}
		\label{fig:flux}
	\end{figure}
	
\begin{figure}[H]
		\centering
		\begin{subfigure}[t]{0.45\textwidth}
			\centering
			\includegraphics[width=\textwidth]{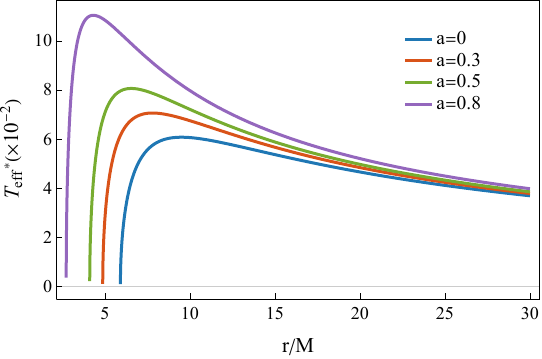}
		\end{subfigure}
		\hfill
		\begin{subfigure}[t]{0.45\textwidth}
			\centering
			\includegraphics[width=\textwidth]{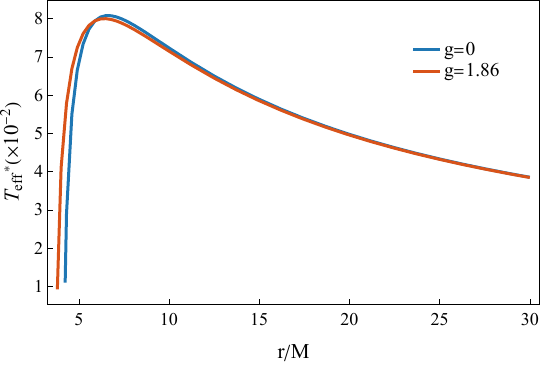}
		\end{subfigure}
		\caption{Dimensionless effective temperature of a thin accretion disk surrounding the rotating SV black hole.
		The top panel shows the dependence on the parameter $a$ for a fixed parameter $g=1$.
		The bottom panel displays the dependence on the parameter $g$ for a fixed parameter $a=0.5$.}
		\label{fig:temperature}
	\end{figure}

	We then show the numerical results of the dimensionless effective temperature in Fig.~\ref{fig:temperature}, where the model parameters are chosen to be the same as those adopted in the previous analysis of the dimensionless radiative flux. 
	It is found that the influence of the parameter $a$ and the parameter $g$ on the effective temperature follows the same qualitative trend as that on the radiative flux, although the overall variation becomes much weaker. 
	This behavior can be understood from the fact that the effective temperature scales with the radiative flux as $T_{\rm eff} \propto \mathcal{F}^{1/4}$, which significantly suppresses the impact of the variation of the parameters.
	We now restore physical units and examine whether  observationally distinguishable features from the Kerr case may arise in the rotating SV black hole. 
	Therefore, we consider two representative astrophysical examples: the supermassive black holes Sgr A* and M87*.
	In our numerical calculations, the physical constants are chosen as: the speed of light 
	$c = 3\times10^{10}\,\mathrm{cm\,s^{-1}}$, the Plank constant 
	$h = 6.625\times10^{-27}\,\mathrm{erg\,s}$, the Boltzmann contant 
	$k = 1.38\times10^{-16}\,\mathrm{erg\,K^{-1}}$,
	the solar mass $M_\odot = 1.989\times10^{33}\,\mathrm{g}$,
	and $1~\mathrm{year} = 3.156\times10^{7}\,\mathrm{s}$. 
	
	For Sgr A* \cite{EventHorizonTelescope:2022wkp,EventHorizonTelescope:2022xqj}, the black hole mass and mass accretion rate are taken to be $M = 4\times10^{6}\,\mathrm{M_\odot}$ and $\dot{M}=10^{-6}\,\mathrm{M_\odot/year}$, respectively. 
	We focus on the peak values of the radiative flux and the effective temperature. 
	For the rotating SV black hole with $a=0.5$ and $g=1.86$, the maximum radiative flux, the maximum effective temperature, and the corresponding radial position are given by
	\begin{equation}
		\begin{aligned}
			\mathcal{F}_{\max} &= 6.67 \times 10^{12}\ \mathrm{erg\,cm^{-2}\,s^{-1}},\\
			T_{\rm eff}^{\max} &= 1.85 \times 10^{4}\ \mathrm{K},\\
			r_{\max} &= 6.42\, M .
		\end{aligned}
	\end{equation}
	For comparison, the corresponding quantities for the Kerr black hole with the same spin parameter $a=0.5$ read
	\begin{equation}
		\begin{aligned}
			\mathcal{F}_{\max} &= 6.95 \times 10^{12}\ \mathrm{erg\,cm^{-2}\,s^{-1}},\\
			T_{\rm eff}^{\max} &= 1.87 \times 10^{4}\ \mathrm{K},\\
			r_{\max} &= 6.62\, M .
		\end{aligned}
	\end{equation}
	
	As a second example, we consider M87* \cite{EventHorizonTelescope:2019dse}, whose black hole mass and  mass accretion rate are taken to be $M = 6.5\times10^{9}\,\mathrm{M_\odot}$ and $\dot{M}=10^{-3}\,\mathrm{M_\odot/year}$ \cite{Drew:2025euq}. 
	For the rotating SV black hole with $a=0.8$ \cite{Drew:2025euq} and $g=1.5$, we obtain
	\begin{equation}
		\begin{aligned}
			\mathcal{F}_{\max} &= 8.85 \times 10^{9}\ \mathrm{erg\,cm^{-2}\,s^{-1}},\\
			T_{\rm eff}^{\max} &= 3.53 \times 10^{3}\ \mathrm{K},\\
			r_{\max} &= 4.22\, M .
		\end{aligned}
	\end{equation}
	while for the Kerr black hole with $a=0.8$, the corresponding results are
	\begin{equation}
		\begin{aligned}
			\mathcal{F}_{\max} &= 9.40 \times 10^{9}\ \mathrm{erg\,cm^{-2}\,s^{-1}},\\
			T_{\rm eff}^{\max} &= 3.59 \times 10^{3}\ \mathrm{K},\\
			r_{\max} &= 4.42\, M .
		\end{aligned}
	\end{equation}
	
It should be emphasized that both the radiative flux and the effective temperature are defined in the local rest frame of the accretion disk and therefore are not directly observable quantities. 
From an observational perspective, a more relevant quantity is the spectral luminosity $\nu \mathcal{L}_{\nu,\infty}$ measured by a distant observer, which encodes the global radiative properties of the disk \cite{Boshkayev:2023fft,Jiang:2024njc,Joshi:2013dva}. 
Assuming that the disk emits as a collection of local blackbody radiators and neglecting the effect of light bending, the spectral luminosity at infinity can be obtained by integrating the local emission over the entire disk surface.
Under the thin-disk approximation, the spectral luminosity takes the form
	\begin{equation}
		\label{eq:nuLnu}
		\nu \mathcal{L}_{\nu,\infty}
		=
		\frac{8\pi h \cos\gamma}{c^{2}}
		\int_{r_{\rm ISCO}}^{r_{\rm out}}
		\int_{0}^{2\pi}
		\frac{\nu \nu_{e}^{3} r\,dr\,d\phi}
		{\exp\!\left(\dfrac{h\nu_{e}}{k\, T_{\rm eff}(r)}\right)-1},
	\end{equation}
where $r_{\rm out}$ denotes the outer edge of the accretion disk and is chosen to be sufficiently large to ensure the convergence of the numerical integration. 
The parameter $\gamma$ represents the inclination angle of the accretion disk. For the equatorial accretion disk we considered here, $\gamma=0$. 
The symbols $\nu$ and $\nu_e$ denote the observed frequency at infinity and the emitted frequency measured in the local rest frame of the accretion disk, respectively, and they are related through $\nu_e = \nu / g_{\rm out}$.
The redshift factor $g_{\rm out}$ is determined by the spacetime geometry and the orbital motion of the disk matter, and is given by
\begin{equation}
	\label{eq:gout}
	g_{\rm out}
	=	\sqrt{- g_{tt}- 2 g_{t\phi}\Omega- g_{\phi\phi}\Omega^{2}}.
\end{equation}
A detailed analysis of the redshift factor and its role in shaping the observed radiation will be presented in the next section.
\begin{figure}[H]
	\centering
	\begin{subfigure}[t]{0.45\textwidth}
		\centering
		\includegraphics[width=\textwidth]{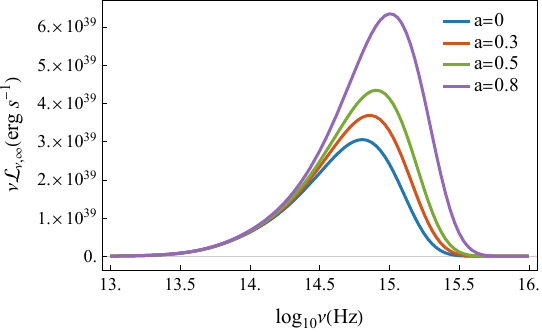}
	\end{subfigure}
	\hfill
	\begin{subfigure}[t]{0.45\textwidth}
		\centering
		\includegraphics[width=\textwidth]{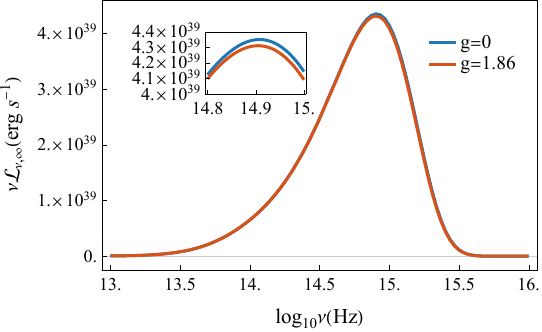}
	\end{subfigure}
	\caption{Spectral luminosity of a thin accretion disk surrounding the rotating SV black hole. The top panel shows the dependence on the parameter $a$ for a fixed parameter $g=1$. The bottom panel displays the dependence on the parameter $g$ for a fixed parameter $a=0.5$.}
	\label{fig:nuLnu}
\end{figure}

For the numerical evaluation of the spectral luminosity, we fix the outer edge of the accretion disk at $r_{\rm out}=1000M$ and set the inclination angle to $\gamma=0$, while the black hole mass and mass accretion rate are chosen to be consistent with observational estimates for Sgr A*.
Under these assumptions, the spectral luminosity $\nu\mathcal{L}_{\nu,\infty}$ of rotating regular black holes in the frequency range $10^{13}$–$10^{16}\mathrm{Hz}$ is computed for different values of the spin parameter $a$ and the regularization parameter $g$, as shown in Fig.~\ref{fig:nuLnu}.
In this frequency band, the spectral luminosity always exhibits a pronounced maximum, which serves as the main quantity of interest in the following discussion.
In the upper panel, for a fixed parameter $g=1$, the spectral luminosity increases significantly with increasing spin. In particular, for $a=0$, corresponding to the spherically symmetric SV black hole, the maximum value is $\nu\mathcal{L}_{\nu,\infty}^{\rm max}=3.049\times10^{39}\mathrm{erg\, s^{-1}}$, occurring at $\nu=6.40\times10^{14}\mathrm{Hz}$, whereas for $a=0.8$ the peak rises to $\nu\mathcal{L}_{\nu,\infty}^{\rm max}=6.339\times10^{39}\mathrm{erg\,s^{-1}}$ and occurs at $\nu=1.02\times10^{15}\mathrm{Hz}$, exceeding two times of the maximum value of the $a=0$ case.
In contrast, the lower panel shows that for rotating SV black hole with fixed spin $a=0.5$, the dependence on the parameter $g$ is much weaker: when $g$ varies from $0$ to $1.86$, the peak value decreases slightly from $4.350\times10^{39}\mathrm{erg\,s^{-1}}$ at $\nu=8.10\times10^{14}\mathrm{Hz}$ to $4.311\times10^{39}\mathrm{erg\,s^{-1}}$ at $\nu=8.03\times10^{14}\mathrm{Hz}$.
Both the peak amplitude and its corresponding frequency are therefore marginally smaller for larger $g$, and the Kerr case yields a maximum spectral luminosity that is approximately $1.01$ times that of the rotating SV black hole.

	\section{Optical Appearance of the Black Hole}
	\label{sec:raytracing}
	
	Having analyzed the radiative properties of thin accretion disks, 
	we now turn to the optical appearance of the rotating SV black hole as detected by a distant observer.
	To investigate the optical appearance of the black hole, we employ a ray-tracing technique to determine which photon emitted from the accretion disk can reach a distant observer \cite{Pu:2016eml,Younsi:2016azx}. 
	Since not all photons originating from the disk are able to escape to infinity, we adopt a backward ray-tracing approach, in which photon trajectories are traced backward from the observer’s screen toward the black hole until it is determined whether they intersect the accretion disk.
	
	The observer’s screen is modeled as a three-dimensional Cartesian coordinate system $(x,y,z)$, with the negative $z$-axis pointing toward the black hole. 
	The observer is located at a fixed position $(r_{\rm obs},\theta_{\rm obs},\phi_{\rm obs})$ in the Boyer--Lindquist coordinate system.
	In the observer frame, the initial position of a photon is taken as $(x,y,0)$, and its initial spatial velocity is chosen as $(0,0,-1)$, corresponding to incident photons moving  perpendicularly to the screen.
	A crucial step in the ray-tracing procedure is to establish the mapping between the observer frame and the black hole coordinate system.
	Following the literature \cite{Younsi:2016azx}, the Cartesian components $(X,Y,Z)$ of the photon’s initial position in the black hole frame are given by
	\begin{align}
		X &= D\cos\phi_{\rm obs}-x\sin\phi_{\rm obs}, \\
		Y &= D\sin\phi_{\rm obs}+x\cos\phi_{\rm obs}, \\
		Z &= r_{\rm obs}\cos\theta_{\rm obs}+y\sin\theta_{\rm obs},
	\end{align}
	where
	\begin{equation}
		D=\sin\theta_{\rm obs}\sqrt{r_{\rm obs}^{2}+a^{2}}-y\cos\theta_{\rm obs}.
	\end{equation}
	The corresponding Boyer--Lindquist coordinates of the photon position are then obtained as
	\begin{align}
		r &=\sqrt{\sigma+\sqrt{\sigma^{2}+a^{2}Z^{2}}}, \\
		\theta &=\arccos\frac{Z}{r}, \\
		\phi &=\arctan(X,Y),
	\end{align}
	where
	\begin{equation}
		\sigma=\frac{X^{2}+Y^{2}+Z^{2}-a^{2}}{2}.
	\end{equation}
	Under the same transformation, the spatial components of the photon’s initial velocity in Boyer--Lindquist coordinates read
	\begin{align}
		\Sigma &= r^{2}+a^{2}\cos^{2}\theta,\qquad 
		\mathcal{R}=\sqrt{r^{2}+a^{2}},\qquad 
		\Phi=\phi-\phi_{\rm obs},\\
		\dot r &= -\frac{1}{\Sigma}\!\left[r\mathcal{R}\sin\theta\sin\theta_{\rm obs}\cos\Phi
		+ \mathcal{R}^{2}\cos\theta\cos\theta_{\rm obs}\right],\\
		\dot\theta &= -\frac{1}{\Sigma}\!\left[\mathcal{R}\cos\theta\sin\theta_{\rm obs}\cos\Phi
		- r\sin\theta\cos\theta_{\rm obs}\right],\\
		\dot\phi &= \frac{1}{\mathcal{R}}\sin\theta_{\rm obs}\sin\Phi\,\csc\theta .
	\end{align}
	Without loss of generality, the initial time coordinate of the photon is set to $t=0$, while its temporal velocity component is determined from the geodesic constraint,
	\begin{equation}
		\beta=-\frac{g_{t\phi}\dot\phi}{g_{tt}},\qquad
		\aleph=\frac{\delta-g_{ij}\dot x^{i}\dot x^{j}}{g_{tt}},\qquad
		\dot t=\beta+\sqrt{\beta^{2}+\aleph},
	\end{equation}
	where $\delta=0$ for photons and the spatial indices run over $i,j=r,\theta,\phi$.
	Once the initial position and four-velocity of a photon are specified, its canonical momentum is obtained via the Legendre transformation
	\begin{equation}
		p_\mu = g_{\mu\nu}\dot x^\nu .
	\end{equation}
	The photon propagation in the rotating regular black hole spacetime is then governed by the Hamiltonian equations
	\begin{equation}
		\dot x^\mu=\frac{\partial \mathcal{H}}{\partial p_\mu},\qquad
		\dot p_\mu=-\frac{\partial \mathcal{H}}{\partial x^\mu},
	\end{equation}
	with the Hamiltonian
	\begin{equation}
		\mathcal{H}=\frac{1}{2}g^{\mu\nu}p_\mu p_\nu=0,
	\end{equation}
	corresponding to the null geodesics of photons.
	By numerically integrating these equations, one can reconstruct the full photon trajectories and determine whether they intersect the accretion disk, thereby producing the optical image observed by a distant observer.

	\begin{figure*}[]
		\centering
		\begin{subfigure}[t]{0.32\textwidth}
			\centering
			\includegraphics[width=\textwidth]{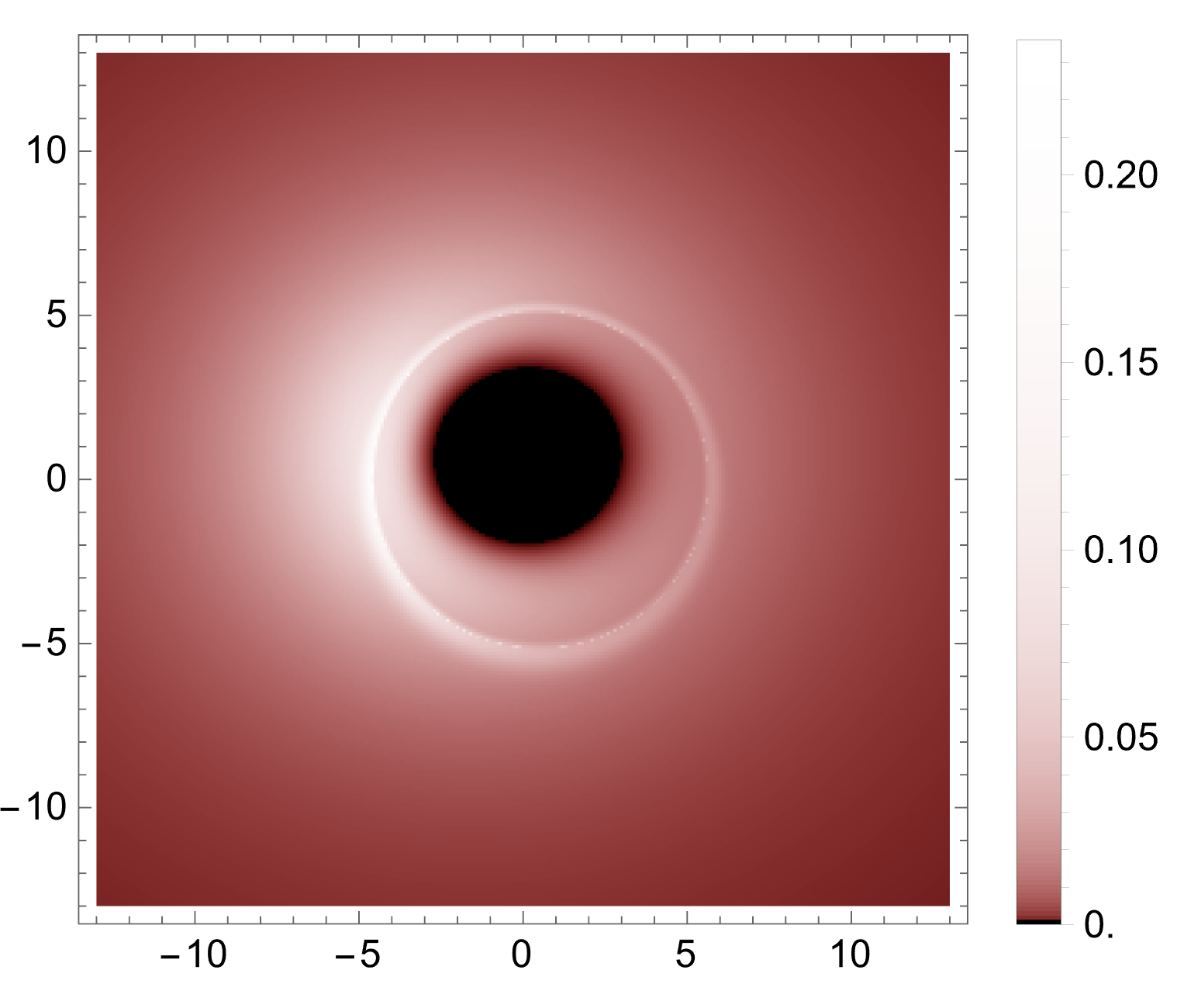}
			\caption{$\theta_{\rm obs}=30^\circ,g=0$}
		\end{subfigure}
		\begin{subfigure}[t]{0.32\textwidth}
			\centering
			\includegraphics[width=\textwidth]{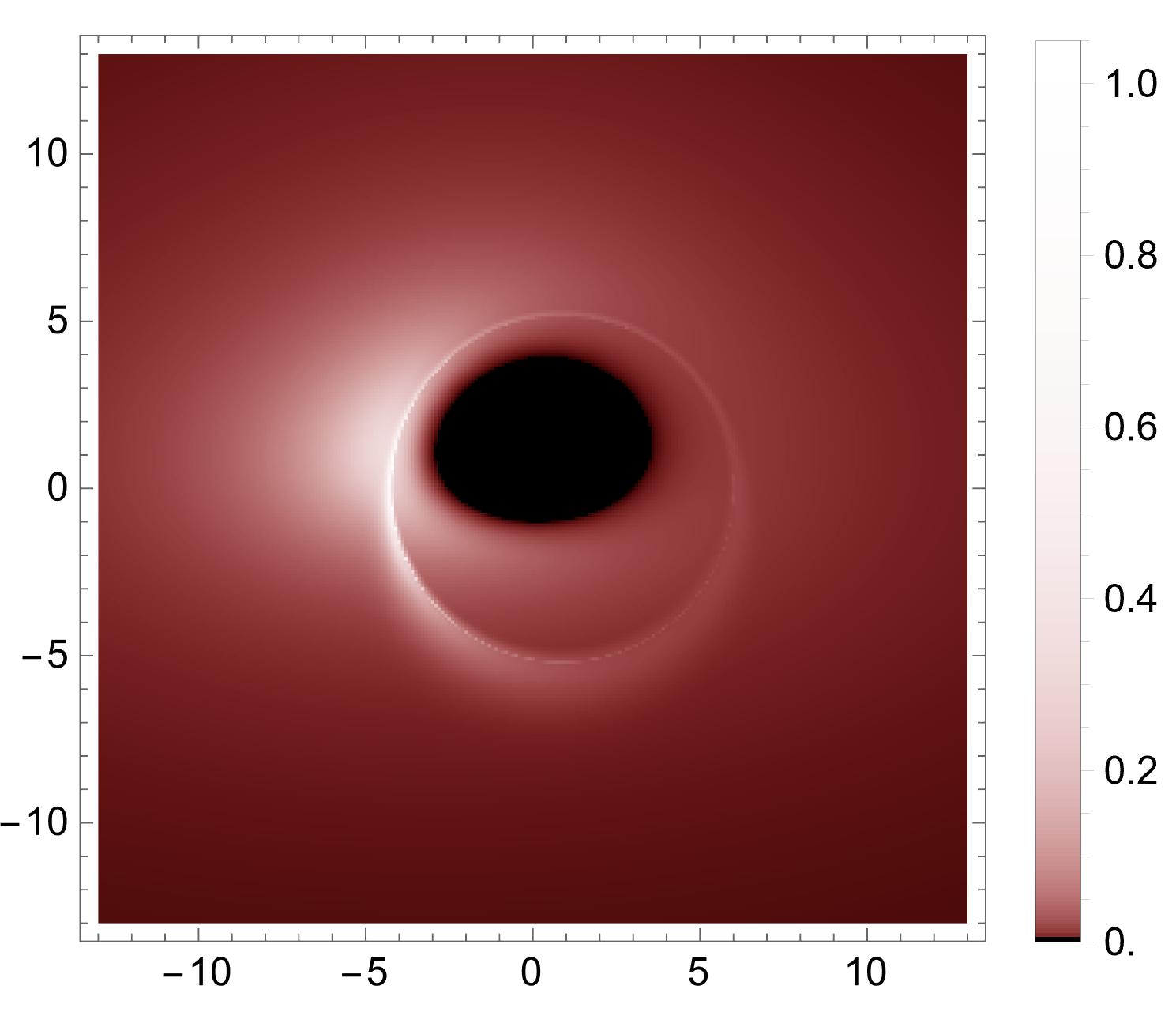}
			\caption{$\theta_{\rm obs}=60^\circ,g=0$}
		\end{subfigure}
		\begin{subfigure}[t]{0.32\textwidth}
			\centering
			\includegraphics[width=\textwidth]{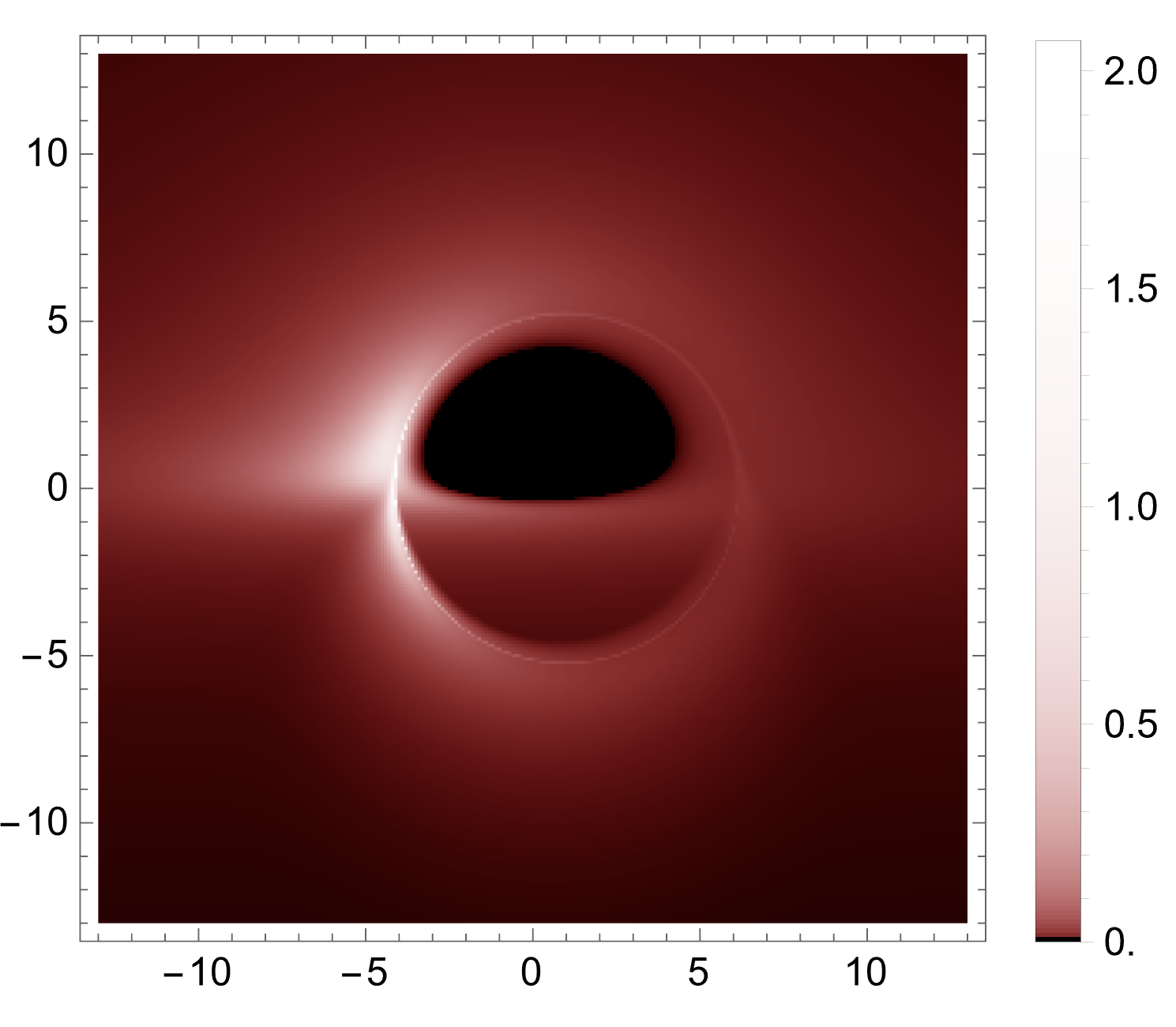}
			\caption{$\theta_{\rm obs}=80^\circ,g=0$}
		\end{subfigure}
		
		\vspace{0.2cm}
		
	\begin{subfigure}[t]{0.32\textwidth}
		\centering
		\includegraphics[width=\textwidth]{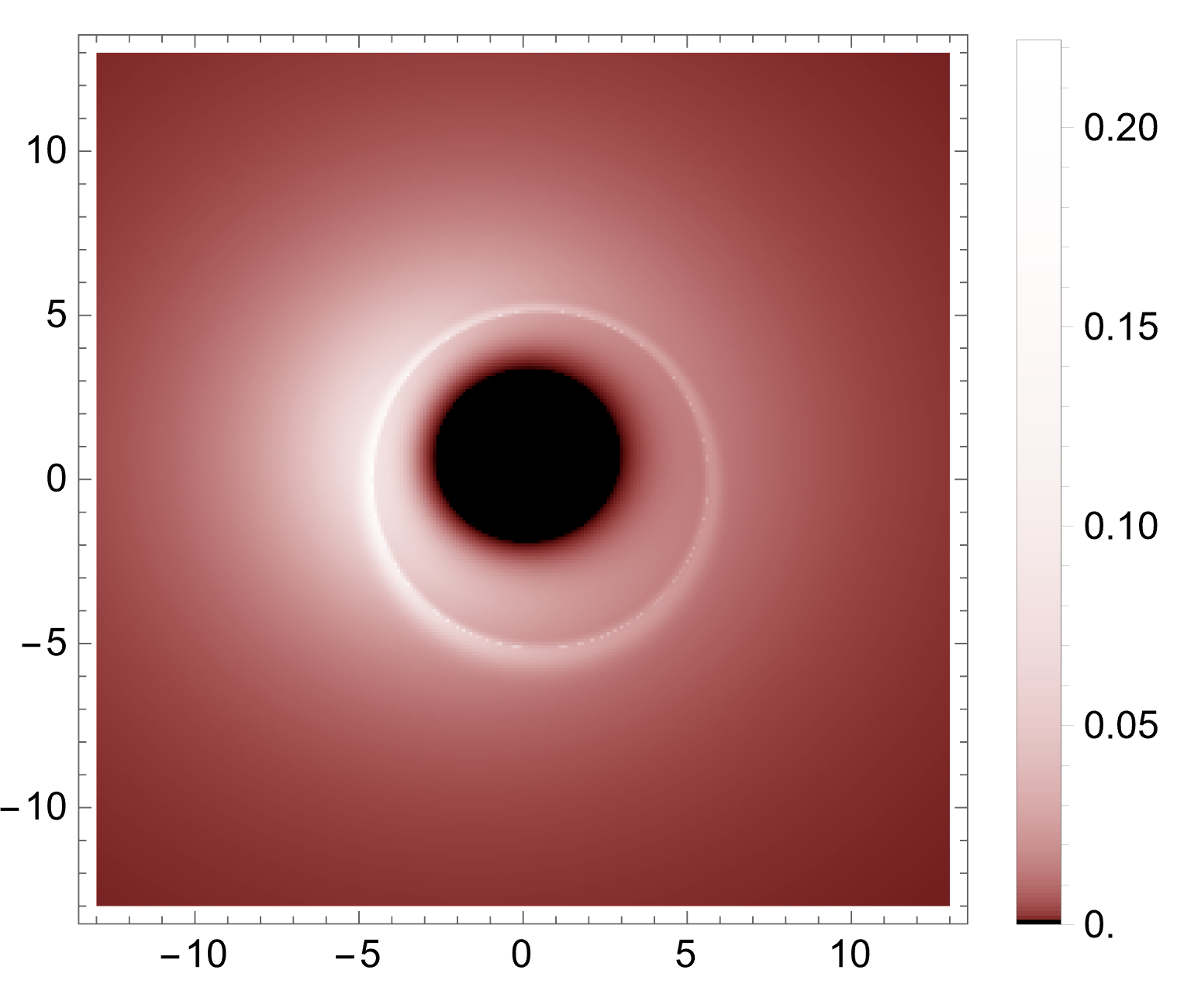}
		\caption{$\theta_{\rm obs}=30^\circ,g=0.5$}
	\end{subfigure}
	\begin{subfigure}[t]{0.32\textwidth}
		\centering
		\includegraphics[width=\textwidth]{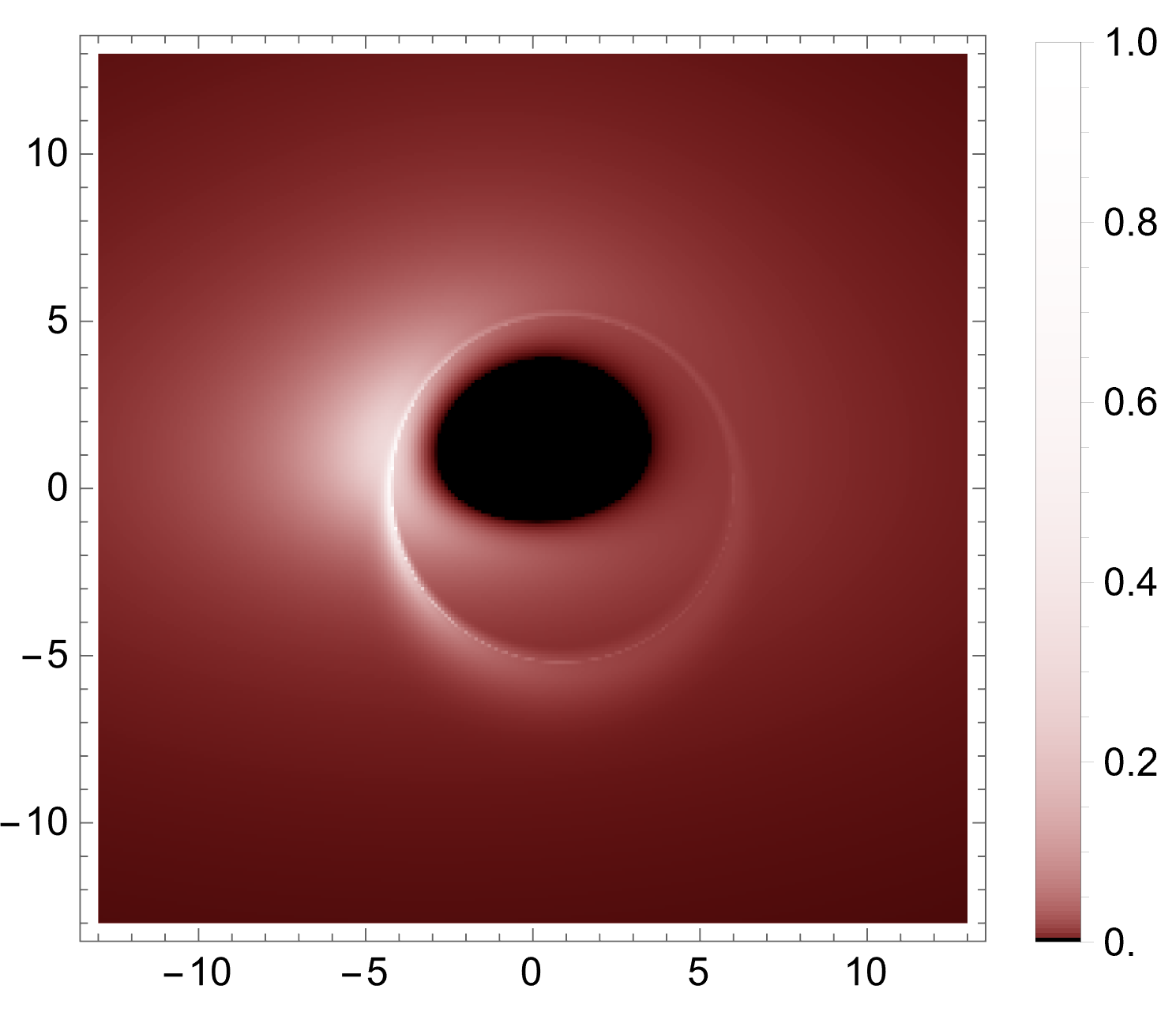}
		\caption{$\theta_{\rm obs}=60^\circ,g=0.5$}
	\end{subfigure}
	\begin{subfigure}[t]{0.32\textwidth}
		\centering
		\includegraphics[width=\textwidth]{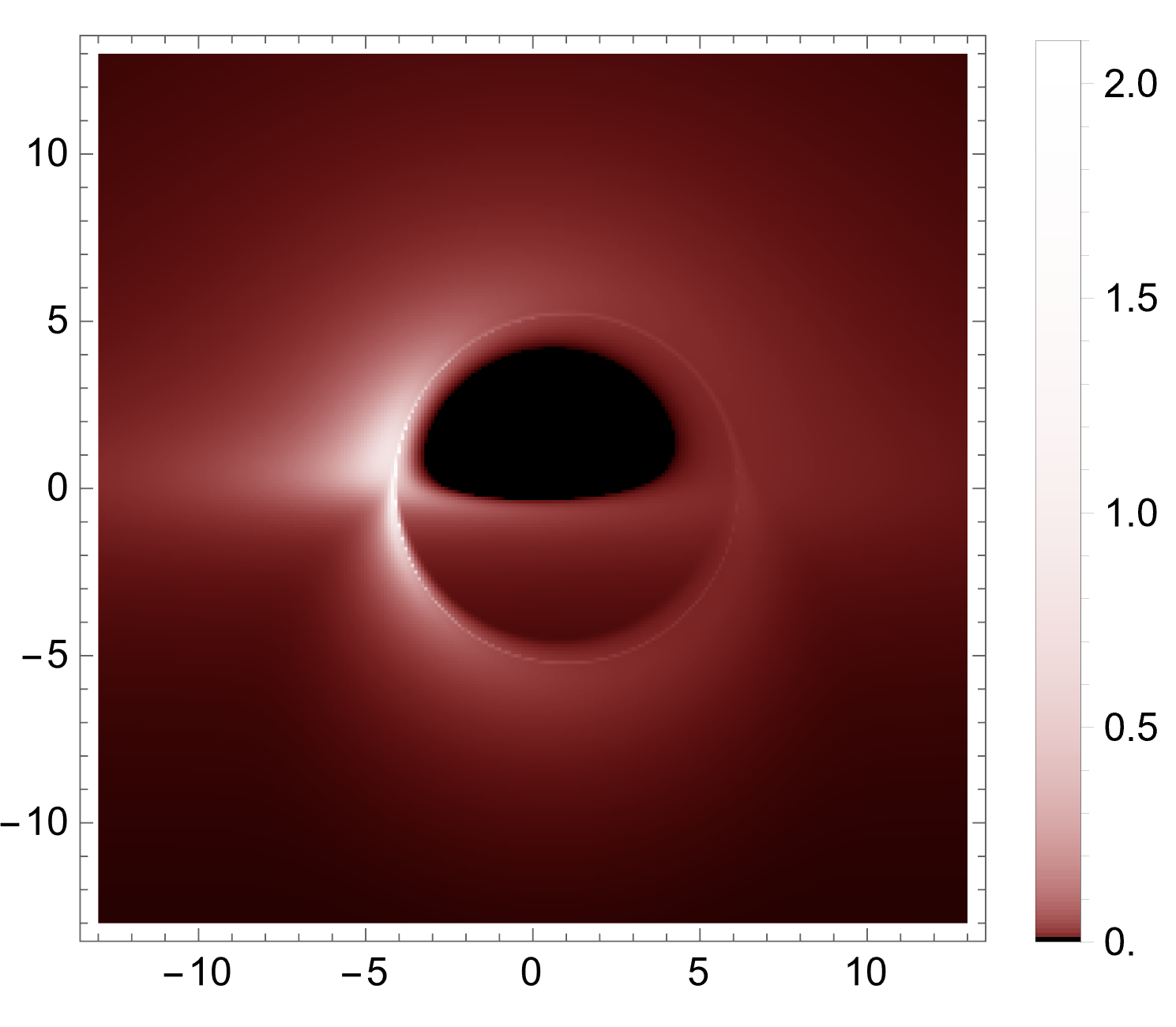}
		\caption{$\theta_{\rm obs}=80^\circ,g=0.5$}
	\end{subfigure}
	
	\vspace{0.2cm}
	
	\begin{subfigure}[t]{0.32\textwidth}
		\centering
		\includegraphics[width=\textwidth]{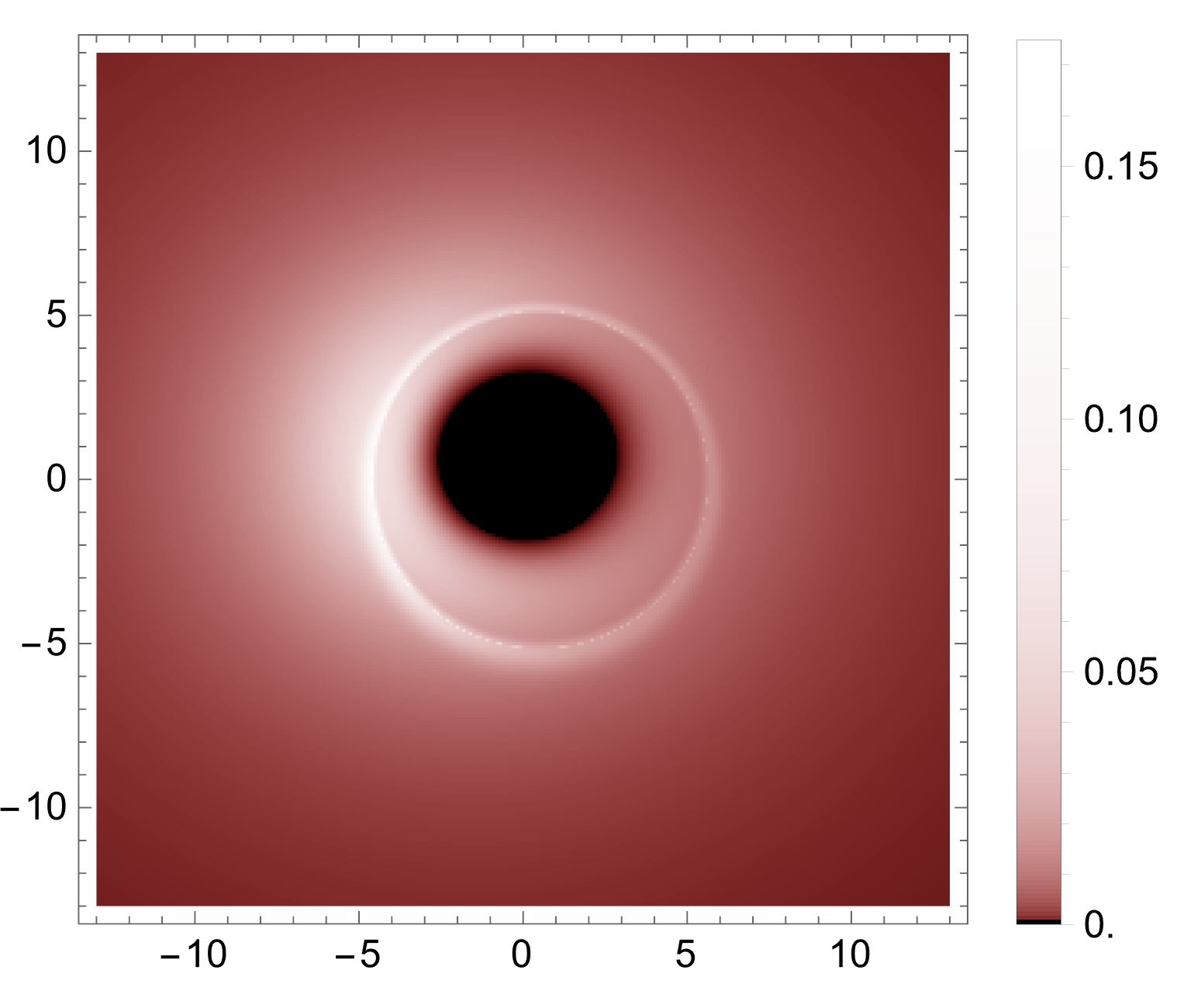}
		\caption{$\theta_{\rm obs}=30^\circ,g=1$}
	\end{subfigure}
	\begin{subfigure}[t]{0.32\textwidth}
		\centering
		\includegraphics[width=\textwidth]{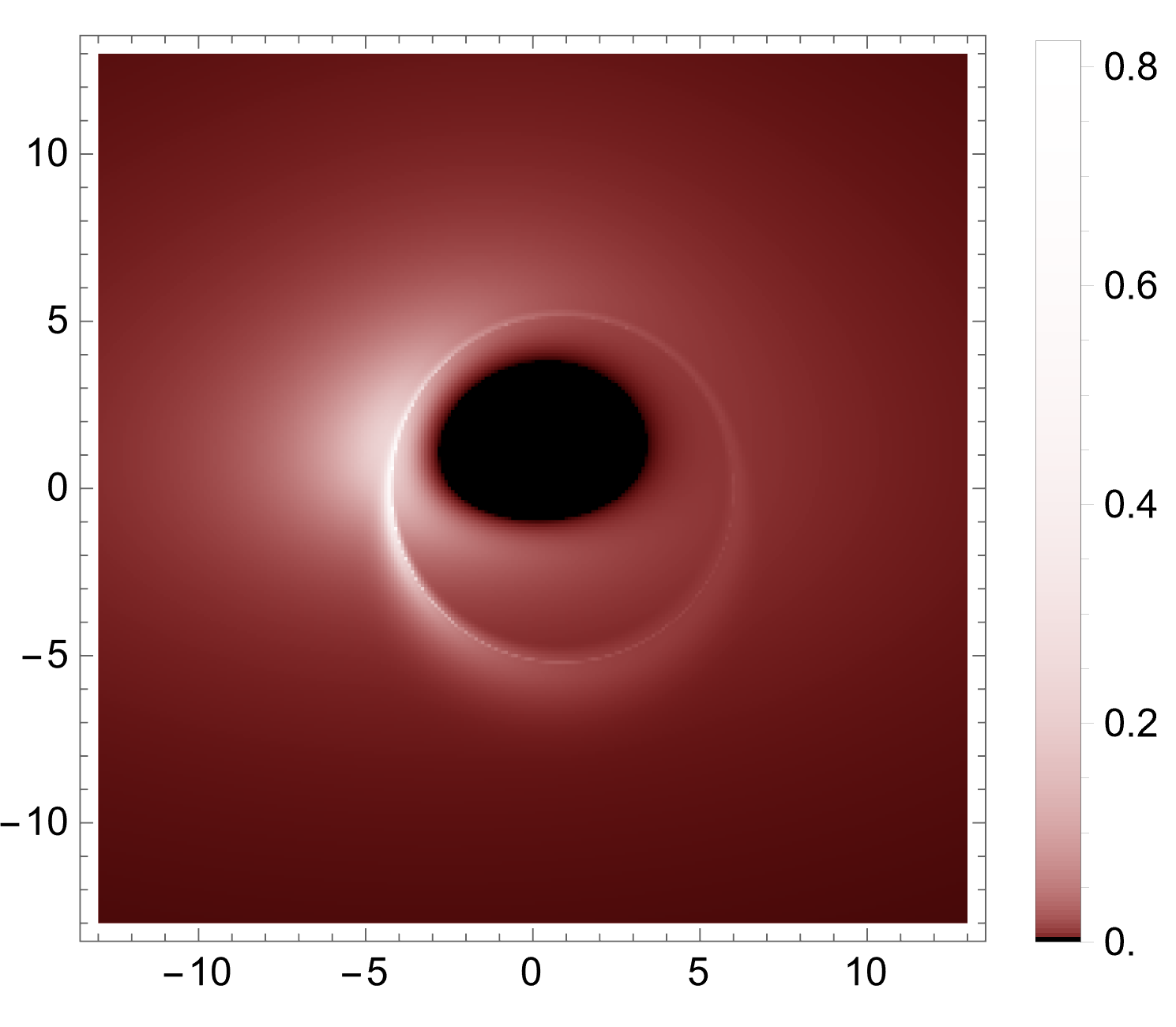}
		\caption{$\theta_{\rm obs}=60^\circ,g=1$}
	\end{subfigure}
	\begin{subfigure}[t]{0.32\textwidth}
		\centering
		\includegraphics[width=\textwidth]{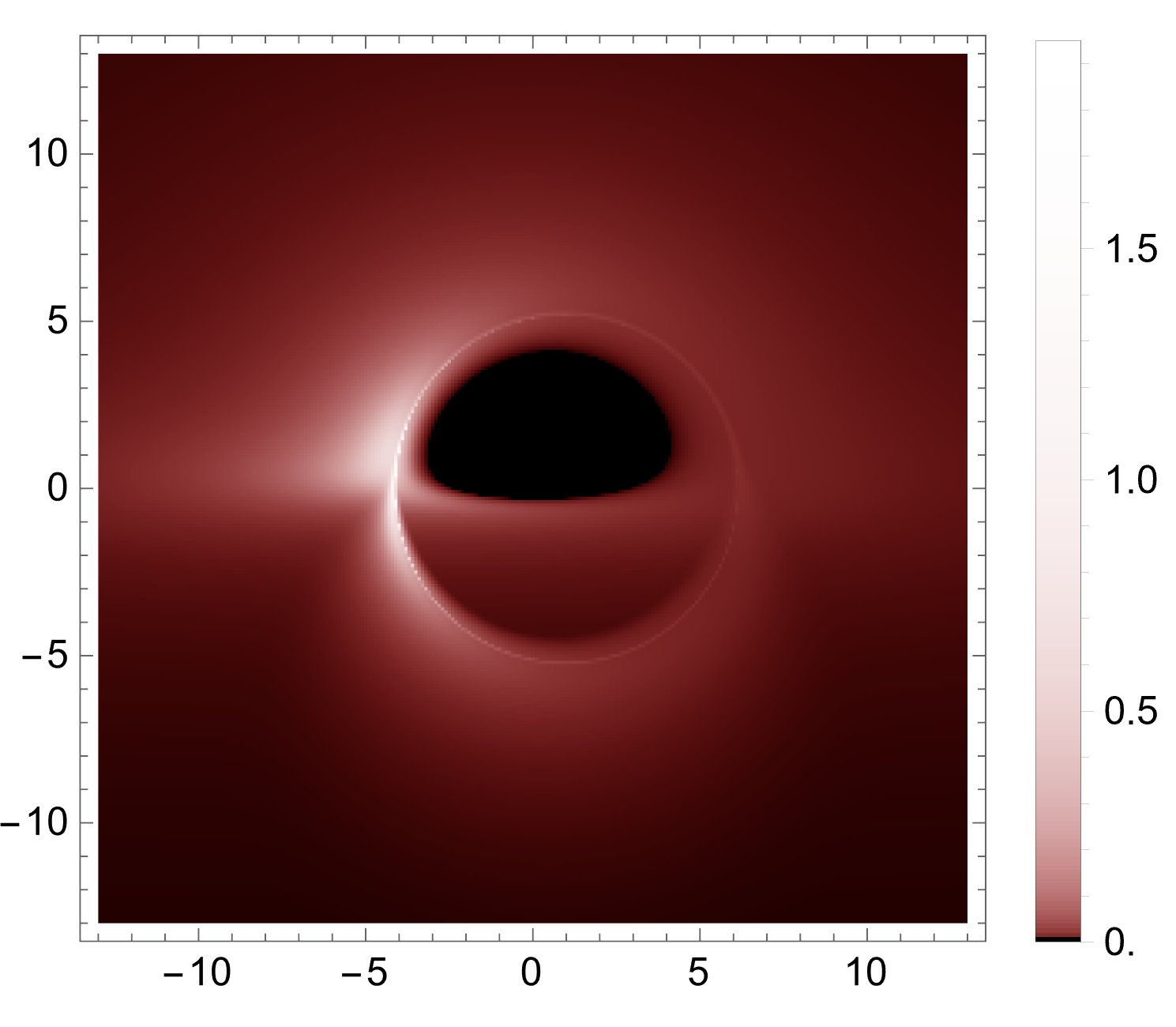}
		\caption{$\theta_{\rm obs}=80^\circ,g=1$}
	\end{subfigure}
	
	\vspace{0.2cm}
	
	\begin{subfigure}[t]{0.32\textwidth}
		\centering
		\includegraphics[width=\textwidth]{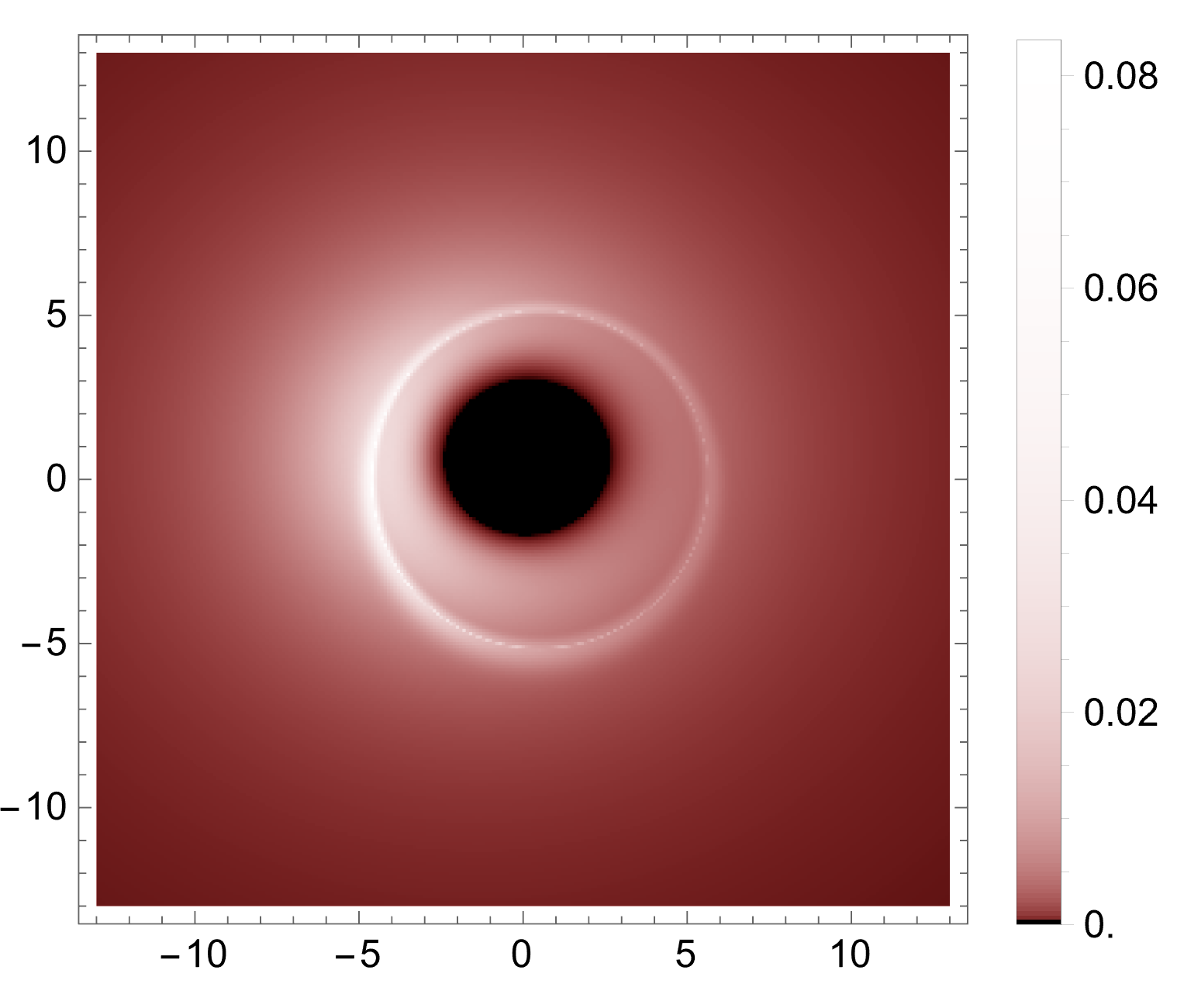}
		\caption{$\theta_{\rm obs}=30^\circ,g=1.5$}
	\end{subfigure}
	\begin{subfigure}[t]{0.32\textwidth}
		\centering
		\includegraphics[width=\textwidth]{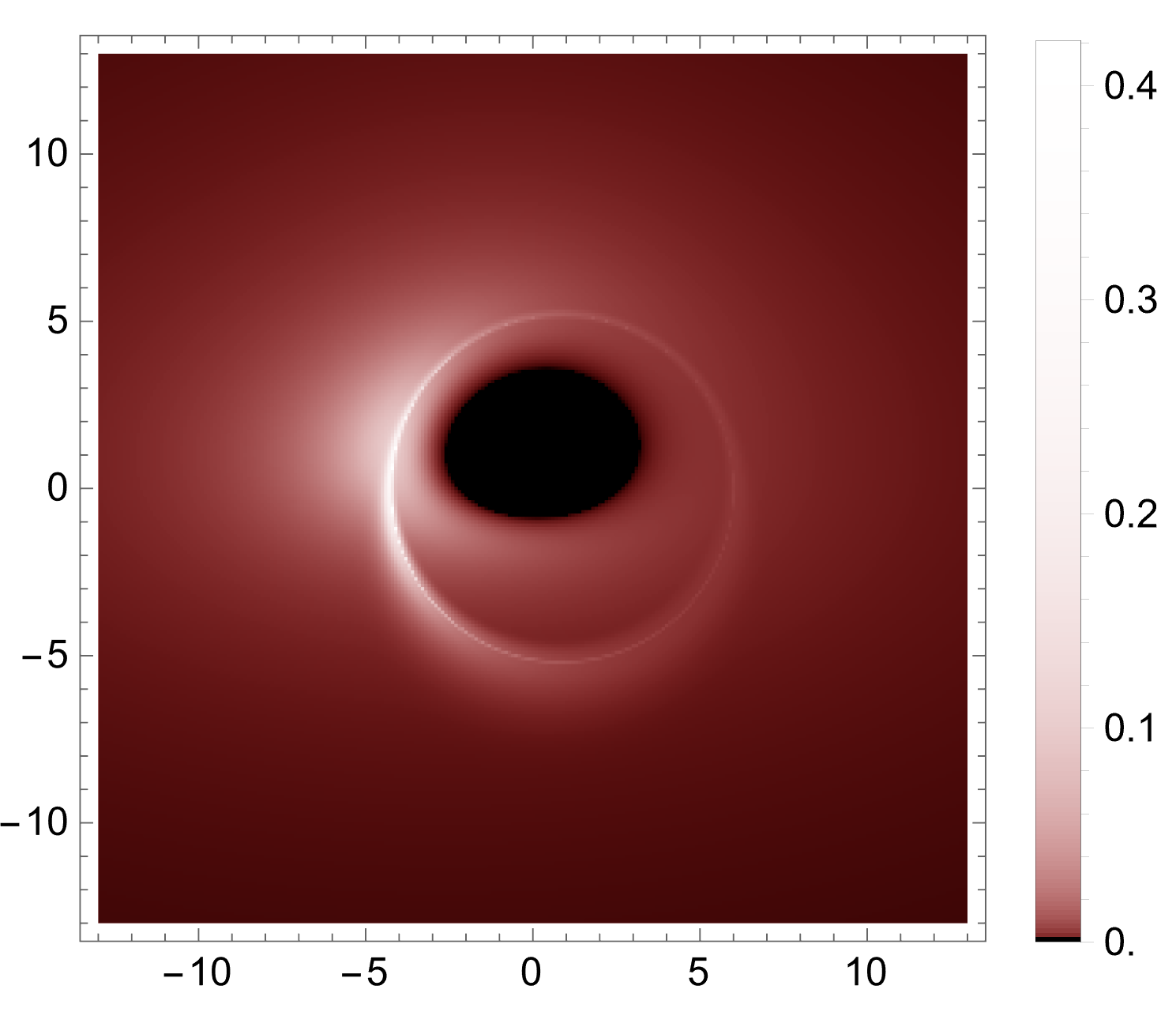}
		\caption{$\theta_{\rm obs}=60^\circ,g=1.5$}
	\end{subfigure}
	\begin{subfigure}[t]{0.32\textwidth}
		\centering
		\includegraphics[width=\textwidth]{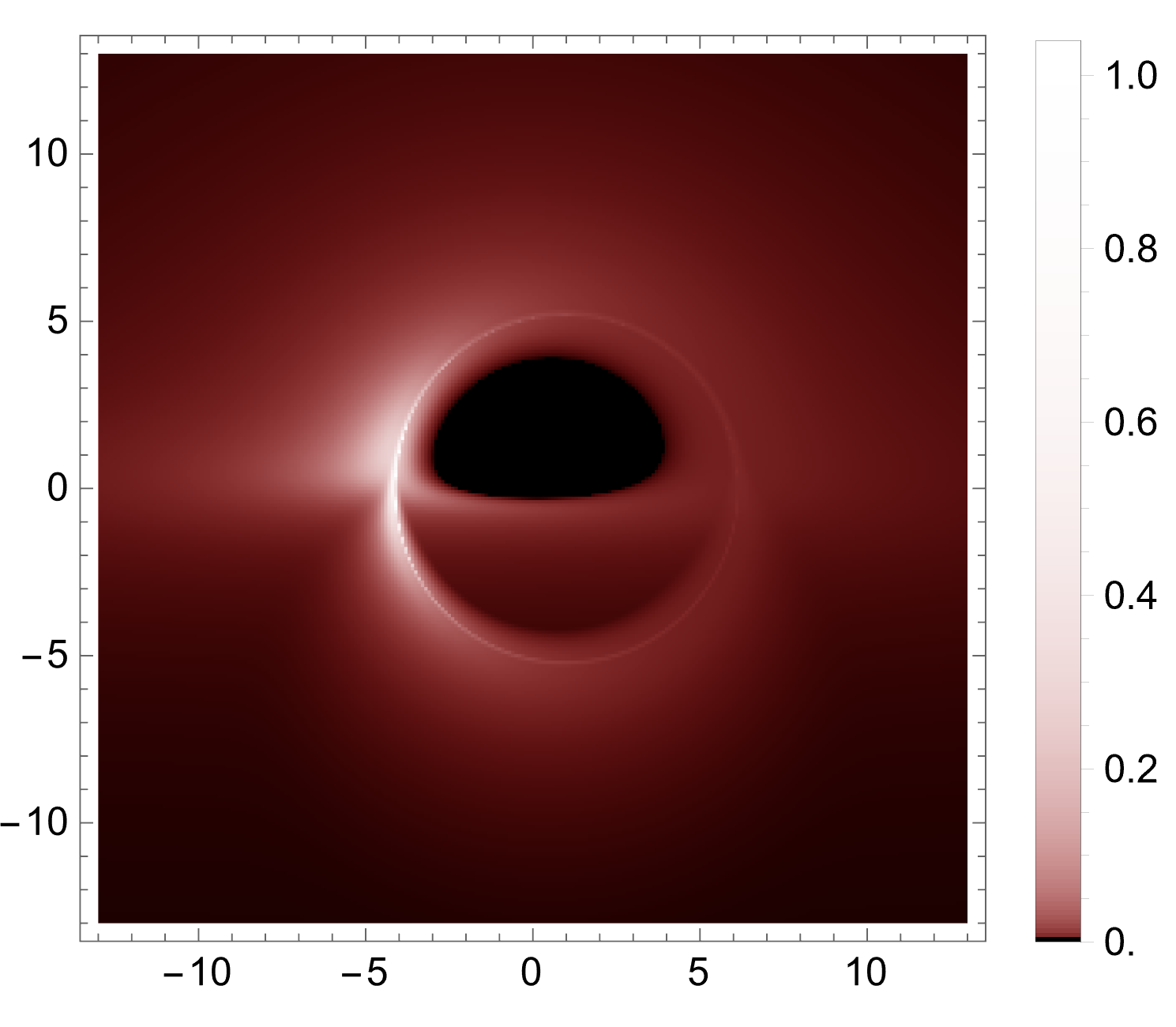}
		\caption{$\theta_{\rm obs}=80^\circ,g=1.5$}
		\end{subfigure}
		
		\caption{Optical appearance of the rotating SV black hole with a fixed parameter $a=0.5$.
			From left to right, the observational inclination angles $\theta_{\rm obs}$ are $30^\circ$, $60^\circ$, and $80^\circ$.
			From top to bottom, the parameter $g$ takes the values $0$ (Kerr case), $0.5$, $1$, and $1.5$.}
		\label{fig:appearance}
	\end{figure*}

	The observed specific intensity $I_{\rm obs}$ can be written as a sum over multiple intersections between photon trajectories and the accretion disk.
	Following the radiative GRMHD imaging prescription, it is given by \cite{Chael:2021rjo,He:2024amh,Hu:2023pyd}
	\begin{equation}
		I_{\rm obs}=\sum_{n=1}^{N_{\max}} f_n\, g_{n}^3\, j(r_e),
	\end{equation}
	where $n$ denotes the number of times a photon trajectory intersects the accretion disk.
	Since higher-order crossings contribute negligibly to the observed intensity, we truncate the sum at $N_{\max}=3$.
	Specifically, $n=1$ corresponds to the direct image, $n=2$ to the lensed image, and $n=3$ to the higher order image.
	The coefficients $f_n$ are phenomenological fudge factors accounting for the relative contributions of different image orders.
	In this work, we adopt $f_1=1, f_2=f_3=\frac{2}{3}$. 
    The function $j(r_e)$ denotes the emissivity of the accretion disk and depends on the radial coordinate $r_e$ of the emission point.
	The emissivity can be expressed as \cite{Chael:2021rjo}
	\begin{equation}
		j(r_e)=\exp\left(p_1 z+p_2 z^2\right),
		\label{j}
	\end{equation}
	where $z = \log(r_{e}/r_{+})$ and for $230\,\mathrm{GHz}$ imaging simulations, the parameters are chosen as $p_1=-2$ and $p_2=-\tfrac{1}{2}$.
	Finally, $g_{n}$ is the redshift factor relating the emitted frequency to the observed frequency,
	\begin{equation}
		g_{n}=\frac{\nu_{\rm obs}}{\nu_e}
		=\frac{(p_\mu u^\mu)_{\rm obs}}{(p_\mu u^\mu)_e},
	\end{equation}
	where $p^\mu$ is the photon four-momentum, and $u^\mu$ denotes the four-velocity of the observer and the emitting matter in the accretion disk, respectively.
	
It should be emphasized that the redshift factor takes different forms inside and outside the innermost stable circular orbit.
	For photon emission originating from the region $r>r_{\rm ISCO}$, we assume that the accretion flow follows stable circular orbits in the equatorial plane. Consequently, the redshift factor is given by
	\begin{equation}
		g_{n} = g_{\rm out}
		=\frac{p_t}{u^t\left(p_t+\Omega p_\phi\right)} .
	\end{equation}
	In the previous section, the redshift factor appearing in the spectral luminosity formula was approximated by $g_{\rm out}$, which is sufficient for computing the spectral luminosity, as the dominant contribution arises from regions outside the ISCO.
	In contrast, for emission arising from the plunging region $r_+<r<r_{\rm ISCO}$, the accretion flow no longer follows circular motion. Following the standard approach in \cite{Liu:2025hhg,Hou:2022eev}, we assume that the specific energy and specific angular momentum of the particles are conserved during the fall, maintaining their values at the ISCO, i.e., $E=E_{\rm ISCO}$ and $L=L_{\rm ISCO}$. In this region, the redshift factor is evaluated as
	\begin{equation}
		g_{n} = g_{\rm in}
		=\frac{p_t}{u^t_{\rm in} p_t + u^r_{\rm in} p_r + u^\phi_{\rm in} p_\phi},
	\end{equation}
	where $u^t_{\rm in}$, $u^r_{\rm in}$, and $u^\phi_{\rm in}$ denote the four-velocity components of the infalling matter, in which the parameters $E$ and $L$ are replaced by $E_{\rm ISCO}$ and $L_{\rm ISCO}$, respectively.
	With these ingredients, the black hole image on the observer’s screen can be constructed.
	
	We fix the parameter to $a=0.5$ and present the optical appearance of the rotating SV black hole for different values of the parameter $g$ and different observer inclination angles in Fig.~\ref{fig:appearance}. 
	In each image, a prominent bright photon ring appears as a narrow luminous curve, which is a direct consequence of strong gravitational lensing in the vicinity of the black hole. 
	In addition, a central dark region is also clearly obvious in each image, corresponding to photons that fall into the event horizon in the backward ray-tracing method. 
	This feature represents the direct imaging of the event horizon and is commonly referred as the shadow of the black hole.  
	The three images in the first row correspond to the Kerr black hole case with $g=0$ and different observation angles of $30^\circ$, $60^\circ$, and $80^\circ$, while the subsequent rows from top to bottom correspond to increasing values of the parameter $g=0.5$, $1$, and $1.5$, respectively.  
	As $g$ increases, the size of the inner shadow decreases noticeably and the overall brightness of the images is suppressed, indicating that the regularization parameter modifies the near-horizon spacetime geometry.

	\begin{figure*}[tb]
		\centering
		\begin{subfigure}[t]{0.32\textwidth}
			\centering
			\includegraphics[width=\textwidth]{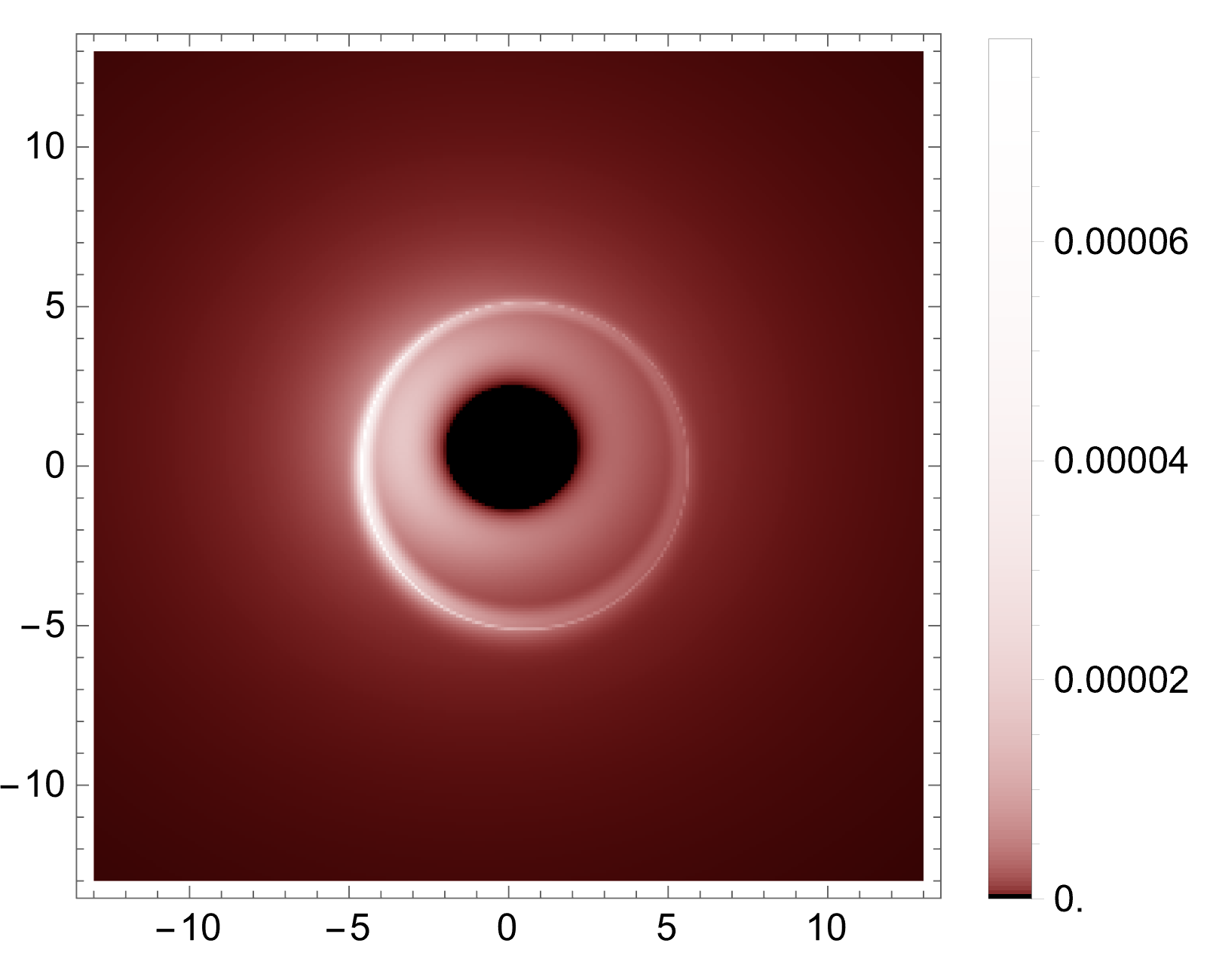}
			\caption{$\theta_{\rm obs}=30^\circ$}
		\end{subfigure}
		\begin{subfigure}[t]{0.32\textwidth}
			\centering
			\includegraphics[width=\textwidth]{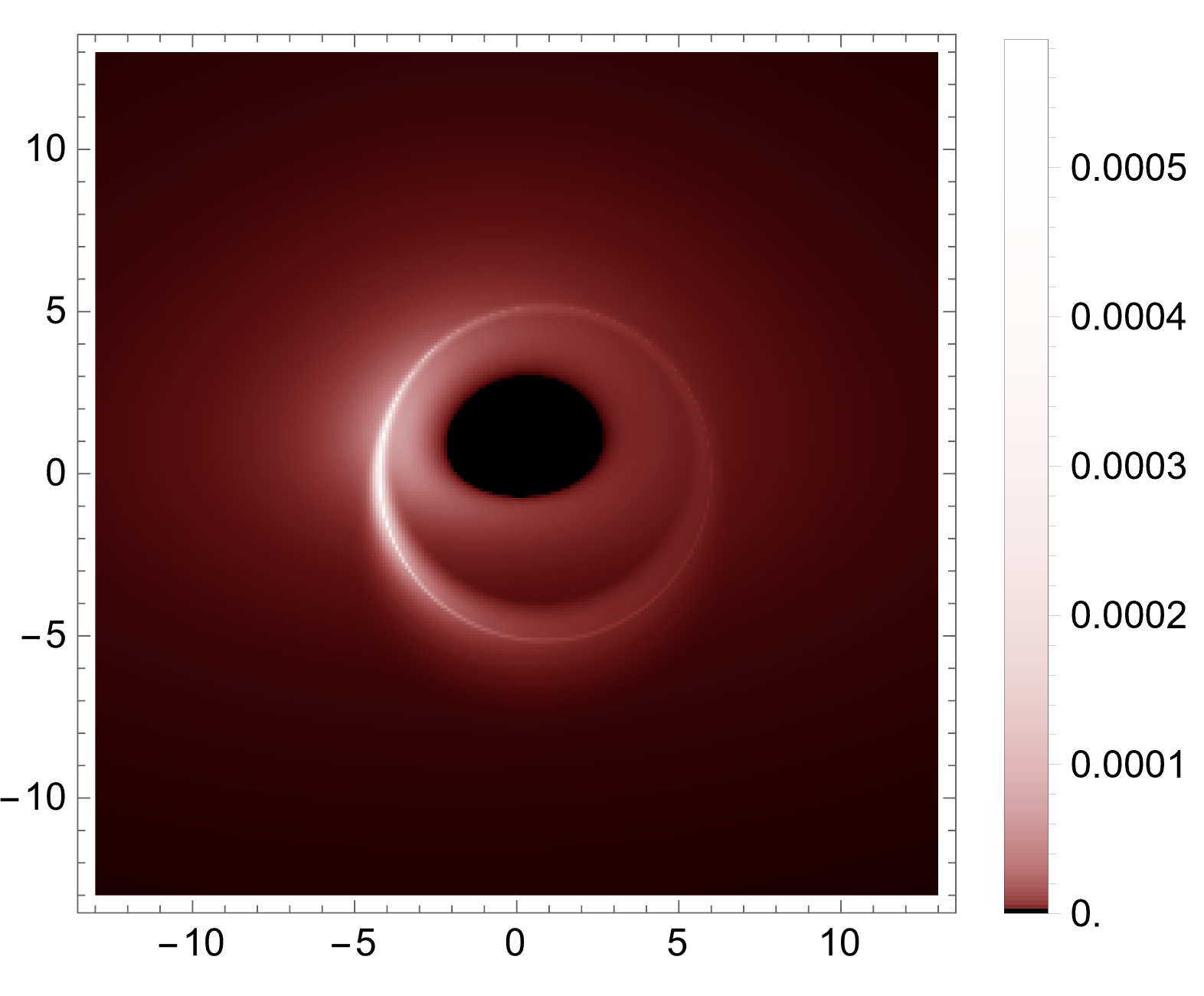}
			\caption{$\theta_{\rm obs}=60^\circ$}
		\end{subfigure}
		\begin{subfigure}[t]{0.32\textwidth}
			\centering
			\includegraphics[width=\textwidth]{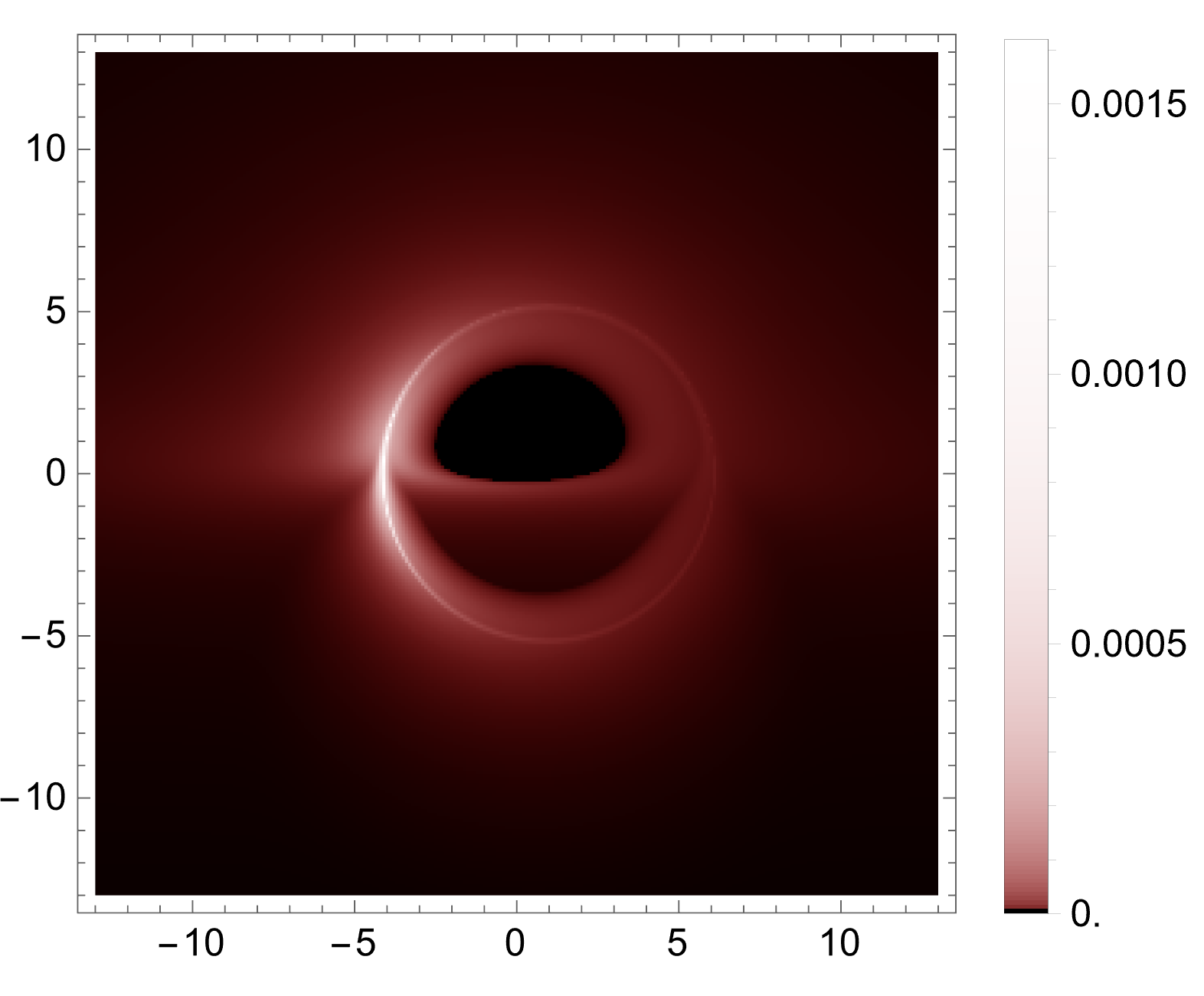}
			\caption{$\theta_{\rm obs}=80^\circ$}
		\end{subfigure}
		
		\caption{Optical appearance of the rotating SV black hole with $a=0.5$ and $g=1.86$ for different observational inclination angle $\theta_{\rm obs}$.
			The overall intensity is significantly suppressed compared with the cases shown in Fig.~\ref{fig:appearance}, while the photon ring and inner shadow structures remain clear.}
		\label{fig:largeg}
	\end{figure*}	

		It is worth emphasizing that the range of the parameter $g$ considered in Fig.~\ref{fig:appearance} is deliberately restricted to $g\leq1.5$.
	As shown in Fig.~\ref{fig:horizon}, for a fixed parameter $a=0.5$, the event horizon radius decreases rapidly once $g\gtrsim1.5$.
	As a consequence, the emission radius relevant for the radiative transfer approaches smaller values, where the emissivity $j(r_{e})$ given by Eq.~(\ref{j}) becomes strongly suppressed. This leads to a substantial reduction of the observed intensity $I_{\rm obs}$.
	
	For completeness, we have computed the black hole images for the case $a=0.5$ and $g=1.86$, which corresponds to the maximal value of $g$ allowed by the regular black hole condition. The resulting images for inclination angles $\theta_{\rm obs}=30^\circ$, $60^\circ$, and $80^\circ$ are shown in Fig.~\ref{fig:largeg}.
	Although the overall intensity is reduced by several orders of magnitude compared with the cases displayed in Fig.~\ref{fig:appearance}, the qualitative image structure, including the photon ring and the inner shadow, remains intact. To avoid confusion caused by the drastic change in the chosen intensity model, we therefore restrict our detailed analysis to $g\leq1.5$ in the following.

\begin{figure*}[tb]
		\centering
		\begin{subfigure}[t]{0.32\textwidth}
			\centering
			\includegraphics[width=\textwidth]{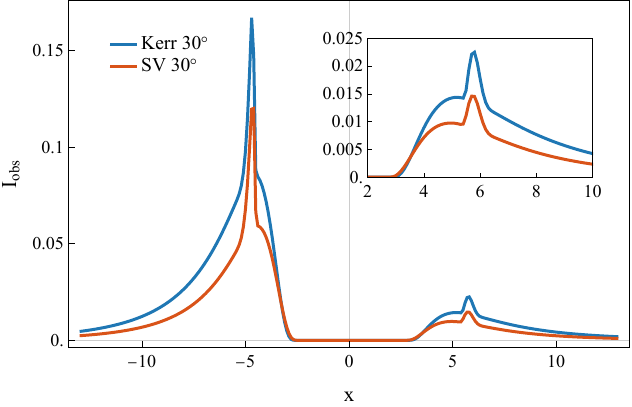}
			\caption{$x$-axis, $\theta_{\rm obs}=30^\circ$}
		\end{subfigure}
		\begin{subfigure}[t]{0.32\textwidth}
			\centering
			\includegraphics[width=\textwidth]{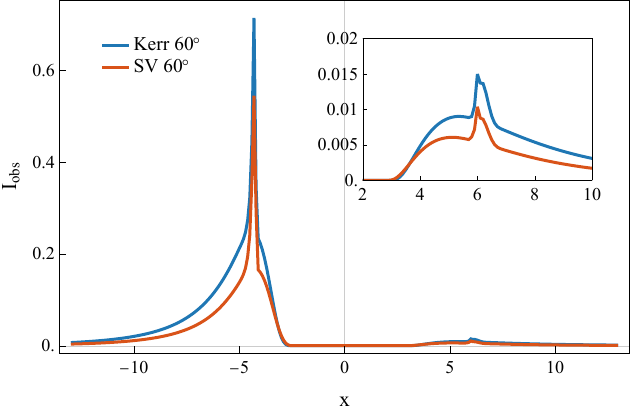}
			\caption{$x$-axis, $\theta_{\rm obs}=60^\circ$}
		\end{subfigure}
		\begin{subfigure}[t]{0.32\textwidth}
			\centering
			\includegraphics[width=\textwidth]{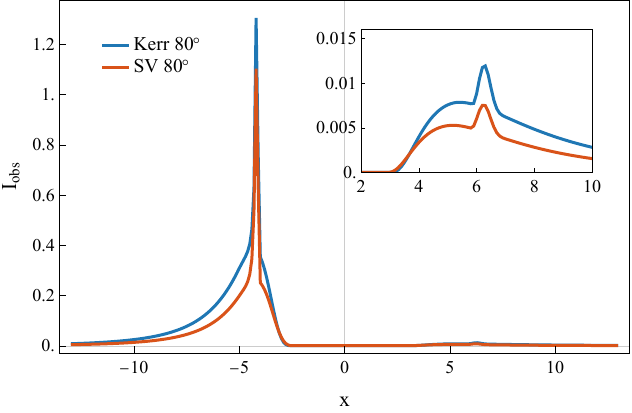}
			\caption{$x$-axis, $\theta_{\rm obs}=80^\circ$}
		\end{subfigure}
		
		\vspace{0.2cm}
		
		\begin{subfigure}[t]{0.32\textwidth}
			\centering
			\includegraphics[width=\textwidth]{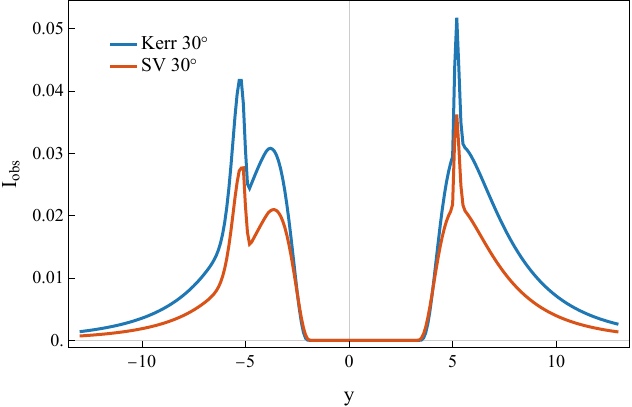}
			\caption{$y$-axis, $\theta_{\rm obs}=30^\circ$}
		\end{subfigure}
		\begin{subfigure}[t]{0.32\textwidth}
			\centering
			\includegraphics[width=\textwidth]{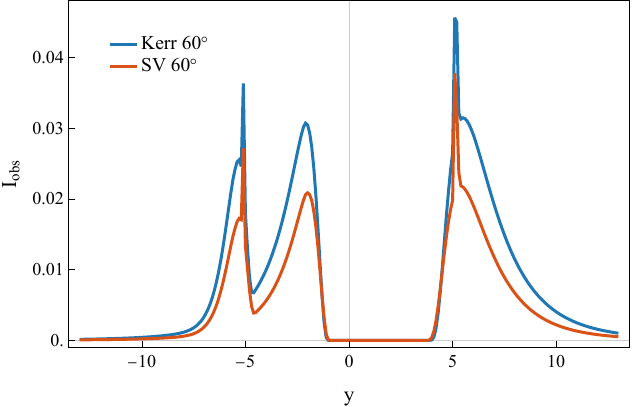}
			\caption{$y$-axis, $\theta_{\rm obs}=60^\circ$}
		\end{subfigure}
		\begin{subfigure}[t]{0.32\textwidth}
			\centering
			\includegraphics[width=\textwidth]{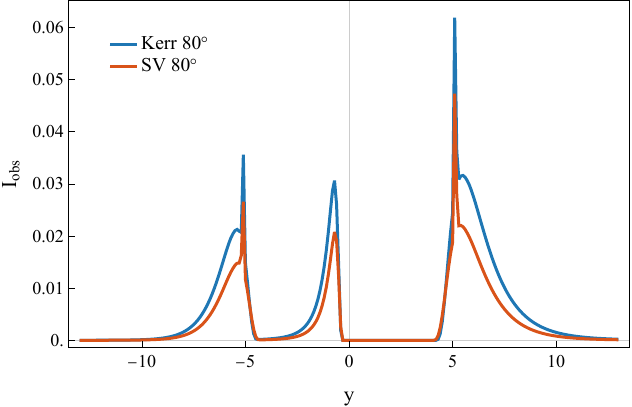}
			\caption{$y$-axis, $\theta_{\rm obs}=80^\circ$}
		\end{subfigure}
		
		\caption{Observed intensity profiles on the observer’s screen along the $x$- and $y$-axes for a Kerr black hole ($a=0.5,\ g=0$) and a rotating SV black hole ($a=0.5,\ g=1$) at different observational inclination angles.
			The top row shows the intensity distributions along the $x$-axis, while the bottom row corresponds to that along the $y$-axis.}
		\label{fig:intensity_profile}
	\end{figure*}

	Since the optical appearance images only provide a qualitative indication of how the parameter $g$ affects the black hole silhouette, we further examine the influence of $g$ on the observed intensity in a more quantitative manner by plotting the intensity profiles on the observer’s screen. In Fig.~\ref{fig:intensity_profile}, 
	we consider the intensity distributions along the $x$- and $y$-axes for a Kerr black hole with $a=0.5$ and for a rotating SV black hole with $a=0.5$ and $g=1$.  
	Along the $x$-axis, the observed intensity always exhibits two prominent peaks. As the inclination angle increases, the left peak becomes enhanced while the right peak is suppressed, leading to an increasing asymmetry between the two peaks. This phenomenon is mainly caused by the Doppler effect, which becomes dominant at large inclination angles. Moreover, the presence of the parameter $g$ reduces the observed intensity at almost all screen points, with significantly suppressed peak values and a modest reduction of the inner shadow size compared with the Kerr case.  
	Along the $y$-axis, three distinct intensity peaks gradually emerge as the inclination angle increases. The peak intensities are substantially reduced relative to the Kerr black hole case by the parameter $g$.

\begin{figure*}[tb]
		\centering
		\begin{subfigure}[t]{0.32\textwidth}
			\centering
			\includegraphics[width=\textwidth]{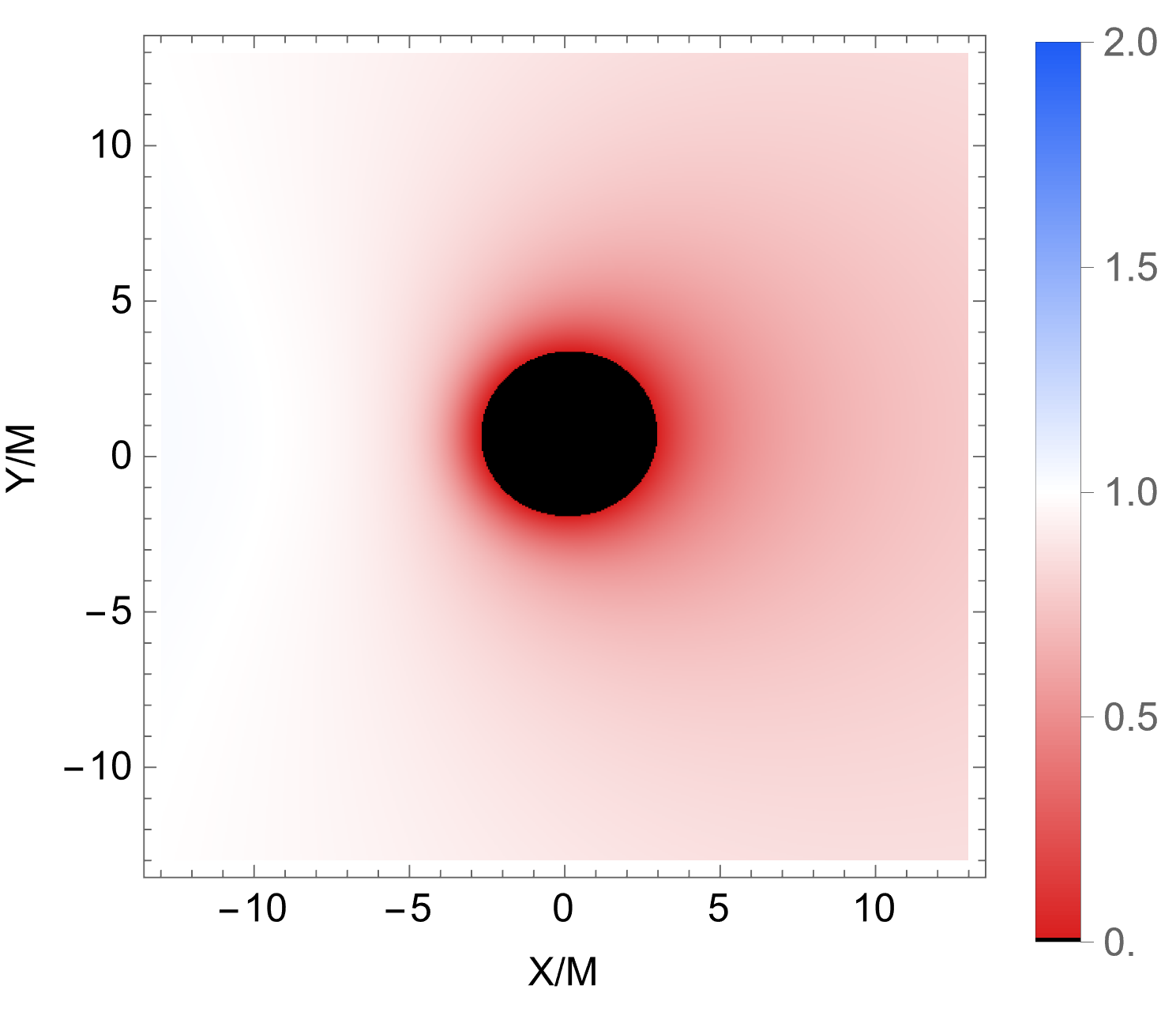}
			\caption{$\theta_{\rm obs}=30^\circ,g=0.5$}
		\end{subfigure}
		\begin{subfigure}[t]{0.32\textwidth}
			\centering
			\includegraphics[width=\textwidth]{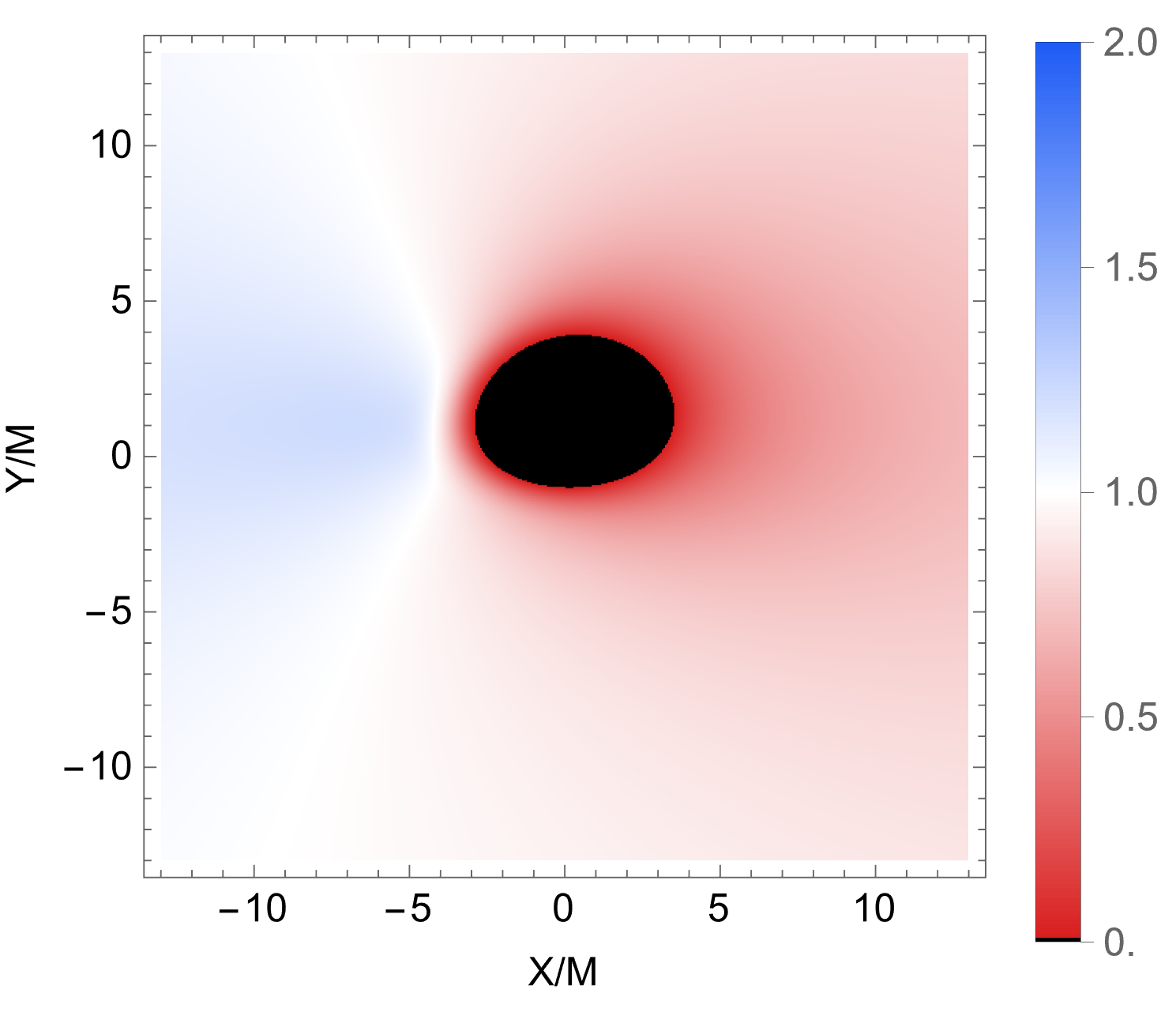}
			\caption{$\theta_{\rm obs}=60^\circ,g=0.5$}
		\end{subfigure}
		\begin{subfigure}[t]{0.32\textwidth}
			\centering
			\includegraphics[width=\textwidth]{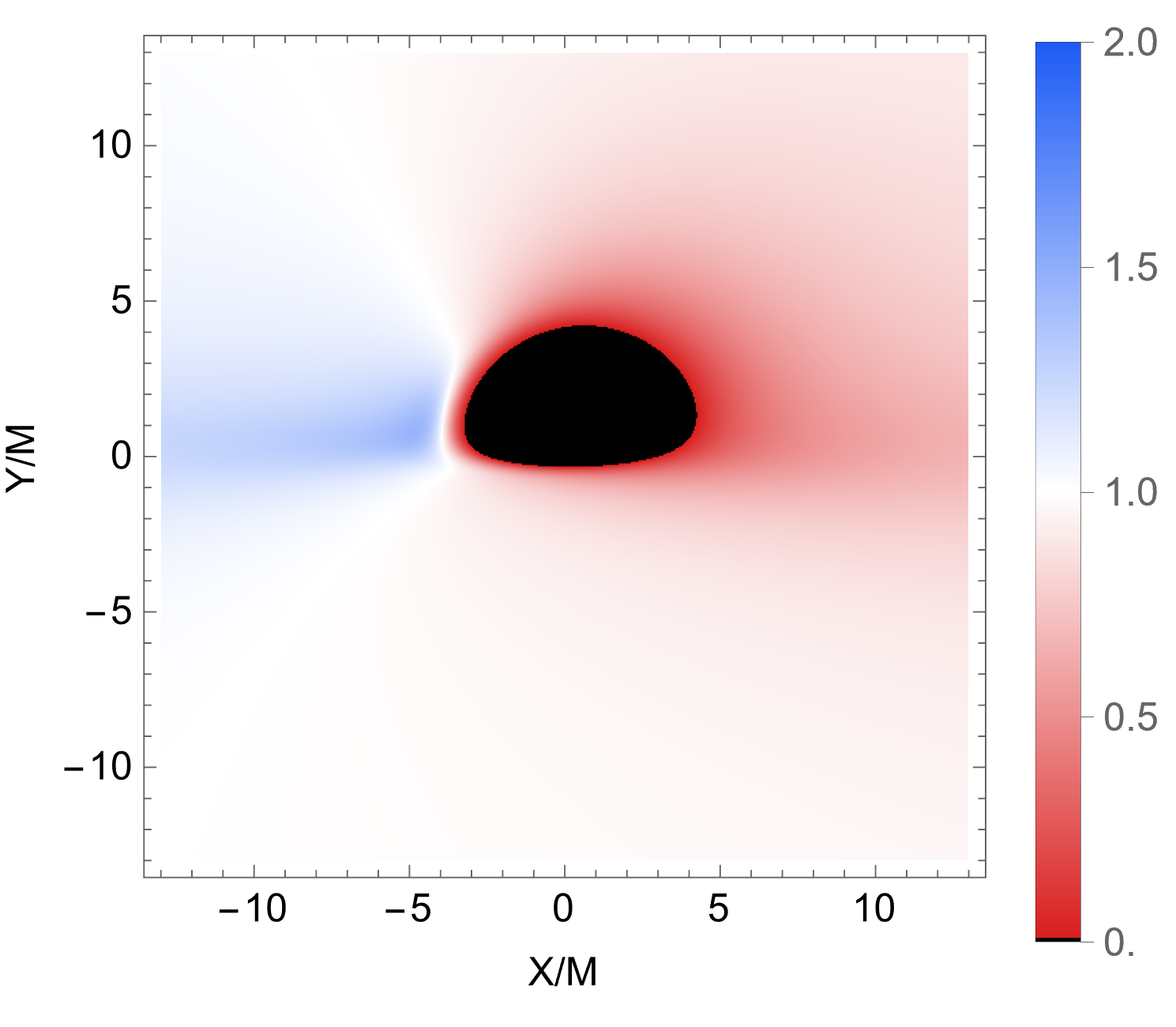}
			\caption{$\theta_{\rm obs}=80^\circ,g=0.5$}
		\end{subfigure}
		
		\vspace{0.2cm}
		
		\begin{subfigure}[t]{0.32\textwidth}
			\centering
			\includegraphics[width=\textwidth]{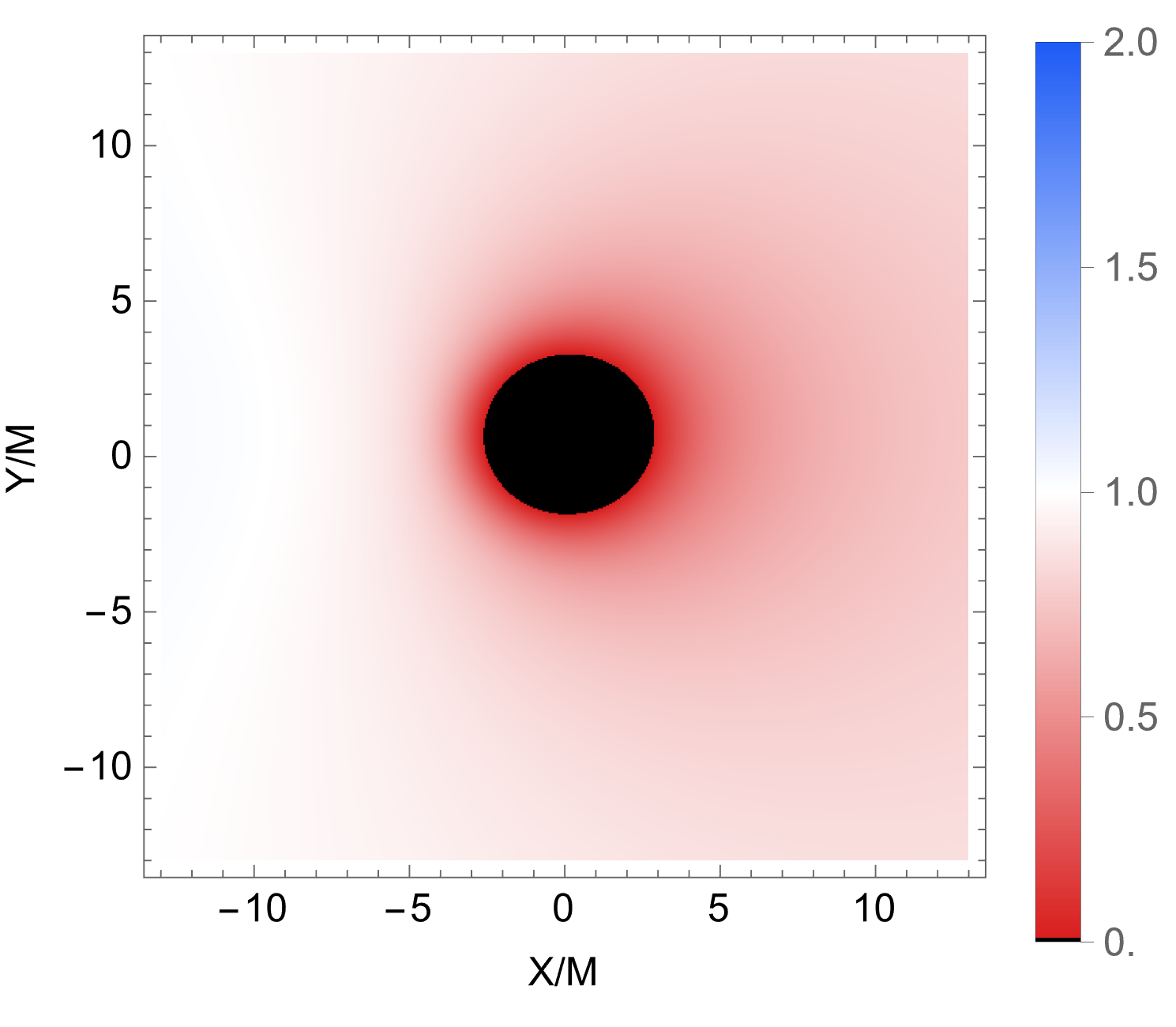}
			\caption{$\theta_{\rm obs}=30^\circ,g=1$}
		\end{subfigure}
		\begin{subfigure}[t]{0.32\textwidth}
			\centering
			\includegraphics[width=\textwidth]{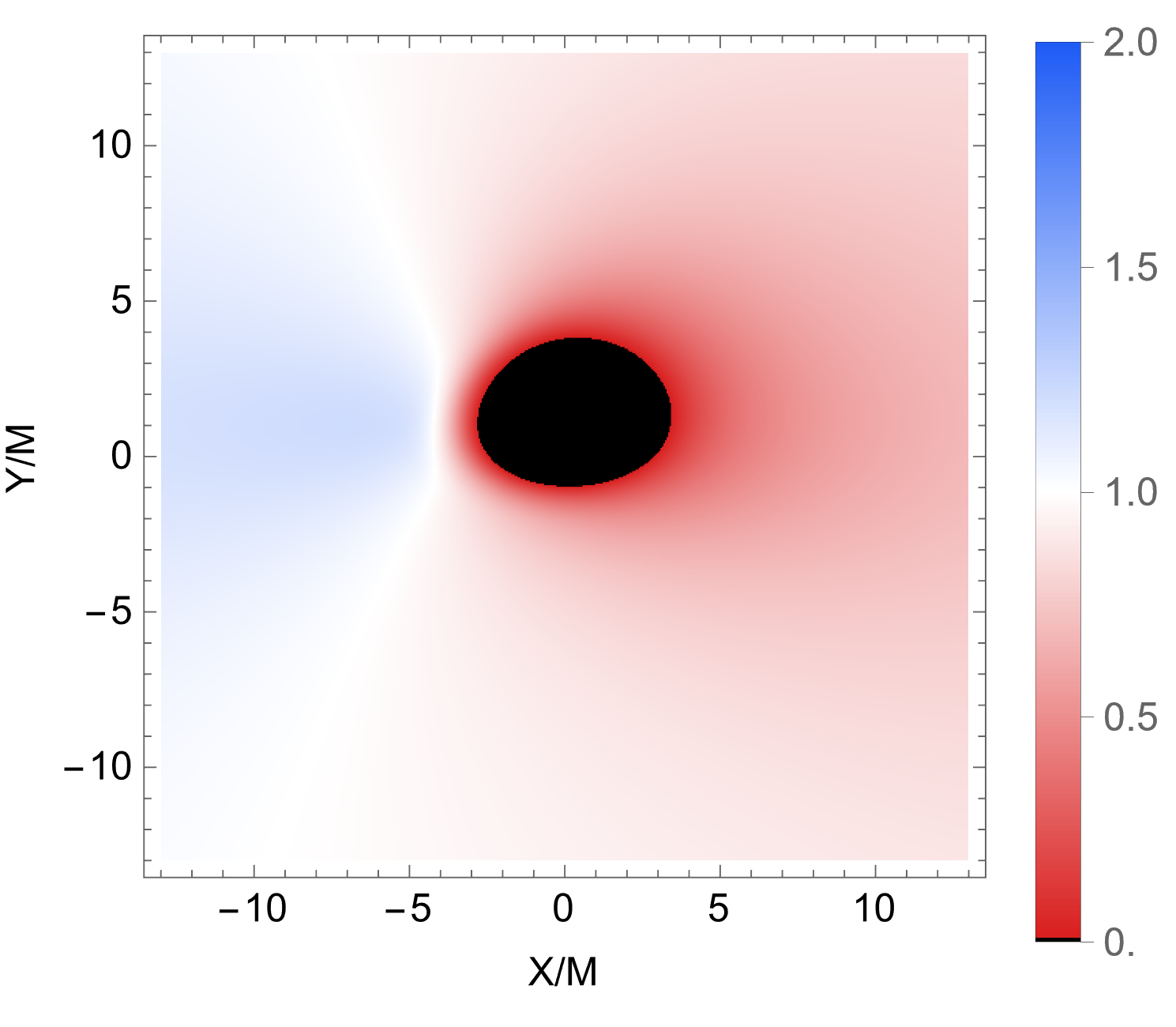}
			\caption{$\theta_{\rm obs}=60^\circ,g=1$}
		\end{subfigure}
		\begin{subfigure}[t]{0.32\textwidth}
			\centering
			\includegraphics[width=\textwidth]{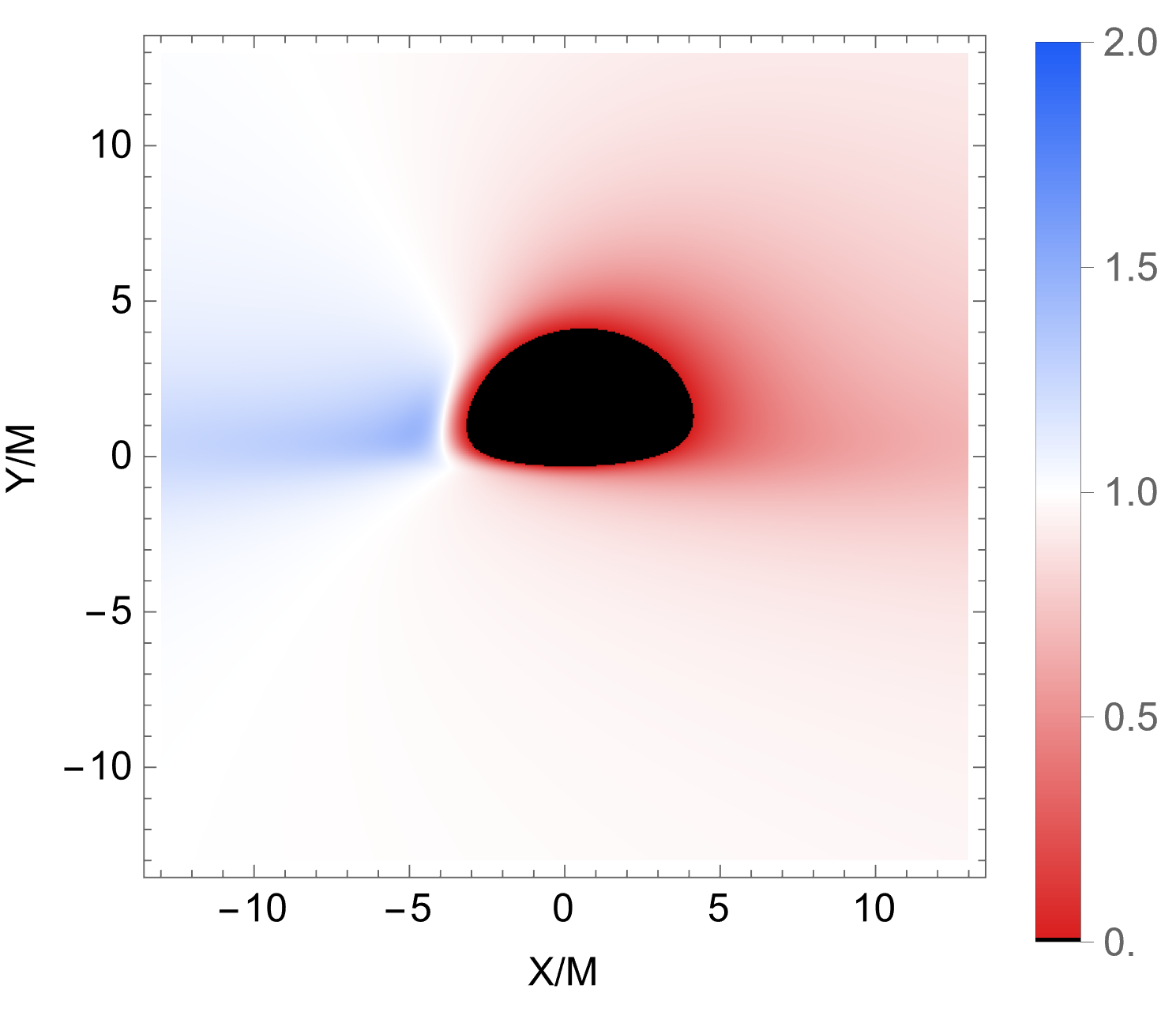}
			\caption{$\theta_{\rm obs}=80^\circ,g=1$}
		\end{subfigure}
		
		\vspace{0.2cm}
		
		\begin{subfigure}[t]{0.32\textwidth}
			\centering
			\includegraphics[width=\textwidth]{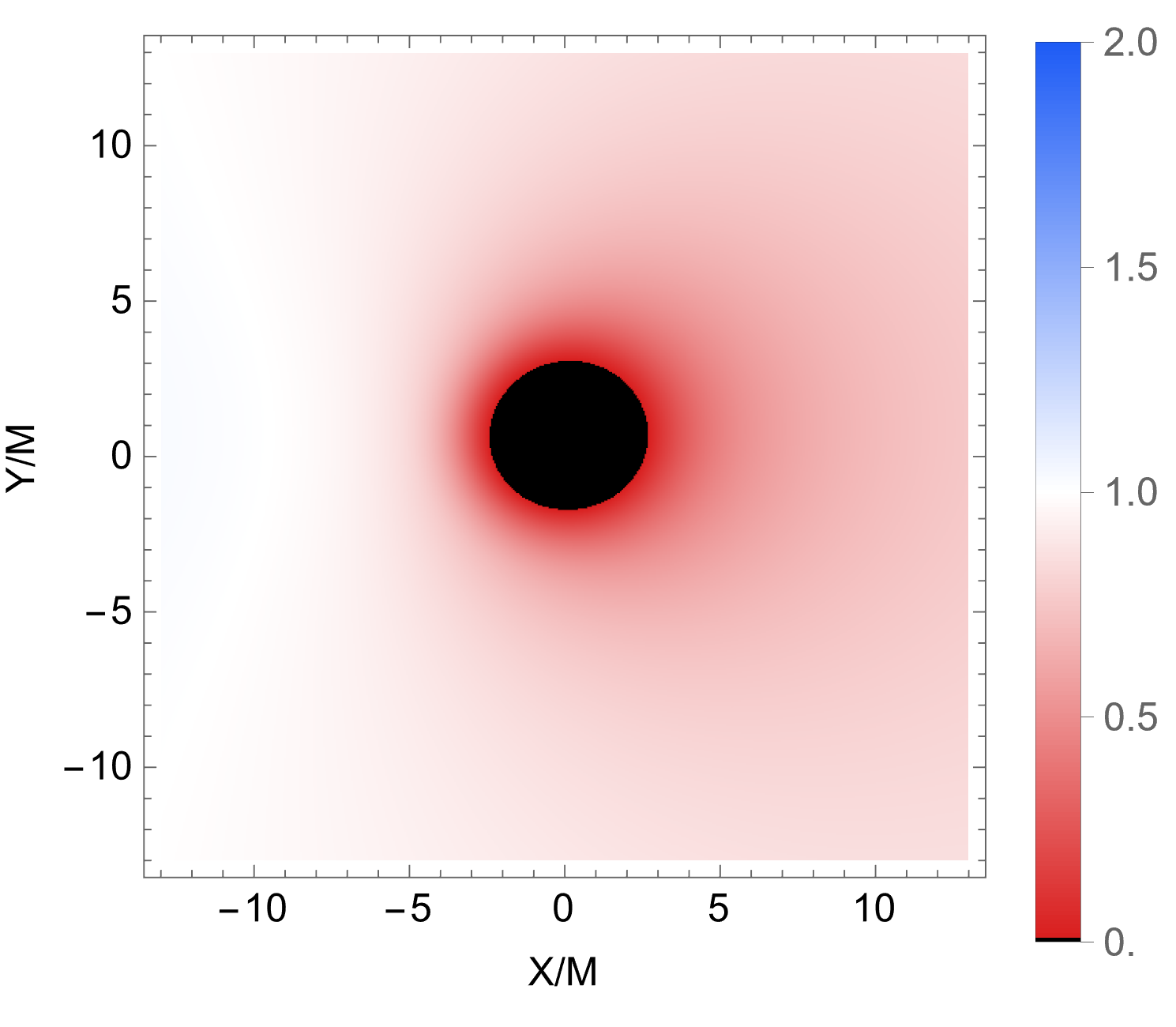}
			\caption{$\theta_{\rm obs}=30^\circ,g=1.5$}
		\end{subfigure}
		\begin{subfigure}[t]{0.32\textwidth}
			\centering
			\includegraphics[width=\textwidth]{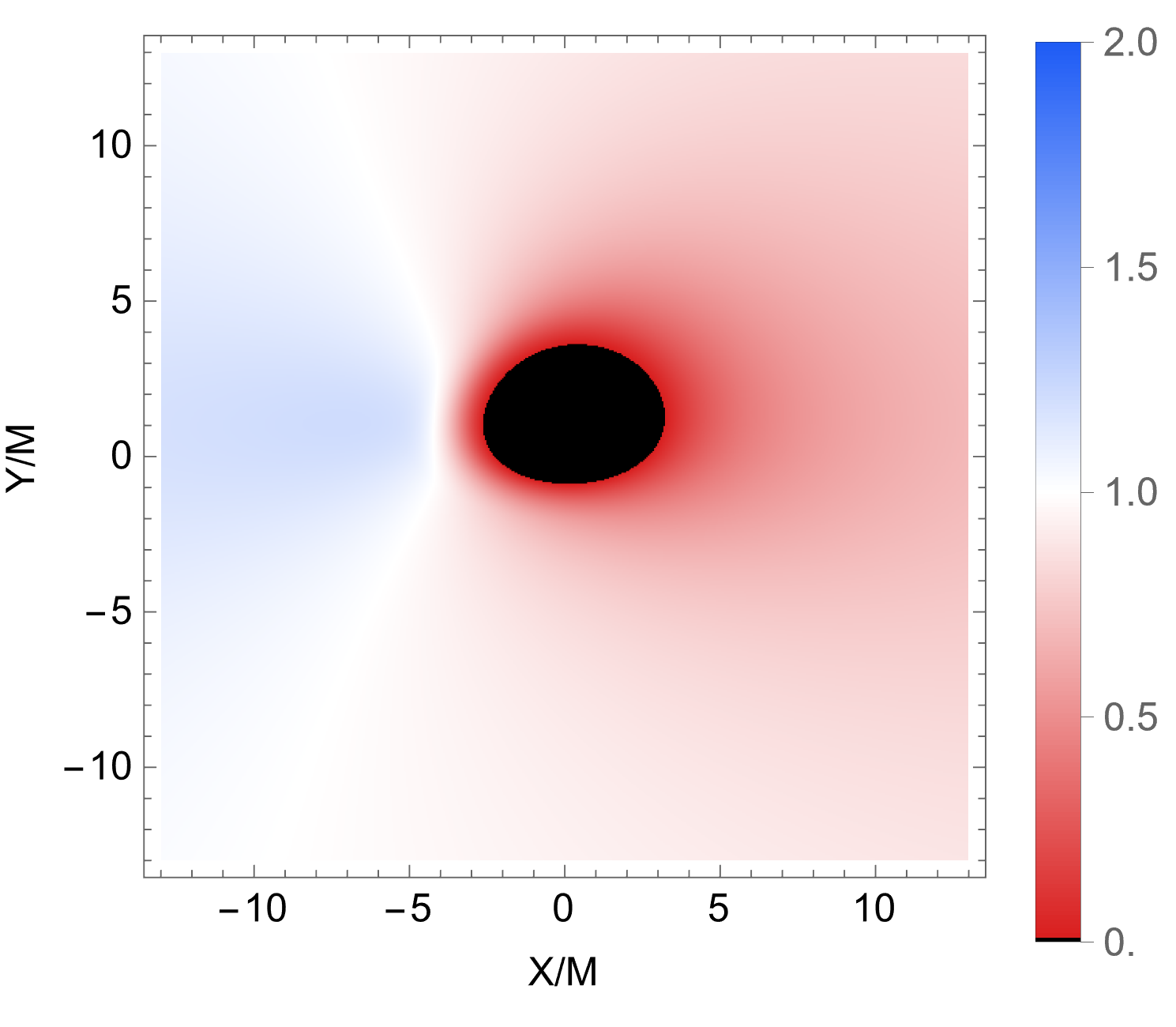}
			\caption{$\theta_{\rm obs}=60^\circ,g=1.5$}
		\end{subfigure}
		\begin{subfigure}[t]{0.32\textwidth}
			\centering
			\includegraphics[width=\textwidth]{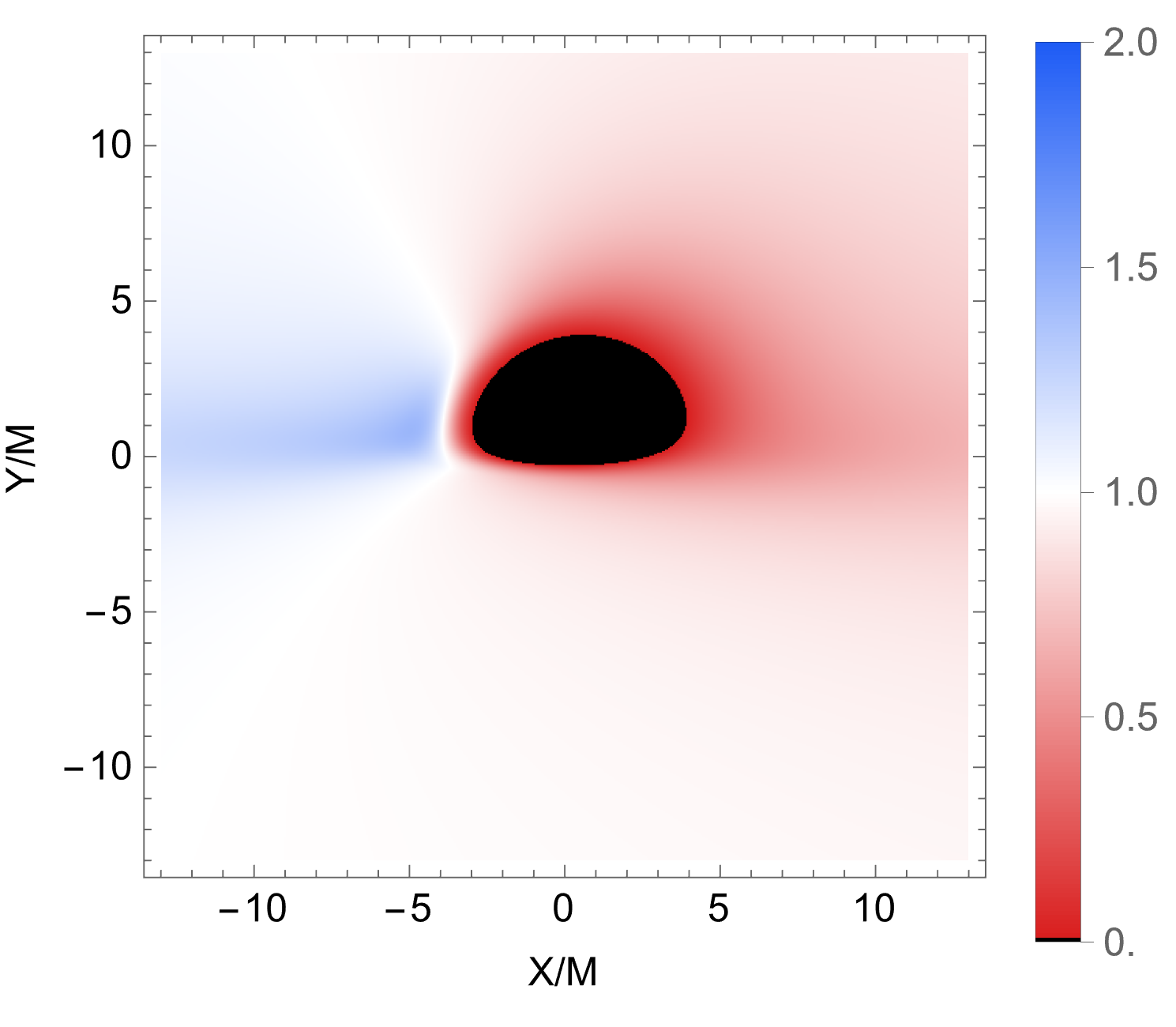}
			\caption{$\theta_{\rm obs}=80^\circ,g=1.5$}
		\end{subfigure}
		
		\caption{Distribution of the redshift factor associated with the direct image of a rotating SV black hole with $a=0.5$ for various observational inclination angles and parameters $g$.}
		\label{fig:redshift}
	\end{figure*}

	Given the crucial role played by the redshift factor in shaping the black hole images, we plot the distribution of the redshift factor associated with the direct image for the same set of parameters as in Fig.~\ref{fig:appearance}, excluding the Kerr case, and present the results in Fig.~\ref{fig:redshift}. 
	To visualize the redshift factor more intuitively, a linear color scale is adopted, where red denotes redshifted regions ($g<1$) and blue denotes blueshifted regions ($g>1$). 
	When the observer inclination angle is $\theta_{\rm obs}=30^\circ$, the entire screen is dominated by redshift, whereas for larger inclination angles $\theta_{\rm obs}=60^\circ$ and $80^\circ$, a blueshifted region emerges on the left side of the image, leading to an enhancement of the observed intensity, while the right side remains redshifted and correspondingly dimmer. 
	This behavior is consistent with the increasing asymmetry between the left and right intensity peaks discussed previously. 
	In addition, as the parameter $g$ increases, the overall distribution pattern of redshifted and blueshifted regions remains unchanged; however, the reduction of the inner shadow causes the redshifted area to expand. 
	Moreover, for $\theta_{\rm obs}=80^\circ$, as the increase of $g$, the blueshifted region becomes slightly stronger, resulting in the increase of the intensity contrast in the image.
	
\begin{figure*}[tb]
		\centering
		\begin{subfigure}[t]{0.32\textwidth}
			\centering
			\includegraphics[width=\textwidth]{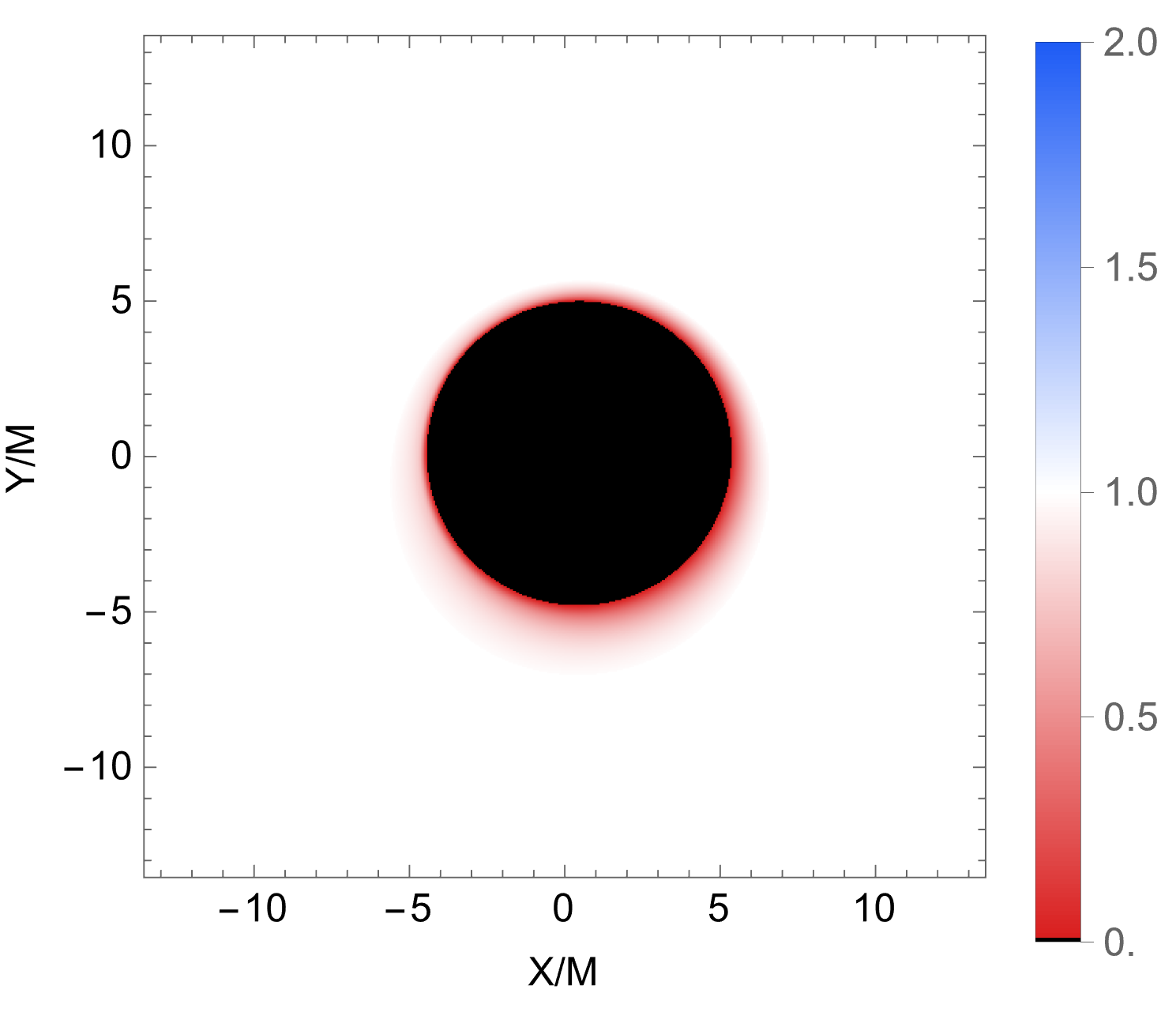}
			\caption{$\theta_{\rm obs}=30^\circ,g=0.5$}
		\end{subfigure}
		\begin{subfigure}[t]{0.32\textwidth}
			\centering
			\includegraphics[width=\textwidth]{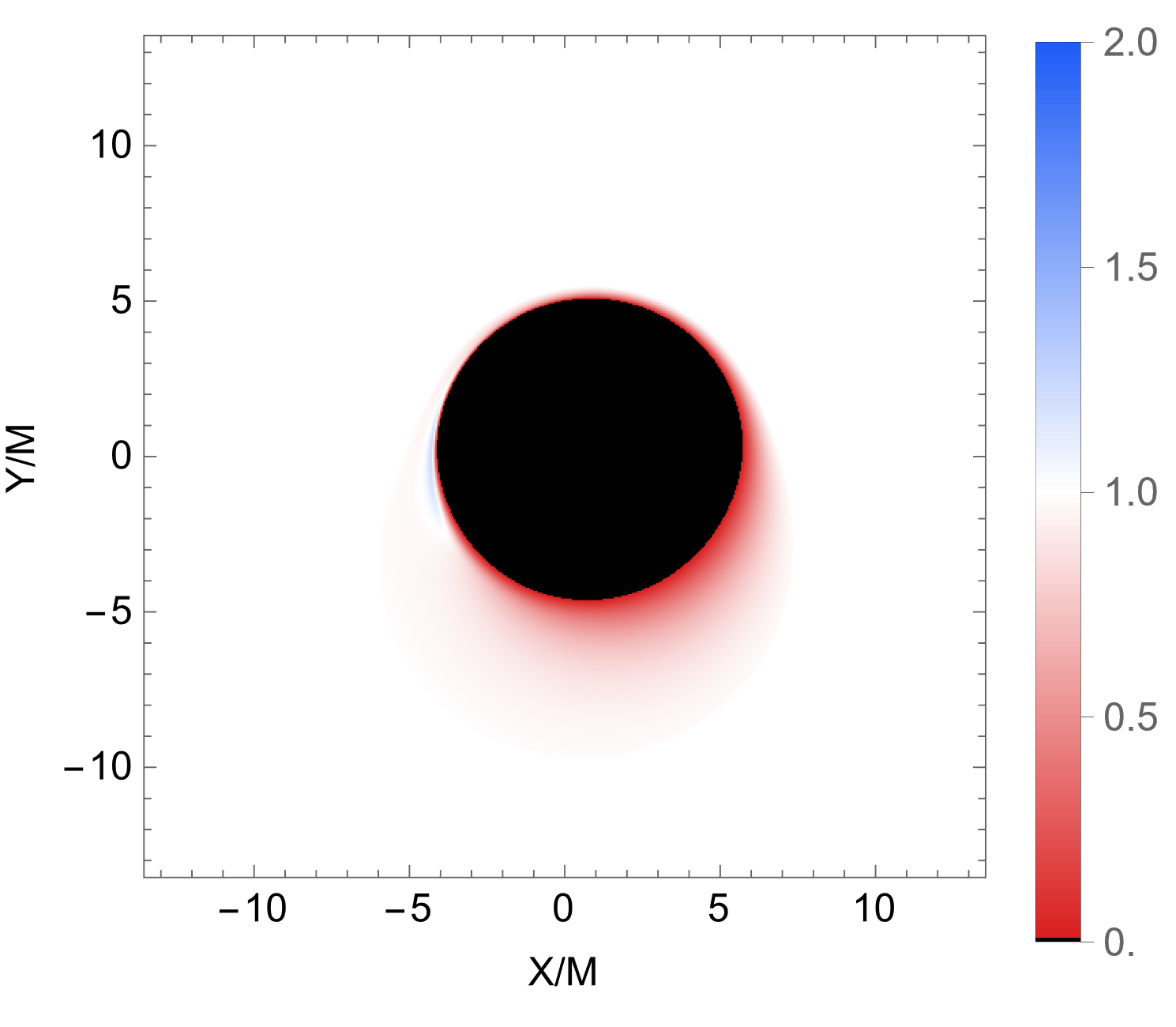}
			\caption{$\theta_{\rm obs}=60^\circ,g=0.5$}
		\end{subfigure}
		\begin{subfigure}[t]{0.32\textwidth}
			\centering
			\includegraphics[width=\textwidth]{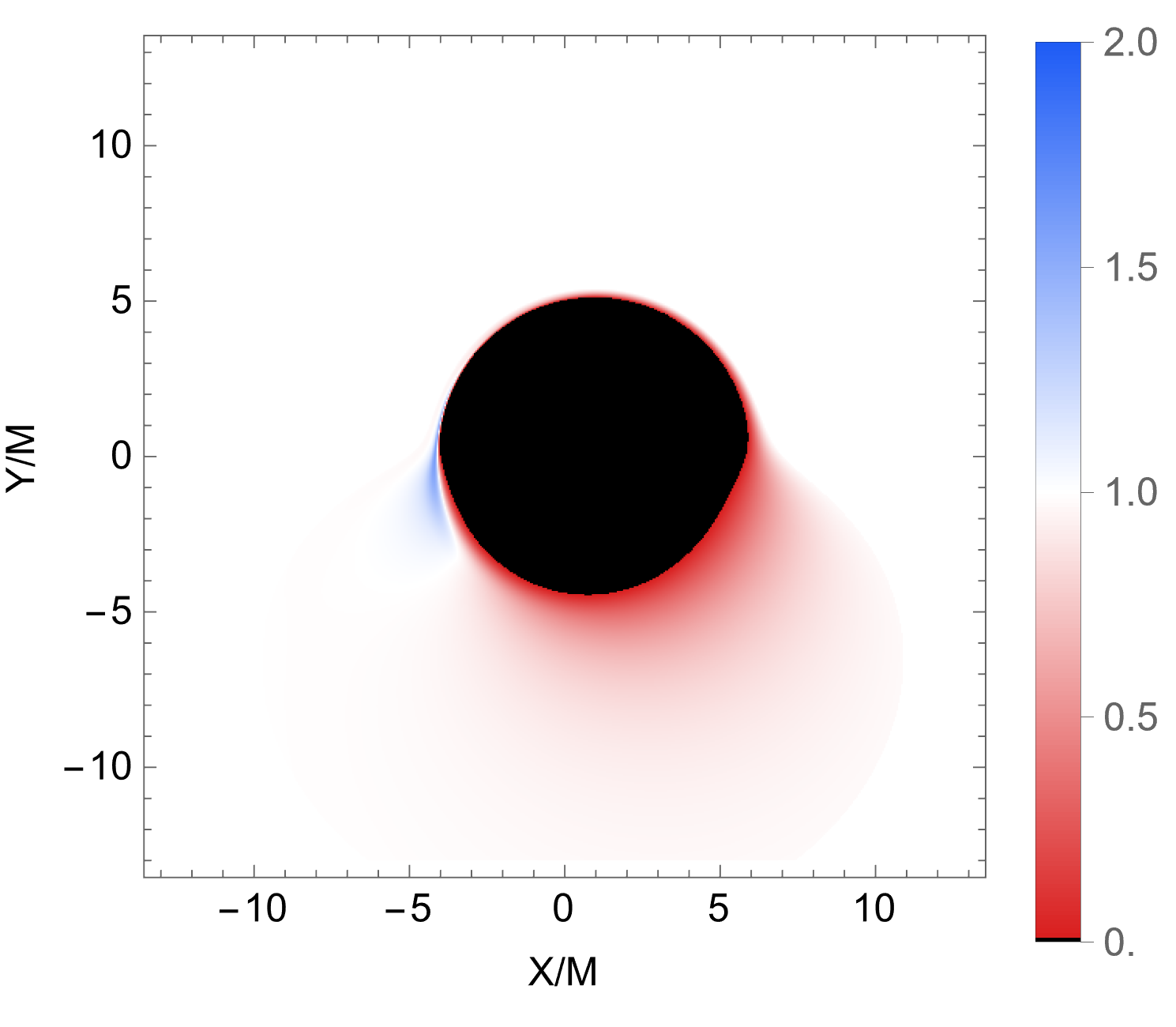}
			\caption{$\theta_{\rm obs}=80^\circ,g=0.5$}
		\end{subfigure}
		
		\vspace{0.2cm}
		
		\begin{subfigure}[t]{0.32\textwidth}
			\centering
			\includegraphics[width=\textwidth]{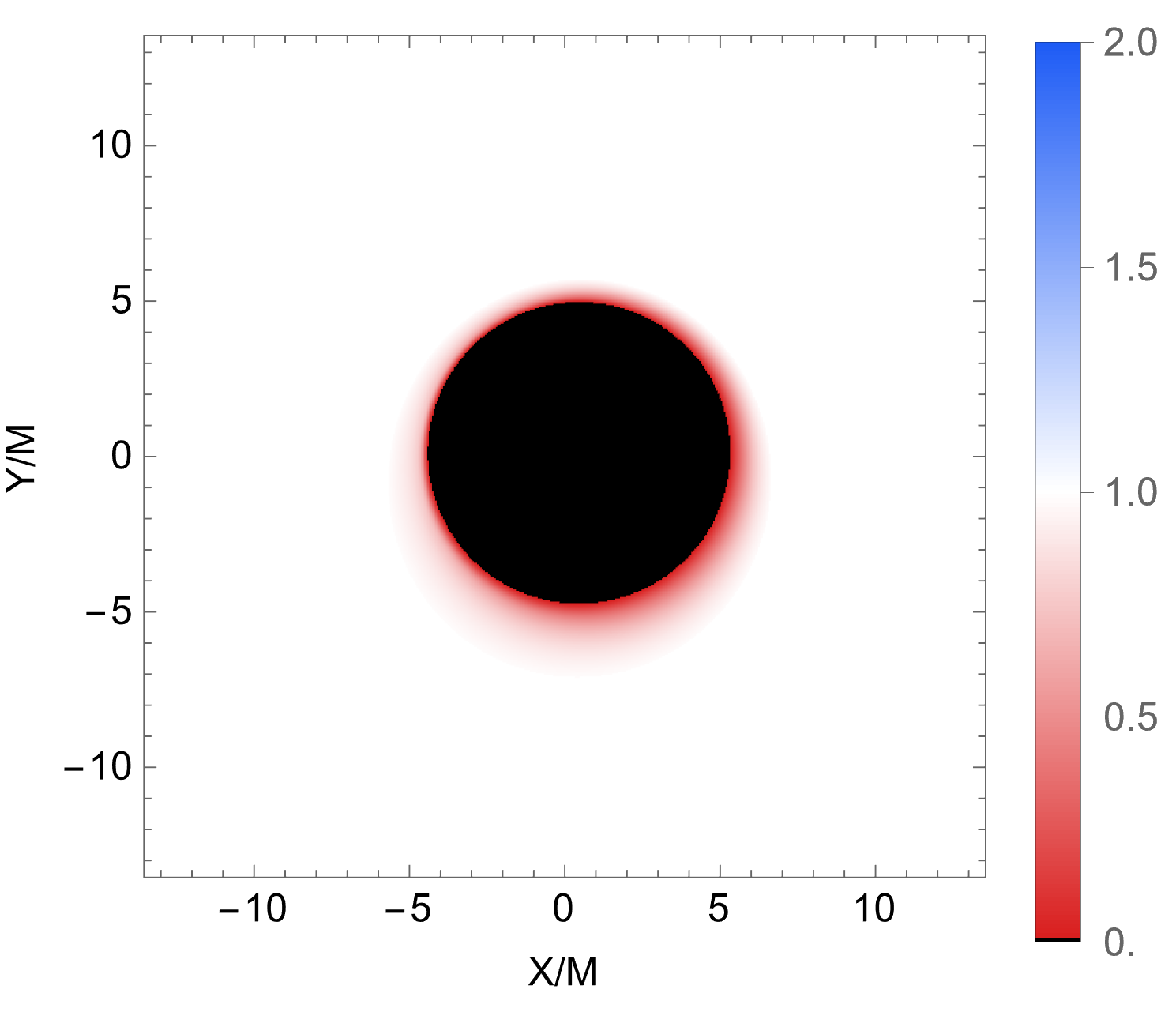}
			\caption{$\theta_{\rm obs}=30^\circ,g=1$}
		\end{subfigure}
		\begin{subfigure}[t]{0.32\textwidth}
			\centering
			\includegraphics[width=\textwidth]{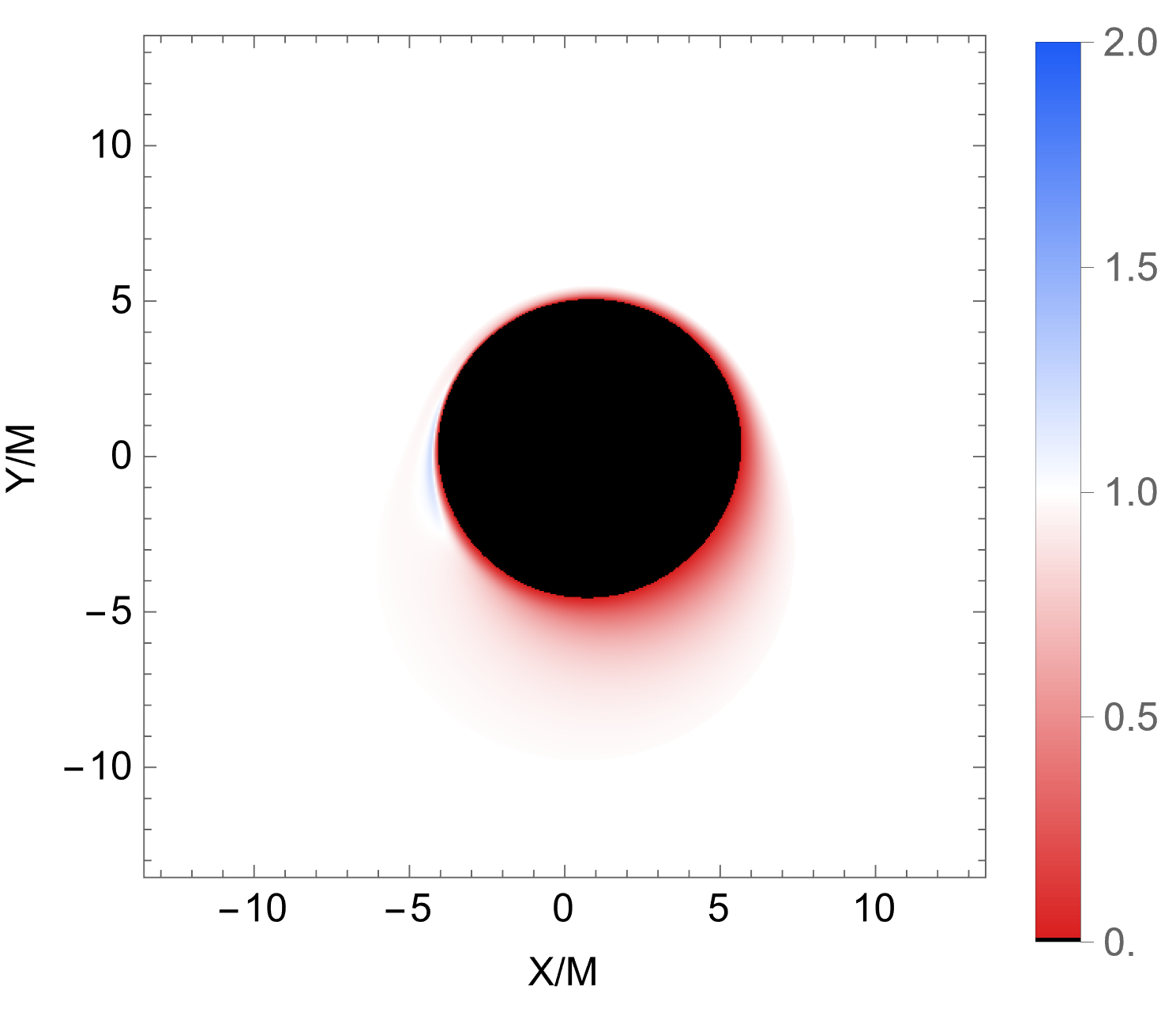}
			\caption{$\theta_{\rm obs}=60^\circ,g=1$}
		\end{subfigure}
		\begin{subfigure}[t]{0.32\textwidth}
			\centering
			\includegraphics[width=\textwidth]{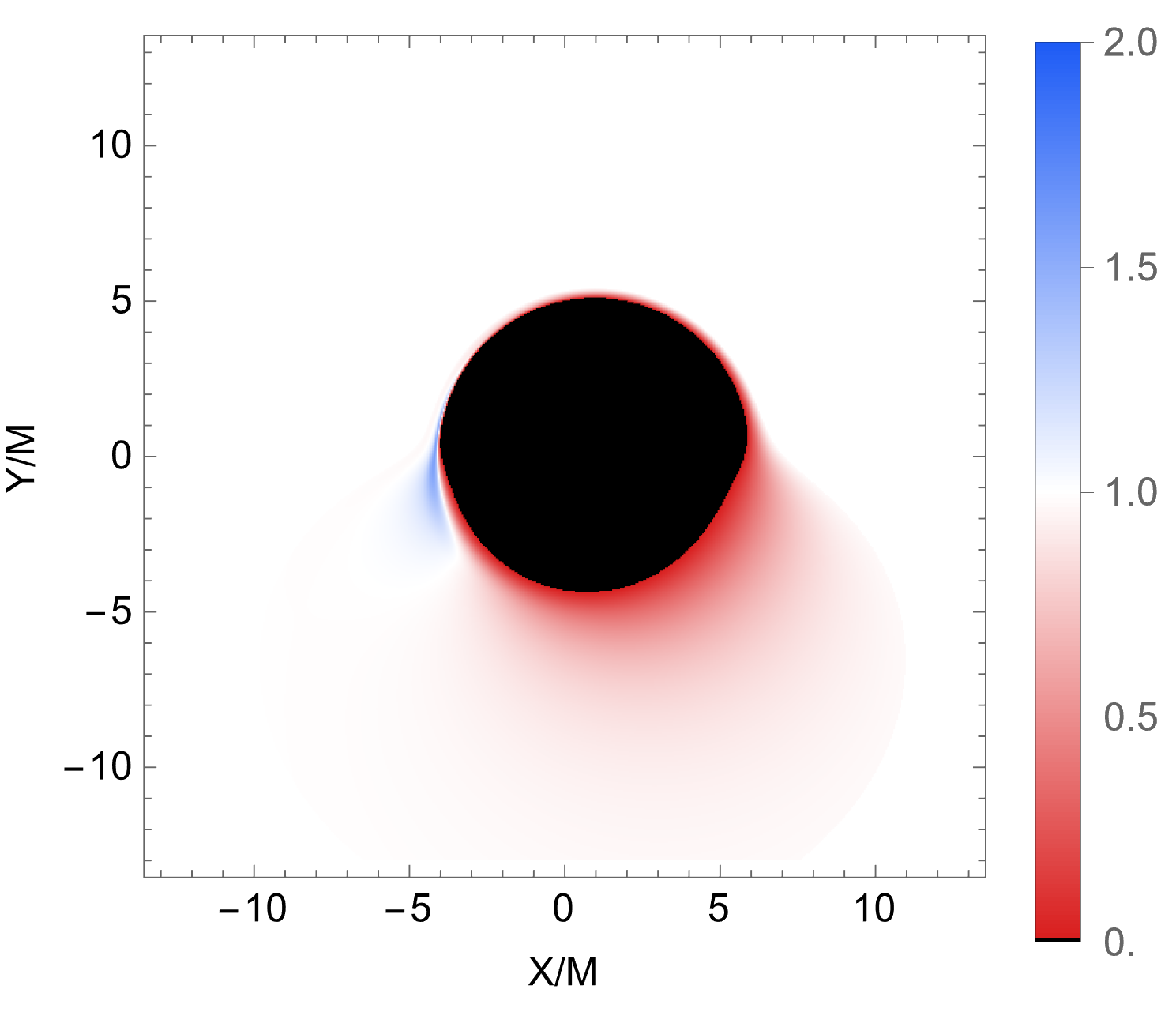}
			\caption{$\theta_{\rm obs}=80^\circ,g=1$}
		\end{subfigure}
		
		\vspace{0.2cm}
		
		\begin{subfigure}[t]{0.32\textwidth}
			\centering
			\includegraphics[width=\textwidth]{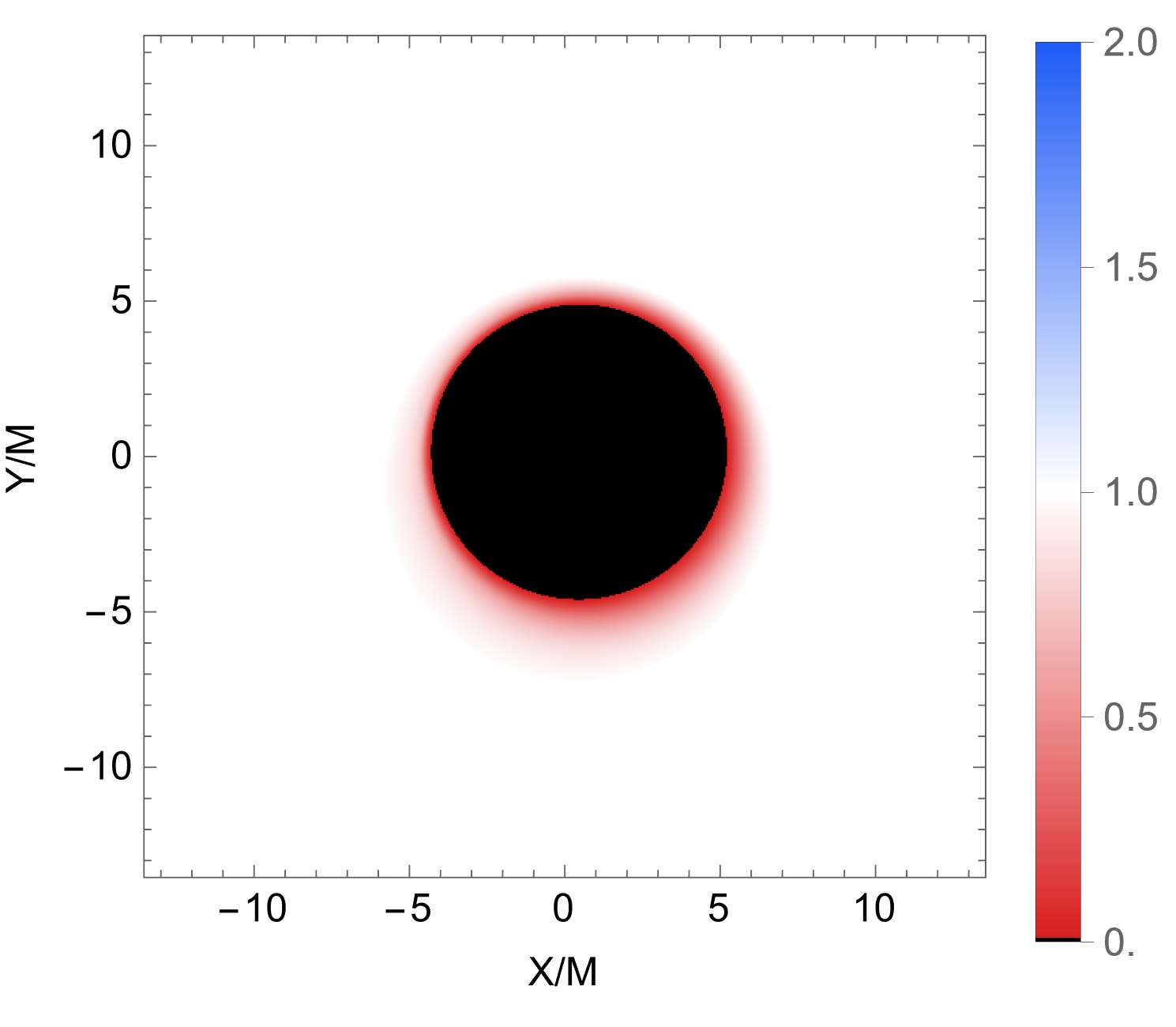}
			\caption{$\theta_{\rm obs}=30^\circ,g=1.5$}
		\end{subfigure}
		\begin{subfigure}[t]{0.32\textwidth}
			\centering
			\includegraphics[width=\textwidth]{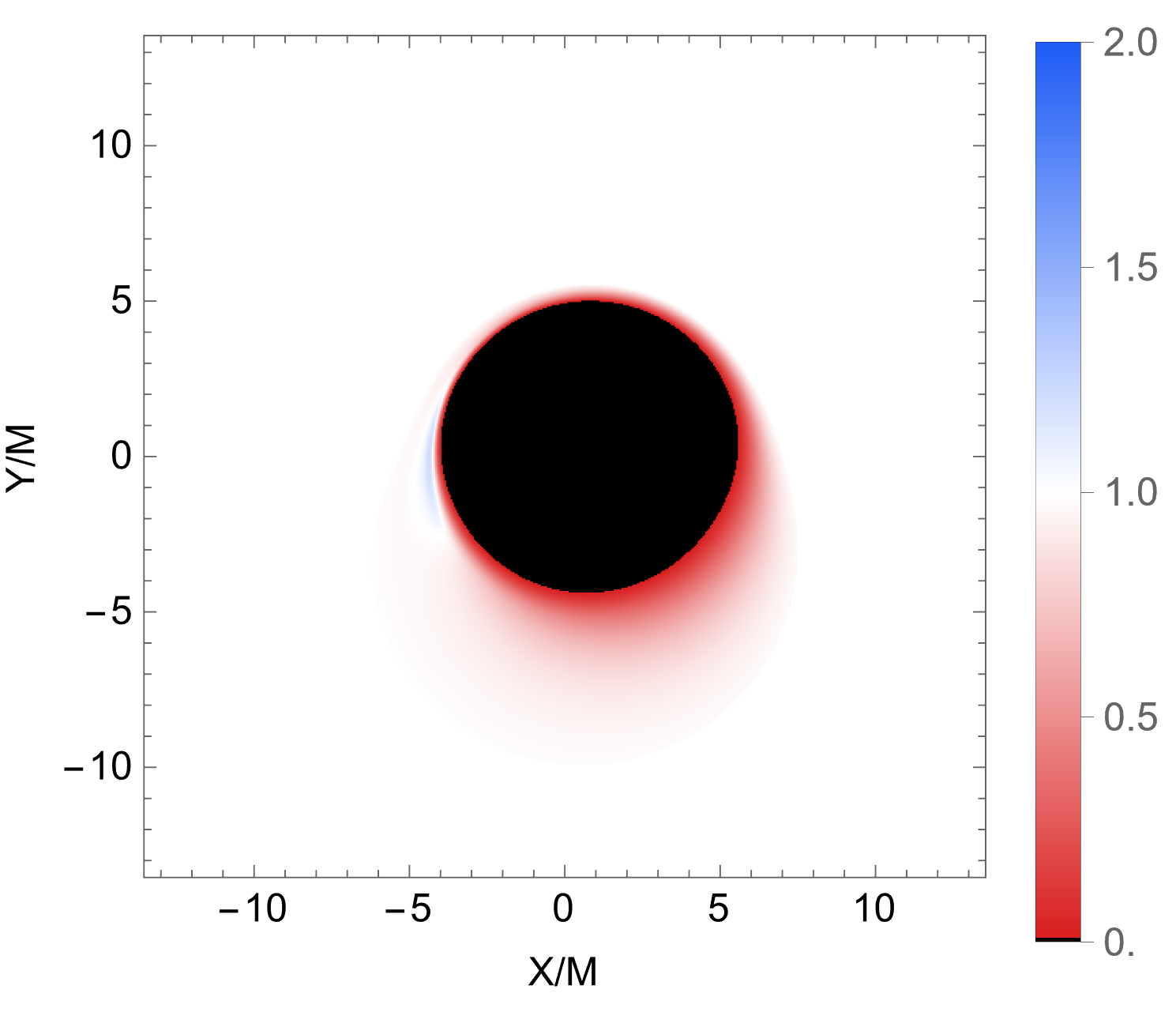}
			\caption{$\theta_{\rm obs}=60^\circ,g=1.5$}
		\end{subfigure}
		\begin{subfigure}[t]{0.32\textwidth}
			\centering
			\includegraphics[width=\textwidth]{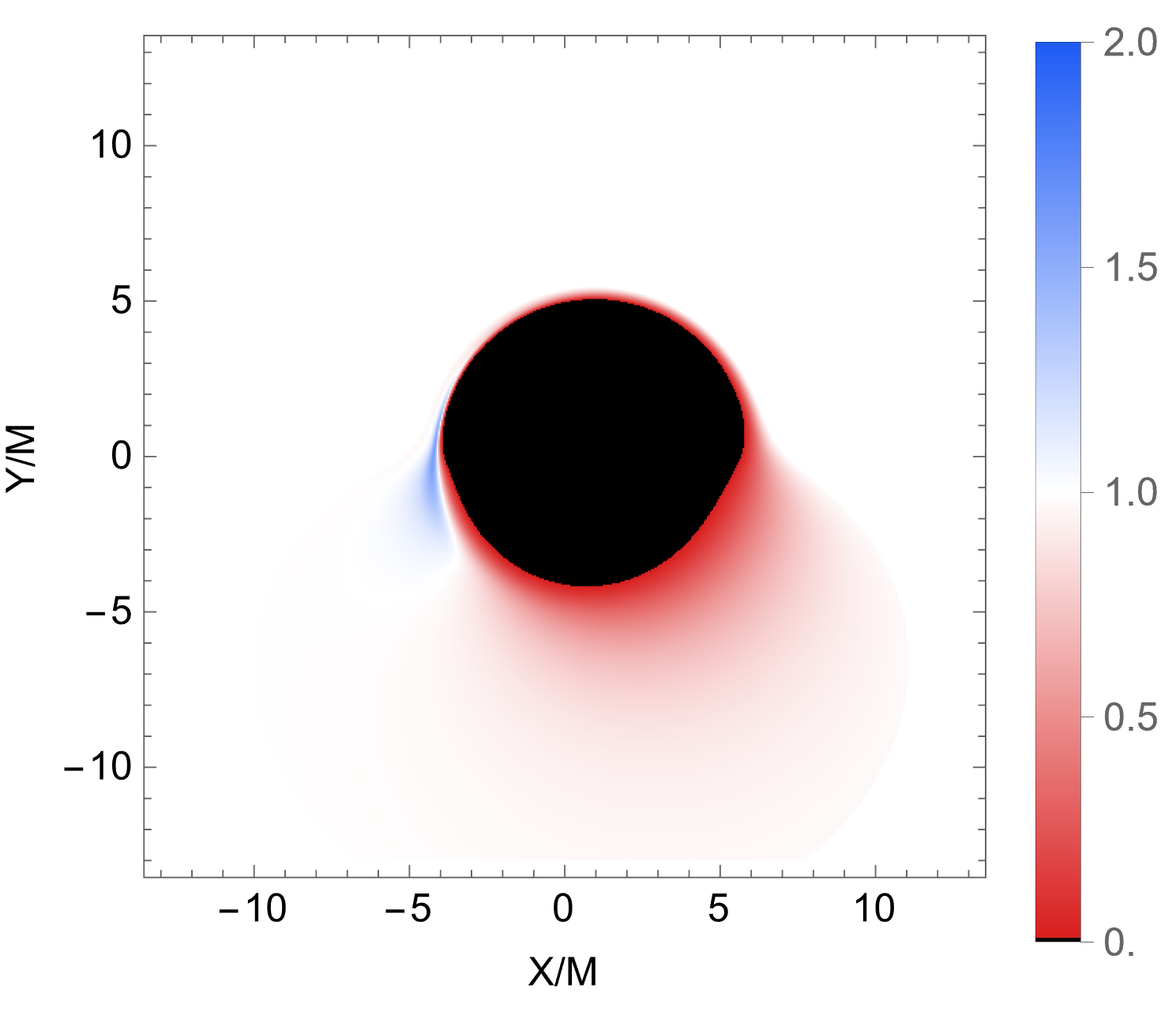}
			\caption{$\theta_{\rm obs}=80^\circ,g=1.5$}
		\end{subfigure}
		
		\caption{Distribution of the redshift factor associated with the lensed image of a rotating SV black hole with $a=0.5$ for various observational inclination angles and parameters $g$.}
		\label{fig:redshift_lensed}
	\end{figure*}

	We then turn to the redshift factor distribution associated with the lensed image, computed for the same set of parameters as that in the direct image case, and the corresponding results are shown in Fig.~\ref{fig:redshift_lensed}. 
	In each panel, the boundary of the inner shadow is surrounded by a prominent red ring, indicating strong redshift effects originating from photon emission in the plunging region inside the ISCO. 
	Overall, the lensed image is dominated by redshift, and similar to the direct image, no blueshifted region appears when the observer inclination angle is $\theta_{\rm obs}=30^\circ$. 
	However, for larger inclination angles $\theta_{\rm obs}=60^\circ$ or $80^\circ$, a small blueshifted region emerges near the left edge of the inner shadow, and this blueshift becomes more pronounced as the parameter $g$ increases. 
	Together with the results for the direct image, these findings indicate that at moderate to high inclination angles, where blueshifted regions are present, the regularization parameter enhances the blueshift effect in the observed black hole images.
	
\begin{figure*}[tb]
		\centering
		\begin{subfigure}[t]{0.32\textwidth}
			\centering
			\includegraphics[width=\textwidth]{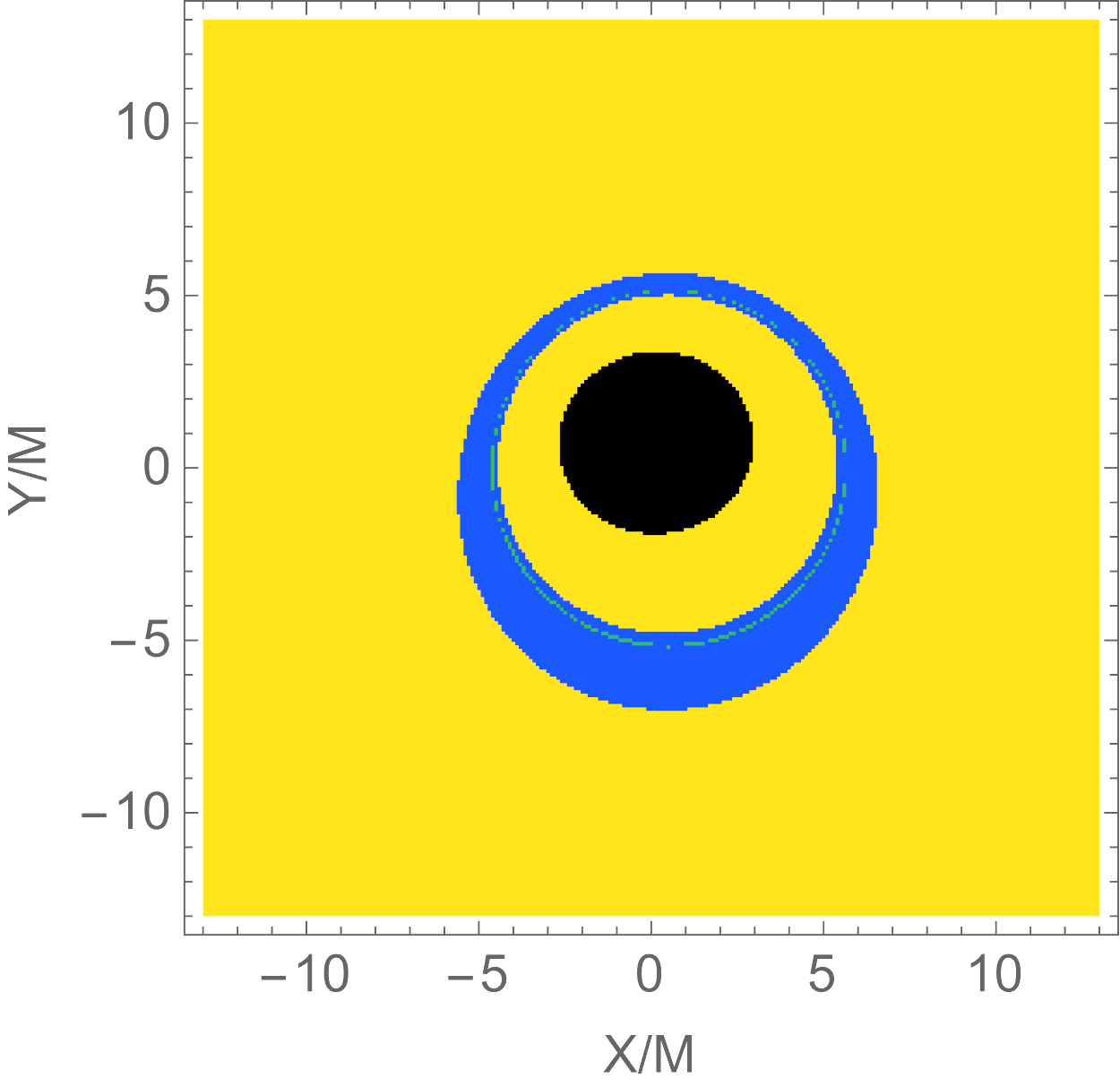}
			\caption{$\theta_{\rm obs}=30^\circ,g=0.5$}
		\end{subfigure}
		\begin{subfigure}[t]{0.32\textwidth}
			\centering
			\includegraphics[width=\textwidth]{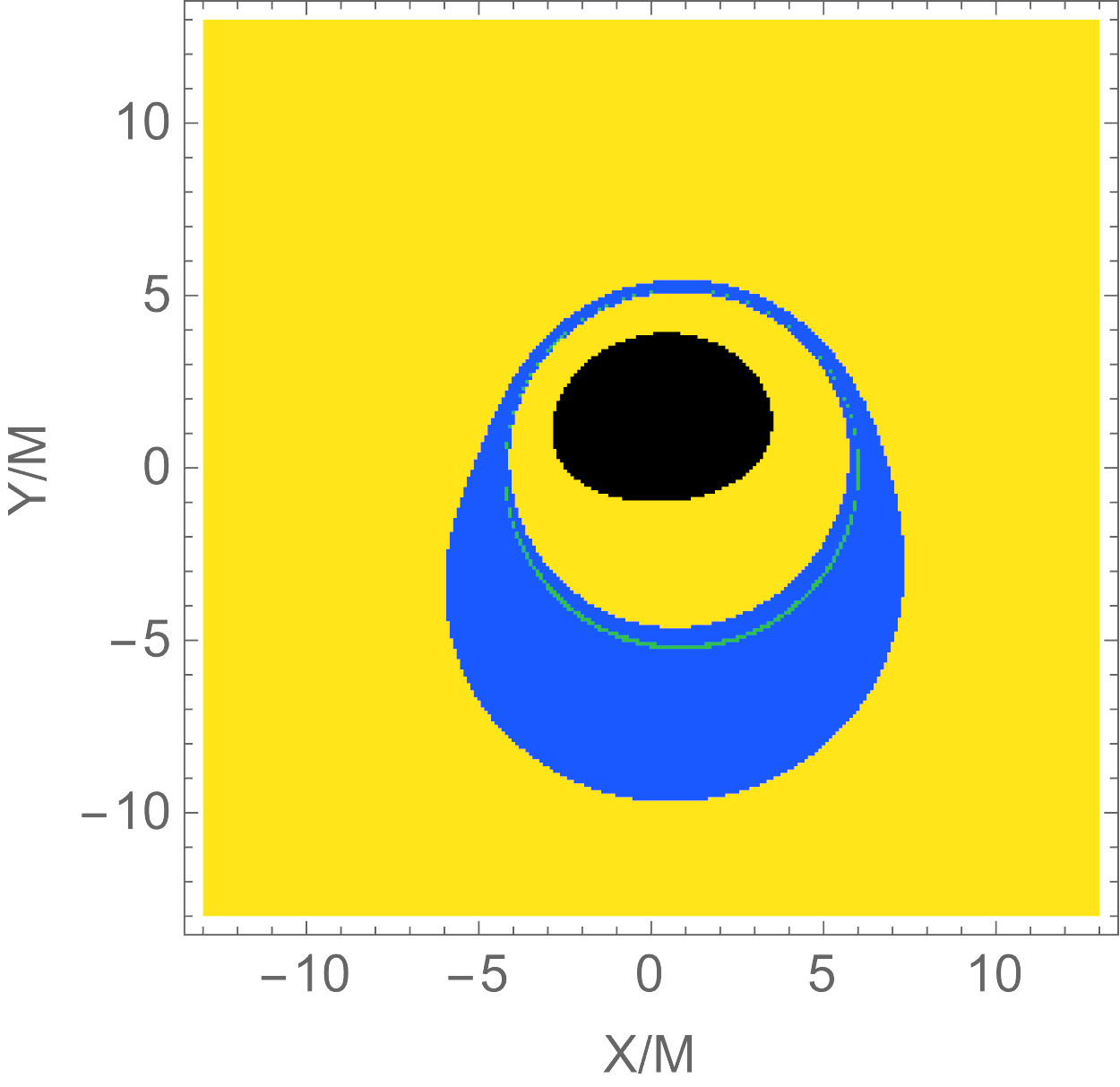}
			\caption{$\theta_{\rm obs}=60^\circ,g=0.5$}
		\end{subfigure}
		\begin{subfigure}[t]{0.32\textwidth}
			\centering
			\includegraphics[width=\textwidth]{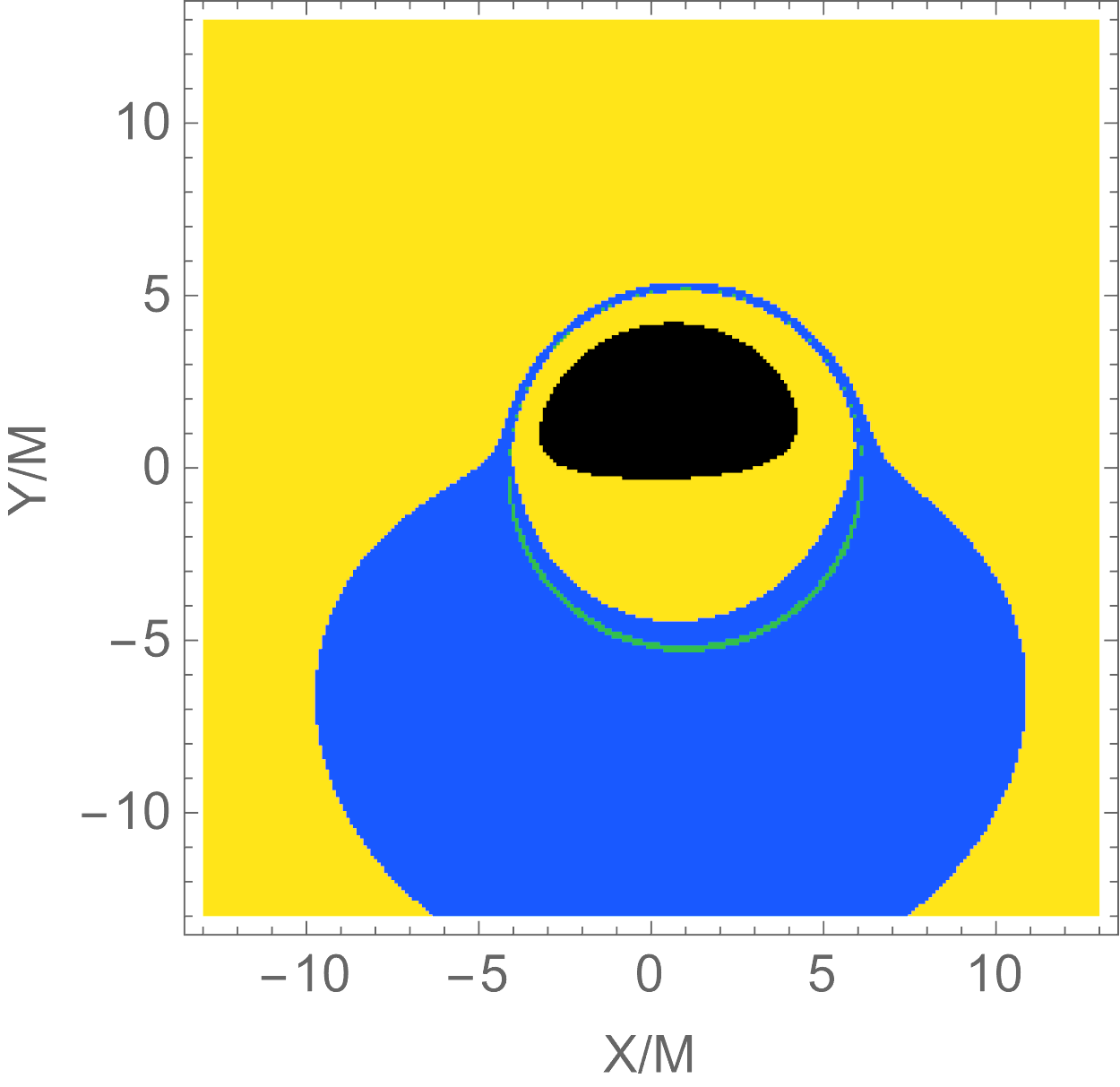}
			\caption{$\theta_{\rm obs}=80^\circ,g=0.5$}
		\end{subfigure}
		
		\vspace{0.2cm}
		
		\begin{subfigure}[t]{0.32\textwidth}
			\centering
			\includegraphics[width=\textwidth]{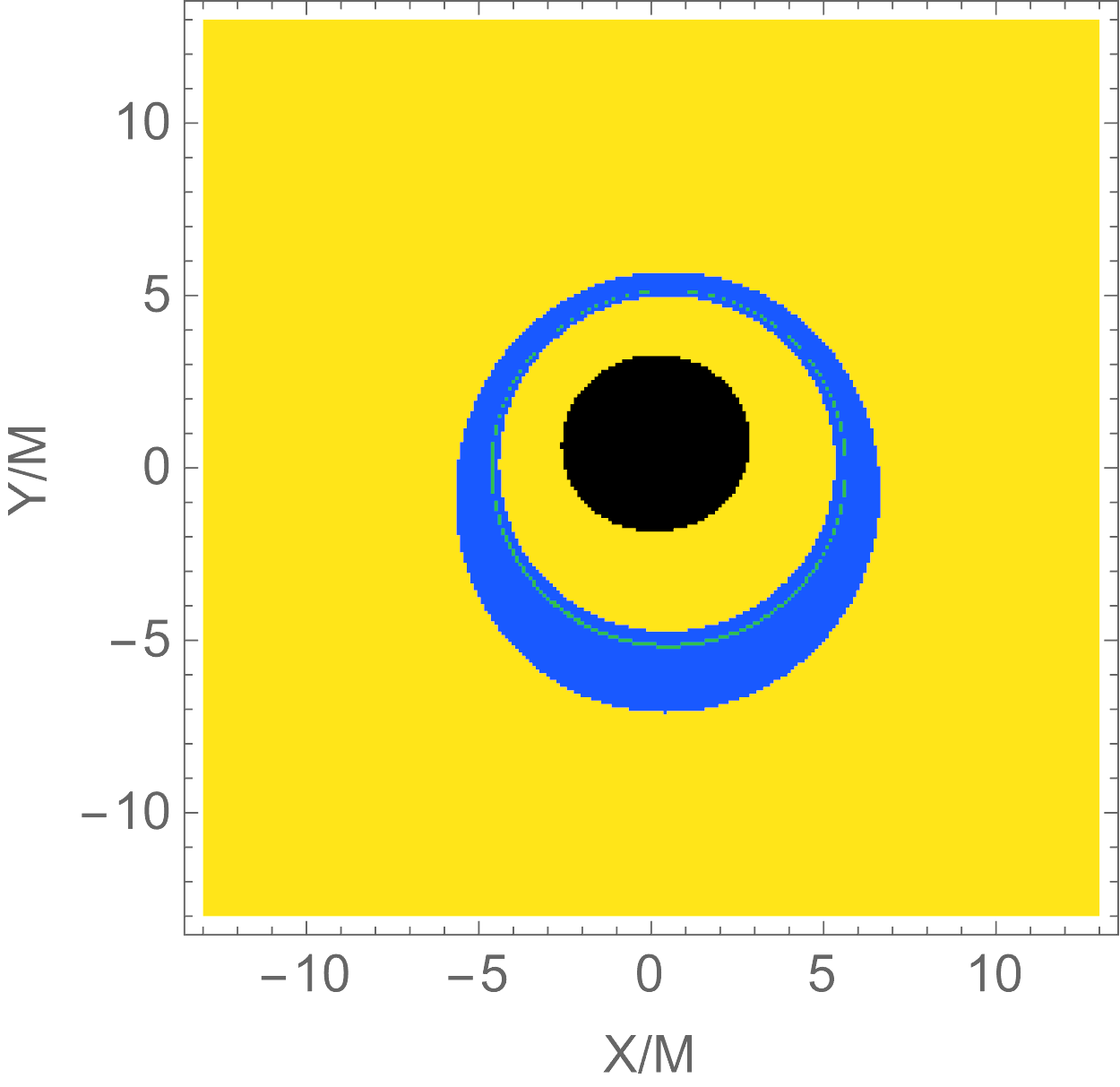}
			\caption{$\theta_{\rm obs}=30^\circ,g=1$}
		\end{subfigure}
		\begin{subfigure}[t]{0.32\textwidth}
			\centering
			\includegraphics[width=\textwidth]{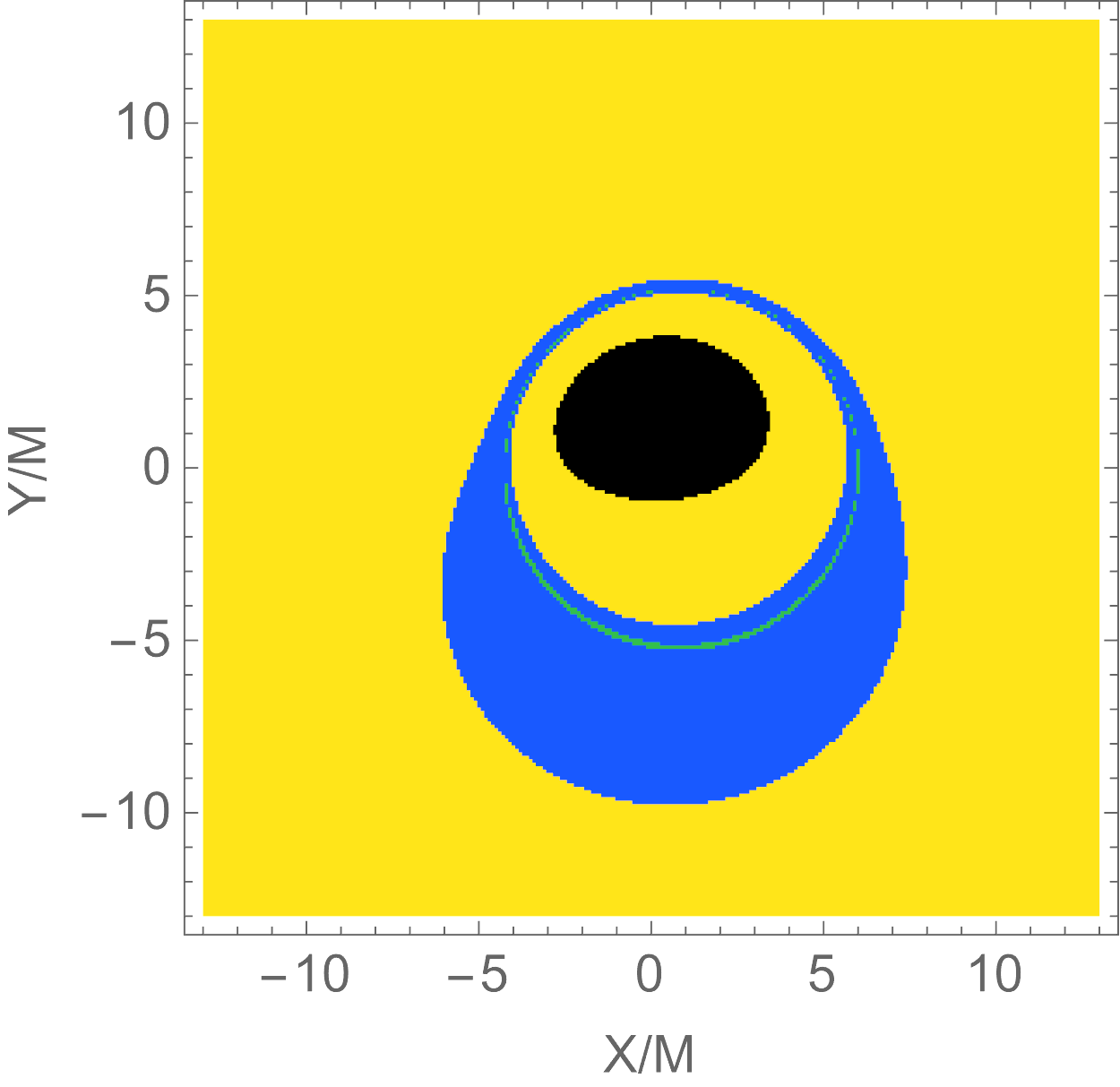}
			\caption{$\theta_{\rm obs}=60^\circ,g=1$}
		\end{subfigure}
		\begin{subfigure}[t]{0.32\textwidth}
			\centering
			\includegraphics[width=\textwidth]{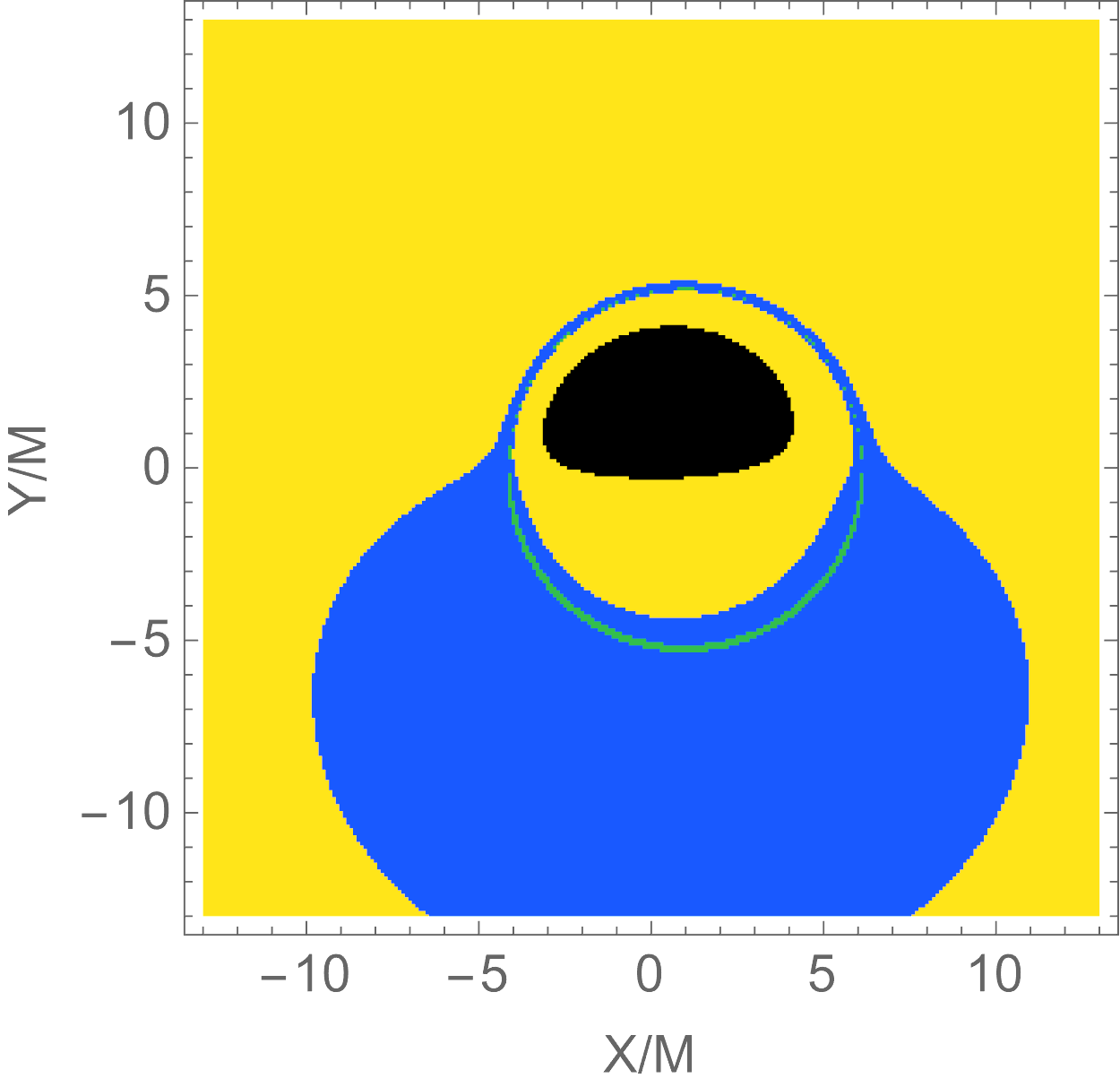}
			\caption{$\theta_{\rm obs}=80^\circ,g=1$}
		\end{subfigure}
		
		\vspace{0.2cm}
		
		\begin{subfigure}[t]{0.32\textwidth}
			\centering
			\includegraphics[width=\textwidth]{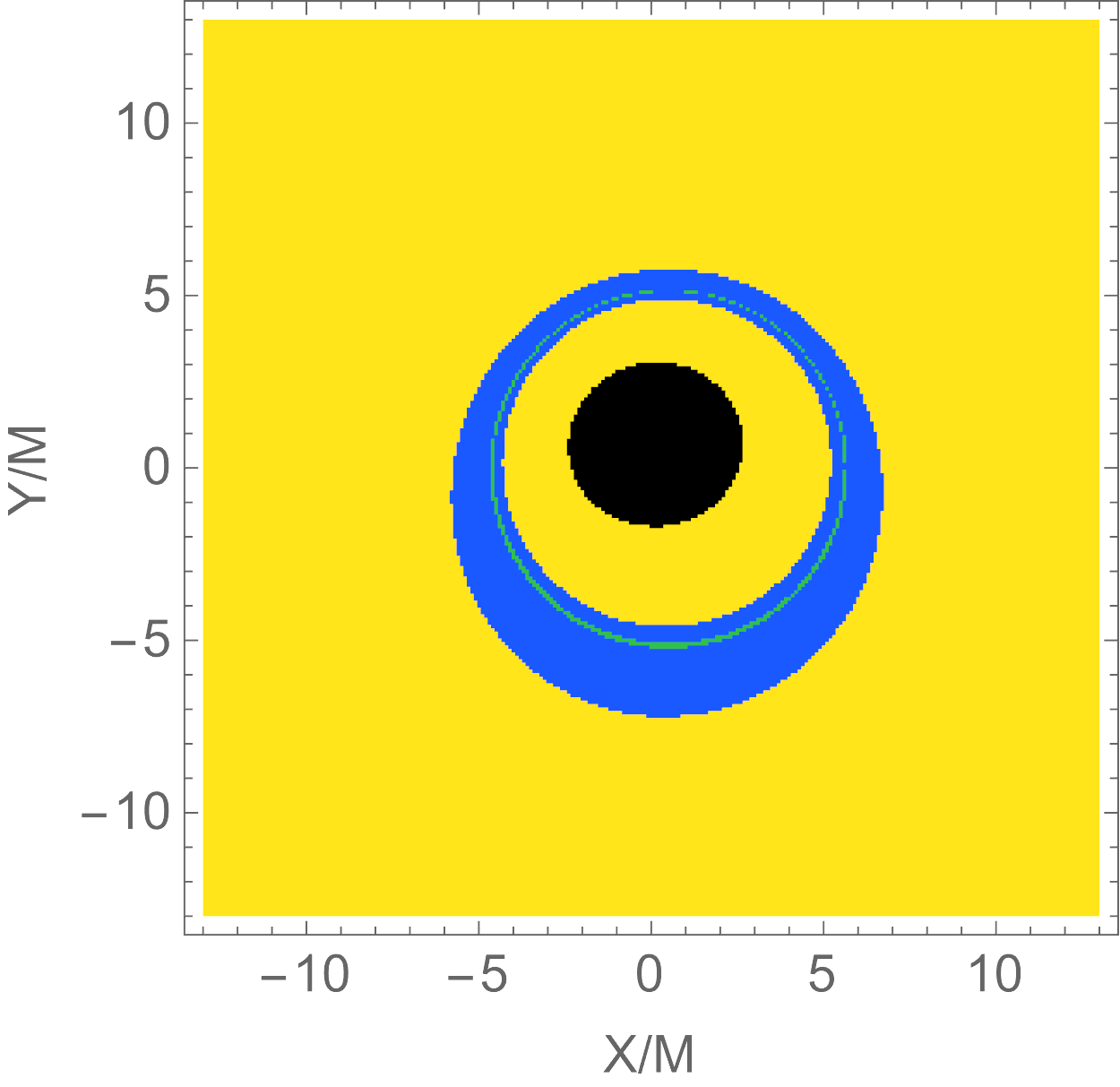}
			\caption{$\theta_{\rm obs}=30^\circ,g=1.5$}
		\end{subfigure}
		\begin{subfigure}[t]{0.32\textwidth}
			\centering
			\includegraphics[width=\textwidth]{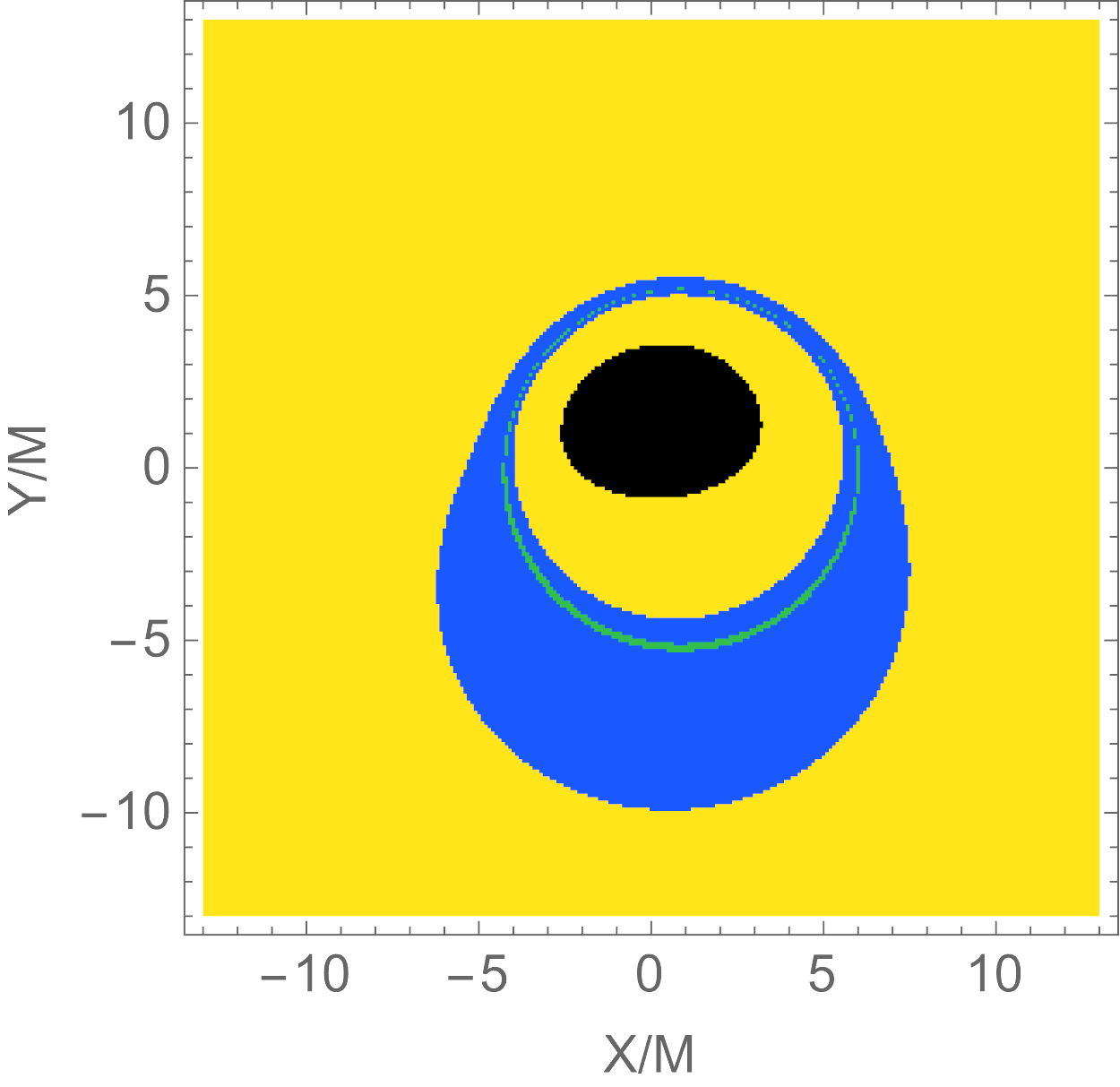}
			\caption{$\theta_{\rm obs}=60^\circ,g=1.5$}
		\end{subfigure}
		\begin{subfigure}[t]{0.32\textwidth}
			\centering
			\includegraphics[width=\textwidth]{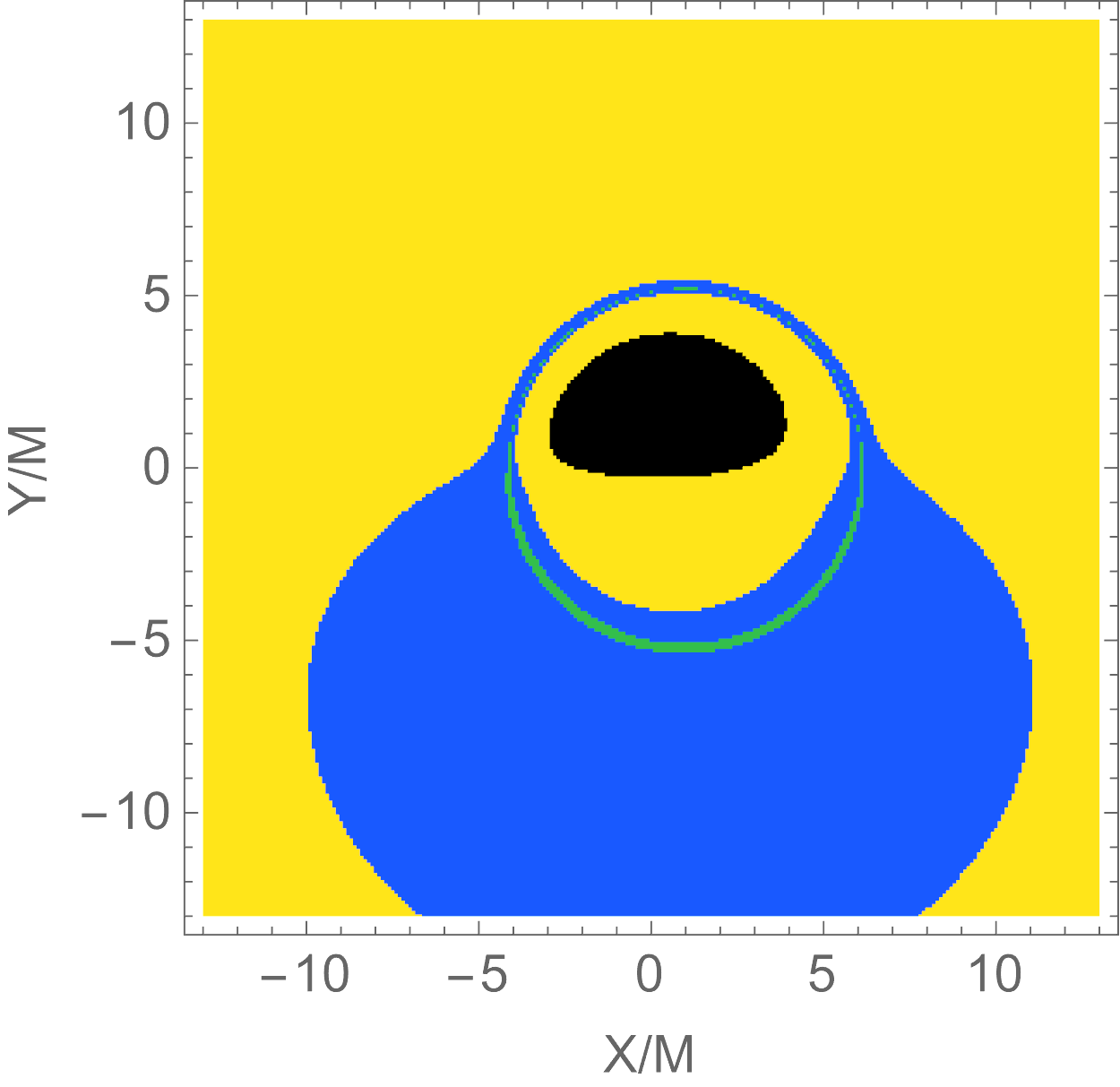}
			\caption{$\theta_{\rm obs}=80^\circ,g=1.5$}
		\end{subfigure}
		
		\caption{Distribution of the observed flux for a rotating SV black hole with $a=0.5$ for various observational inclination angles and parameters $g$.}
		\label{fig:obs_flux}
	\end{figure*}

	To gain a clearer understanding of the differences between the direct image, the lensed image, and the higher order image, we plot the observed flux for the same set of parameters as considered above, which is shown in Fig.~\ref{fig:obs_flux}. 
	In these subfigures, the thin green bands correspond to the higher order images, the blue regions represent the lensed images, and the yellow regions denote the direct images, which cover most of the observer’s screen, while the central black area indicates the inner shadow of the black hole. 
	When the observer inclination angle is $\theta_{\rm obs}=60^\circ$ or $80^\circ$, the inner shadow develops a characteristic hat-like shape, with this feature becoming more pronounced at $\theta_{\rm obs}=80^\circ$, and the increase of the inclination angle also leads to a significant enlargement of the lensed-image region. 
	Moreover, as the parameter $g$ increases, the inner shadow region shrinks while the lensed-image region expands; interestingly, the higher order image also becomes broader. 
	
	\section{Conclusion}
	\label{sec:conclusion}
	\par

In this work, we carry out a study the electromagnetic properties  of the rotating SV black hole with an equatorial thin accretion disk, focusing on the radiative properties and optical appearance of the black hole system. 
By combining analytical calculations and extensive numerical simulations, we systematically explore how the spin parameter $a$ and the regularization parameter $g$ affect both the emission of the accretion disk and the observed images.

For test particles moving on circular orbits in the equatorial plane of the rotating SV black hole, the ISCO radius is obtained numerically and it decreases as the regularization parameter $g$ increases. Interestingly, although the parameter $g$ affects the radial profiles of the specific energy $E$,  angular momentum $L$, angular velocity $\Omega$, and the ISCO radius, it is found that the values of $E$, $L$, and $\Omega$ evaluated at the ISCO are independent of $g$. As a direct consequence, 
the radiative efficiency of the rotating SV black hole is the same as that of the corresponding Kerr black hole.

Although the radiative efficiency of the rotating SV black hole is not affected by the regularization parameter $g$, other radiative properties of a thin accretion disk around the rotating SV black hole depend on the parameter $g$. The dimensionless radiative flux and effective temperature of the accretion disk are computed numerically and presented for different values of the parameters $g$ and $a$. It is found that, for SV black hole with fixed $g$, the influence of parameter $a$ on these local quantities is qualitatively the same as that of the Kerr case, and both quantities are  enhanced as $a$ increases. In contrast, the regularization parameter $g$ suppresses the peak values of the radiative flux and effective temperature. Using the observationally motivated parameters for Sgr~A$^{*}$ and M87$^{*}$, we present two concrete examples to show the influence of $g$ on the peak values of the radiative flux and effective temperature. The peak value of the radiative flux decreases about 4\% and 6\% for $g=1.86,~a=0.5$ and $g=1.5,~a=0.8$, respectively. We can also observe that the suppression effect of the regularization parameter $g$ is more obvious for higher spin black holes.
Besides these local quantities, we also incorporate the redshift factor for a distant observer with zero inclination and compute the spectral luminosity for different parameters.
Our results indicate that the spin parameter $a$ significantly enhances the spectral luminosity, whereas the effect of $g$ is comparatively much weaker.
In particular, for $a=0.5$, the spectral luminosity of the Kerr black hole is approximately $1.01$ times that of the rotating SV black hole with $g=1.86$.

By employing a backward ray-tracing technique, we numerically simulate the optical appearance of rotating SV black holes for different inclination angles and model parameters.
We also compare the intensity profiles along the $x$- and $y$-axes on the observer's screen between a Kerr black hole with $a=0.5$ and a rotating SV black hole with $a=0.5$ and $g=1$.
The increase of $g$ leads to an overall suppression of the observed intensity and a noticeable reduction of the inner shadow size.

In addition, we plot the distribution of the redshift factor for both direct and lensed images. The regularization parameter $g$ slightly enhances the blueshifted regions in both images, and this enhancement is more obvious for observers at large inclination angles (e.g. $60^\circ$ or $80^\circ$).  Furthermore, the direct, lensed and higher-order images of the observed flux distribution are also considered. Although the inclination angle affects this distribution significantly as expected, it is found that the increase of $g$ leads to a noticeable increase in the width of the photon ring region. This implies more refined structures may be more sensitive to the regularization parameter $g$. 
Thus, in order to test the possible deviation from the Kerr hypothesis for the astrophysical black holes, a more advantageous observational object is an astrophysical black hole with higher spin and larger inclination angle, and the observations of its higher-order structures are also crucial.

The results presented in \cite{KumarWalia:2022aop} highlight a degeneracy between rotating SV black holes and Kerr black holes at the level of shadow size.
Our analysis demonstrates that accretion-disk-related observables, including radiative fluxes, spectral luminosities, redshift and observed flux distributions, provide complementary information to break this degeneracy.
These findings offer valuable theoretical guidance for future high-resolution observations and may help to distinguish rotating SV black holes from Kerr black holes, thereby shedding light on the possible existence of regular black holes and deepening our understanding of singularity theorems in strong-gravity regimes.

There are several directions for future research. In this work, we focus on the SV black hole with a thin accretion disk, and one may further consider the SV black hole surrounded by dark matter halo or thick disk models \cite{Chen:2024nua,Li:2025ver,Li:2025knj,Aslam:2025hgl,Wang:2025qpv,Wang:2025gbj,Zeng:2025kyv,Yang:2025whw,Hou:2023bep,Battista:2023iyu,Wang:2025fmz} and investigate the influence of the regularization parameter on related observables. 
Another interesting topic is to consider the polarization images and patterns of the accretion flow around the SV black hole, especially the near horizon one, which can be used to probe the black hole geometry \cite{Zhang:2021hit,Chen:2022scf,Hou:2024qqo,Chen:2024jkm,Zhang:2023bzv,Hou:2022eev,Liu:2022ruc,Qin:2023nog}.
These further research may shed more light on our understanding of distinguishing the rotating SV black hole from the corresponding Kerr black hole.

	\clearpage
	\bibliography{refRSV1}  
	
\end{document}